\newif\ifdraft
\newif\ifprintpagenumbers 
\newif\ifhyperdoc   
\newif\ifhyperdebug   
\newif\iftimes
\newif\ifdistribution  
\newif\ifmynote
\newif\ifconfsub   
\newif\ifmywarnings 
\newif\ifmyidwarnings 
\newif\ifmyidundefined 
\newif\ifNotEOF   
\newif\ifprintmyids 
\newif\ifUseHardcopyIds 
\newif\iftwocolumn   
\newif\ifWideMargins      
\newif\ifgenstyle    
\newif\ifspringer    
\newif\ifkcp         
\newif\ifkcpomit     
\newif\ifkcpinclude  
\def\kcpbreak{\ifkcp\break\else\ \fi} 
\newif\ifmydoublespaceflag 
\newif\ifCoRR        

\mywarningsfalse    
\myidwarningsfalse 
\myidundefinedfalse 

\def\confcode{0}  
\ifnum\confcode<1 \confsubfalse\else\confsubtrue\fi
\draftfalse
\printmyidstrue
\printpagenumbersfalse
\hyperdocfalse
\UseHardcopyIdstrue
\timesfalse
\distributiontrue
\mynotetrue
\springerfalse
\kcpfalse
\iffalse\kcpomitfalse\kcpincludetrue\else\kcpomittrue\kcpincludefalse\fi
\mydoublespaceflagfalse
\ifdraft
   \printpagenumberstrue
\fi
\ifdistribution
   \printpagenumberstrue
\fi
\twocolumnfalse

\CoRRtrue

\def\styleindex{0} %
\ifspringer \def\styleindex{1} \fi
\ifkcp \def\styleindex{2} \fi
\ifnum\styleindex=0 \genstyletrue \else \genstylefalse \fi
\WideMarginstrue

\newbox\tempbox
\newbox\tempboxa
\newdimen\tempdimen
\newcount\tempcount
\newtoks\temptoks
\newtoks\temptoksa
\newtoks\temptoksb
\newtoks\temptoksc

\ifkcp
  \documentclass{book}
  \usepackage{amssymb}
  \usepackage[leqno,fleqn]{amsmath}
  \usepackage{theorem}
  \usepackage{latexsym,kcp}
  \usepackage{makeidx,named}
  \makeindex
  \collection

  \def\Number{\refstepcounter{equation}\begin{list}{\rm(\theequation)\hfill}%
  {\setlength{\labelwidth}{6mm}%
  \setlength{\leftmargin}{8mm}%
  \setlength{\labelsep}{2mm}}
  \item $\displaystyle}

\fi
\ifgenstyle
\documentclass{article}
\fi

\usepackage{url}
\usepackage{keyval}

\ifkcp
  \usepackage{amsmath}
\fi

\ifgenstyle
  \usepackage[reqno,centertags]{amsmath}

  \usepackage[amsmath,thmmarks]{ntheorem}

  \usepackage{setspace} \ifmydoublespaceflag  \fi
\fi

\usepackage{exscale}

\usepackage{array}

\ifkcp
\else
  \usepackage{cite}
\fi

\usepackage{enumerate}
\usepackage{longtable}
\usepackage{version}
\iffalse
   \includeversion{mycomment}

   \let\endmycommentA=\endmycomment
   \def\endmycomment{\endmycommentA ***{\tt\bf\normalsize end of mycomment.}}
\else
   \excludeversion{mycomment}
\fi
\ifkcpomit
   \excludeversion{kcpomit}
   \includeversion{kcpinclude}
\else
   \includeversion{kcpomit}
   \excludeversion{kcpinclude}
\fi

\usepackage[dvips]{graphicx,color}
\usepackage[center,hang]{subfigure}

{\makeatletter
}

\iftimes
   \usepackage{t1enc}
   \usepackage{times,pifont}
   \usepackage{mathptm}
   \usepackage[times]{itlsym}
\else
   \usepackage{itlsym}

\fi

\ifdraft
   \usepackage{ps-timestamp}
   \pstimestampextra{MANNA JOUR. SUB.}
\fi

\ifhyperdoc
\usepackage[  
        pdftitle="An Automata-Theoretic Completeness Proof for Interval Temporal Logic",
        bookmarks=true,
        bookmarksopen=true,
        bookmarksopenlevel=1,
        bookmarksnumbered=true,
        colorlinks=true,
        pagebackref=true
    ]{hyperref}

\else

\fi

\ifhyperdebug

\fi

\ifspringer
\fi
\ifkcp
\fi
\ifgenstyle
\fi

\let\theoremStandardDef=\theorem

\def\ifundefined#1{\expandafter\ifx\csname#1\endcsname\relax}  

\newlinechar=`\^^J

\def\twodollarsigns{$$}
\let\lparen=(
\let\rparen=)

\newbox\tempbox

\newenvironment{pgm*}{
\displaymath\array{@{}l}
}{
\endarray\enddisplaymath\mbox{}}

\newenvironment{defpgm*}{
\displaymath\array{@{}l}
}{
\endarray\enddisplaymath\mbox{}}

\newenvironment{proplist*}{
\displaymath\array{@{}lrcl}
}{
\endarray\enddisplaymath\mbox{}}

\def\useslashslash{%
\def\\{\futurelet\a\starnobreak}%
\def\starnobreak{\if*\a\def\a*{\cr\noalign{\nobreak}}%
\else\def\a{\cr}\fi\a}}

\newenvironment{theorems*}{
\displaymath\array{@{}lrcl}
}{
\endarray\enddisplaymath\mbox{}}

{\list{}{\setlength{\leftmargin}{\mathindent}}\raggedright\item[]}%
{\endlist}
 
\makeatletter
\expandafter \def\csname array*\endcsname #1{%
  \let\@acol\@arrayacol \let\@classz\@arrayclassz
  \let\@classiv\@arrayclassiv \let\\\@arraycr\def\@halignto{to #1}\@tabarray}
\makeatother
\expandafter \let \csname endarray*\endcsname = \endarray
 
\chardef\other=12
\def\ttverbatim{\begingroup
  \catcode`\{=\other \catcode`\}=\other \catcode`\$=\other
  \catcode`\&=\other \catcode`\#=\other \catcode`\%=\other
  \catcode`\~=\other \catcode`\_=\other \catcode`\^=\other
  \catcode`\<=\other
  \obeyspaces\def\par{\leavevmode\endgraf}\obeylines\tt}
{\obeyspaces\gdef {\ }}
\outer\def\begintt{
  \begingroup\small $$\let\par=\endgraf \ttverbatim\parskip=0pt
  \rightskip=-5pc \ttfinish}
{\obeylines\gdef\ttfinish#1^^M#2\endtt{#1\vbox{#2}\endgroup$$\endgroup}}
 
\def\m@th{\mathsurround=0pt}
\def\Rightarrowfill{$\m@th\mathord=\mkern-6mu
  \cleaders\hbox{$\mkern-2mu\mathord=\mkern-2mu$}\hfill
  \mkern-6mu\mathord\Rightarrow$}

\def\lt{<}
\def\gt{>}

\def\lbox#1\rbox{\bgroup\tabskip=0pt
\left(\begin{array}{l}#1\end{array}\right)\egroup}
\def\makeltother{\catcode`\<=\other}
\def\makeltactive{\catcode`\<=\active}
\makeltactive  
\def<#1>{\relax\ifmmode \mathit{#1} \else $\mathit{#1}$\fi{}}

\def\ldotss{\ldots\,}   
 
\begingroup
   \catcode`@=11  
   \global\def\usemybiglandbigr{\relax
      \ifnum\@ptsize=1   
         \typeout{Using my adjusted versions of bigl and bigr for 11pt fonts}
         \font\mybigfont=cmr12 scaled 1095
         \def\bigl##1{\mathopen{\hbox{\mybigfont ##1}}}
         \def\bigr##1{\mathclose{\hbox{\mybigfont ##1}}}
      \fi}
\endgroup

\let\imp=\supset
\def\implies{\;\supset\;}
\def\Implies{\quad\supset\quad}

\def\thm{\vdash}
\def\theorem{\vdash\;}
\def\Theorem{\vdash\quad}

\let\vld=\models
\def\valid{\vld\;}
\def\Valid{\vld\quad}

\let\dvd=/

\def\defeq{\stackrel{\rm def}{=}}
\def\Defeq{\quad\stackrel{\rm def}{=}\quad}
\def\defeqv{\stackrel{\rm def}{\equiv}}
\def\Defeqv{\quad\stackrel{\rm def}{\equiv}\quad}
\def\iff{\quad \mbox{iff}\quad}
\def\citelemma#1/{#1}

\let\union=\cup
\let\intersection=\cap
 
\def\tabb{\quad\qquad}

\def\tempop#1{\mathop{{\it #1}\,}}
\def\tempopsubchk#1{\textstyle  
   \ifcat_\temptoken \mathop{\it #1\!} \else \mathop{\it #1\/}  \fi}

\def\tempord#1{\mathord{{\it #1}}}
\def\tempinner#1{\mathinner{{\it #1}}}
\def\True{\tempord{true}}
\def\False{\tempord{false}}

\let\Not=\neg
\def\And{\mathrel{\scriptstyle\wedge}}

\def\AND{\quad\And\quad}

\def\Or{\mathrel{\scriptstyle\vee}}

\def\Equiv{\quad\equiv\quad}
\def\EQUIV{\;\equiv\;}

\def\Exists#1{{\exists #1}\mathpunct{:}}

\def\First{\tempop{extend}}
\def\Keep{\tempop{keep}}

\def\Halt{\tempop{halt}}

\def\Assert{\tempop{assert}}

\def\Fin{\tempop{fin}}
\def\SFin{\tempop{sfin}}

\def\Uproj{\mathrel{\raise 0.2em\hbox{$\bigtriangledown$}}}

\def\Chop{\mathbin{;}}
\def\Chopstar{^*}

\def\Chopplus{^+}

\def\Chopomega{^\omega}

\let\Tassign=\leftarrow

\def\Skip{\tempord{skip}}
\def\Empty{\tempord{empty}}
\def\More{\tempord{more}}
\def\Inf{\tempord{inf}}
\def\Finite{\tempord{finite}}

\def\Deloc{\raise 0.17em\hbox{$\uparrow$}}   

\def\ReadOnly{\tempop{readonly}}

\def\Var{\tempop{local}}
\def\For{\futurelet\temptoken\Fora}
\def\Fora{\tempopsubchk{for}}
\def\Forever{\futurelet\temptoken\Forevera}
\def\Forevera{\tempopsubchk{forever}}
\def\ChopstarKW{\futurelet\temptoken\ChopstarKWa}
\def\ChopstarKWa{\tempopsubchk{chopstar}}
\def\ChopinfKW{\futurelet\temptoken\ChopinfKWa}
\def\ChopinfKWa{\tempopsubchk{chopinf}}
\def\While{\futurelet\temptoken\Whilea}
\def\Whilea{\tempopsubchk{while}}
\def\Repeat{\futurelet\temptoken\Repeata}
\def\Repeata{\tempopsubchk{repeat}}

\def\Until{\tempop{until}}
\def\Times{\tempinner{times}}

\let\EOD=\#
\let\NIL=-

\def\size#1{\vert #1\vert}      

\def\concat{%
   \mathbin{\raise 0.50em\hbox{$\mskip-1.5\thinmuskip\scriptstyle\frown
      \mskip-2\thinmuskip$}}}

\def\C#1{\ifhmode {\it #1\/}\else \hbox{\it #1\/}\fi}
\def\meta#1{\ifhmode {\it #1\/}\else \hbox{\it #1\/}\fi}
 
\def\intlen#1{{|#1|}}    
 
\def\newrule#1#2{}
\newrule\Assump{Assump.}
\newrule\AssumeContra{Assume-C}
\newrule\Axiom{Axiom}
\newrule\Given{Given}
\newrule\Prop{Prop.}
\newrule\Contra{Contra.}
\newrule\PTL{PTL}
\newrule\FOL{FOL}
\newrule\Quant{Quantifiers}
\newrule\Arith{Arith.}
\newrule\Lists{Lists}
\newrule\Static{Static}
\newrule\MP{MP}
\newrule\AndIntro{$\And$-intro}
\newrule\AndElim{$\And$-elim}
\newrule\OrElim{$\Or$-elim}
\newrule\ExMiddle{excl.-middle}
\newrule\ImpIntro{$\imp$-intro}
\newrule\ImpChain{$\imp$-chain}
\newrule\EqvChain{$\equiv$-chain}
\newrule\EqvElim{$\equiv$-elim}
\newrule\ForallElim{$\forall$-elim}
\newrule\ForallGen{$\forall$-gen}
\newrule\ForallImp{$\forall$-$\imp$}
\newrule\ExistsIntro{$\exists$-intro}
\newrule\ExistsImp{$\exists$-$\imp$}
\newrule\InitElim{init-elim}
\newrule\NextImp{$\Next$-$\imp$}
\newrule\NextLoop{$\Next$-loop}
\newrule\NextNe{$\Next$-$\ne$}
\newrule\BoxIntro{$\Box$-intro}
\newrule\DiamondImp{$\Diamond$-$\imp$}
\newrule\DiamondIntro{$\Diamond$-intro}
\newrule\FinImp{$\Fin$-$\imp$}
\newrule\KeepIntro{$\Keep$-intro}
\newrule\ChopAssoc{$\Chop$-assoc}
\newrule\ImpChop{$\imp$-$\Chop$}
\newrule\ChopImp{$\Chop$-$\imp$}
\newrule\EqvChop{$\equiv$-$\Chop$}
\newrule\ChopEqv{$\Chop$-$\equiv$}
\newrule\OrChop{$\Or$-$\Chop$}
\newrule\ChopOr{$\Chop$-$\Or$}
\newrule\AndChop{$\And$-$\Chop$}
\newrule\ChopAnd{$\Chop$-$\And$}
\newrule\ChopAndA{$\Chop$-$\And$-1}
\newrule\ChopAndB{$\Chop$-$\And$-2}
\newrule\NextChop{$\Next$-$\Chop$}
\newrule\EmptyChop{$\Empty$-$\Chop$}
\newrule\FirstChop{first-$\Chop$}
\newrule\StateChop{state-$\Chop$}
\newrule\FinChop{$\Fin$-$\Chop$}
\newrule\TrueChop{$\True$-$\Chop$}
\newrule\ChopDiamond{$\Chop$-$\Diamond$}
\newrule\AssertChop{$\Assert$-$\Chop$}
\newrule\ChopAssert{$\Chop$-$\Assert$}
\newrule\EmptyChop{$\Empty$-$\Chop$}
\newrule\ChopEmpty{$\Chop$-$\Empty$}
\newrule\ChopChain{$\Chop$-chain}
\newrule\FinAssert{$\Fin$-$\Assert$}
\newrule\BoxGen{$\Box$-gen}
\newrule\BoxChopImp{$\Box\Chop$-$\imp$}
\newrule\BoxChopAnd{$\Box\Chop\And$}
\newrule\WeakNextBoxChop{$\WeakNext\Box\Chop$}
\newrule\HaltTrue{$\Halt$-$\True$}
\newrule\HaltFalse{$\Halt$-$\False$}
\newrule\HaltTrueChop{$\Halt$-$\True$-$\Chop$}
\newrule\HaltFalseChop{$\Halt$-$\False$-$\Chop$}
\newrule\HaltIntro{$\Halt$-intro}
\newrule\HaltChopIntro{$\Halt$-$\Chop$-intro}
\newrule\WhileFalse{$\While$-$\False$}
\newrule\WhileTrue{$\While$-$\True$}
\newrule\WhileFin{$\While$-$\Fin$}
\newrule\WhileIntro{$\While$-intro}
\newrule\WhileElim{$\While$-elim}
\newrule\WhileInv{$\While$-inv}
\newrule\WhileBodyRep{$\While$-body-rep}
\newrule\ForTimesEmpty{$\For$-$\Times$-$\Empty$}
\newrule\ForTimesMore{$\For$-$\Times$-$\More$}
\newrule\ForTimesIntro{$\For$-$\Times$-intro}
\newrule\ForTimesElim{$\For$-$\Times$-elim}
\newrule\ForTimesBodyRep{$\For$-$\Times$-body-rep}
\newrule\FrAssignFin{$:=$-$\Fin$}
\newrule\TAssignFin{$\Tassign$-$\Fin$}
\newrule\FrChopLemma{fr-$\Chop$-lemma}
\newrule\FrChopChain{fr-$\Chop$-chain}
\newrule\FrChopAnd{fr-$\Chop$-$\And$}
\newrule\FrWhileAnd{fr-$\While$-$\And$}
\newrule\FrWhileElim{fr-$\While$-elim}
\newrule\FrWhileInv{fr-$\While$-inv}
\newrule\AndParA{$\And$-$\parallel$-1}
\newrule\ParAndA{$\parallel$-$\And$-1}
\newrule\AndParB{$\And$-$\parallel$-2}
\newrule\ParAndB{$\parallel$-$\And$-2}
\newrule\OrPar{$\Or$-$\parallel$}
\newrule\ParOr{$\parallel$-$\Or$}
\newrule\ImpPar{$\imp$-$\parallel$}
\newrule\ImpOrPar{$\imp$-$\Or$-$\parallel$}
\newrule\ParImp{$\parallel$-$\imp$}
\newrule\ParImpOr{$\parallel$-$\imp$-$\Or$}
\newrule\EqvPar{$\equiv$-$\parallel$}
\newrule\ParEqv{$\parallel$-$\equiv$}
\newrule\DiamondPar{$\Diamond$-$\parallel$}
\newrule\DiamondIntro{$\Diamond$-intro}
\newrule\DiamondParRead{$\Diamond$-$\parallel$-$\ReadOnly$}
\newrule\DiamondParKeep{$\Diamond$-$\parallel$-$\Keep$}
\newrule\StatePar{state-$\parallel$}
\newrule\ParElim{$parallel$-elim}
\newrule\NextParEmpty{$\Next$-$\parallel$-$\Empty$}
\newrule\EmptyParNext{$\Empty$-$\parallel$-$\Next$}
\newrule\NextParNext{$\Next$-$\parallel$-$\Next$}
\newrule\NextParKeep{$\Next$-$\parallel$-$\Keep$}
\newrule\ReadParRead{read-$\parallel$-read}
\newrule\NextParRead{$\Next$-$\parallel$-read}
\newrule\EmptyParRead{$\Empty$-$\parallel$-read}
 
\def\today{\number\day\space
\ifcase\month\or
January\or February\or March\or April\or May\or June\or
July\or August\or September\or October\or November\or December\fi
\space\number\year}

\makeltother

\let\theorem=\theoremStandardDef

\newenvironment{myfigure}{\begin{figure}}
   {\end{figure}\myignoretrue}
\newenvironment{mytable}{\begin{table}}
   {\end{table}\myignoretrue}

\let\D=\displaystyle
\newenvironment{myarray}{\begin{array}[t]{@{}>{\D}l@{}}}
   {\end{array}\myignoretrue}

\ifspringer
   \spnewtheorem{mytheorem}[theorem]{Theorem}{\bfseries}{\itshape}
   \spnewtheorem{myclaim}[theorem]{Claim}{\bfseries}{\itshape}
   \spnewtheorem{mycorollary}[theorem]{Corollary}{\bfseries}{\itshape}
   \spnewtheorem{mylemma}[theorem]{Lemma}{\bfseries}{\itshape}
   \spnewtheorem{mydefin}[theorem]{Definition}{\bfseries}{\itshape}
   \spnewtheorem*{myremark}{Remark}{\itshape}{\rmfamily}
   \spnewtheorem*{myproof}{Proof}{\itshape}{\rmfamily}
\fi

\ifkcp

\fi

\ifgenstyle
   \gdef\mysquareforqed{{\hbox{\rlap{$\sqcap$}$\sqcup$}}}
   \newtheorem{mytheorem}{Theorem}

   \newtheorem{mycorollary}[mytheorem]{Corollary}
   \newtheorem{mylemma}[mytheorem]{Lemma}
   \newtheorem{mydefin}[mytheorem]{Definition}
   \newtheorem{myremark}[mytheorem]{Remark}
   {\theoremstyle{nonumberplain}
   \theoremheaderfont{\itshape}
   \theorembodyfont{\rmfamily}
   \theoremsymbol{\ensuremath{_\mysquareforqed}}
   \qedsymbol{\ensuremath{_\mysquareforqed}}
   \newtheorem{myproof}{\scshape Proof}
   }
\fi
\let\ProofStep=\paragraph

\newdimen\dispskipdimen
\dispskipdimen=\parindent

\newenvironment{widedisplaymath}{%
\begingroup\displayindent=0pt\twodollarsigns}{
\twodollarsigns\endgroup\myignoretrue}

\newif\ifWithinProveIt  

\newbox\proveitformulabox
\setbox\proveitformulabox=
    \hbox to 4.5in{}

{\makeatletter
\global\let\myignoretrue=\@ignoretrue
}

\newcolumntype{C}{>{$}c<{$}}
\newcolumntype{L}{>{$}l<{$}}
\newcolumntype{R}{>{$}r<{$}}

\def\Exists#1{\exists #1\mathpunct{.}}

\def\SDiamond{\mathop{{\Diamond}^{+}}}
\def\SBox{\mathop{{\Box}^{+}}}

\def\Atoms{\ifmmode \mathit{Atoms}\else $\mathit{Atoms}$\fi}
\def\shift{\mathbin{\uparrow}}
\def\DD#1{\langle #1\rangle}
\def\BB#1{[#1]}
\def\eh{\ifmmode \mathord{\mathit{eh}}\else $\mathord{\mathit{eh}}$\fi}
\def\En{\ifmmode \mathit{En} \else $\mathit{En}$\fi}
\def\ETL{\ifmmode \mathrm{ETL} \else $\mathrm{ETL}$\fi}
\def\FE{\ifmmode \mathrm{FE} \else $\mathrm{FE}$\fi}
\def\FEV{\ifmmode \mathrm{FE}_V \else $\mathrm{FE}_V$\fi}
\def\FL{\ifmmode \mathrm{FL} \else $\mathrm{FL}$\fi}
\def\FLV{\ifmmode \mathrm{FL}_V \else $\mathrm{FL}_V$\fi}

\def\H{{\mathcal H}}
\def\ITL{\ifmmode \mathrm{ITL}\else $\mathrm{ITL}$\fi}
\def\LangI{\mathord{\mathcal{L}_I}}
\def\LangOmegaI{\mathord{\mathcal{L}_I^\omega}}

\def\LangPlusI{\mathord{\mathcal{L}_I^+}}
\def\NL{\ifmmode \mathrm{NL}\else $\mathrm{NL}$\fi}
\def\NLone{\ifmmode \NL^{\!1}\else $\NL^{\!1}$\fi}
\def\NLoneV{\ifmmode \NL^{\!1}_V\else $\NL^{\!1}_V$\fi}
\def\PDL{\ifmmode \mathrm{PDL}\else $\mathrm{PDL}$\fi}
\def\PITL{\ifmmode \mathrm{PITL} \else $\mathrm{PITL}$\fi}
\def\PITLV{\ifmmode \mathrm{PITL}_V \else $\mathrm{PITL}_V$\fi}
\def\PrevL{\ifmmode \mathrm{PrevL}\else $\mathrm{PrevL}$\fi}
\def\PrevLone{\ifmmode \PrevL^{\!1}\else $\PrevL^{\!1}$\fi}
\def\PrevLoneV{\ifmmode \PrevL^{\!1}_V\else $\PrevL^{\!1}_V$\fi}
\def\PROP{\ifmmode \mathrm{PROP}\else $\mathrm{PROP}$\fi}
\def\PTL{\ifmmode \mathrm{PTL}\else $\mathrm{PTL}$\fi}
\def\PTLP{\ifmmode \mathrm{PTL^P}\else $\mathrm{PTL^P}$\fi}
\def\PTLV{\ifmmode \mathrm{PTL}_V \else $\mathrm{PTL}_V$\fi}
\def\QPTL{\ifmmode \mathrm{QPTL}\else $\mathrm{QPTL}$\fi}

\def\Reg{\ifmmode \mathrm{Reg} \else $\mathrm{Reg}$\fi}
\def\sh{\ifmmode \mathord{\mathit{sh}} \else $\mathord{\mathit{sh}}$\fi}
\def\Since{\tempop{since}}

\def\First{\tempop{first}}
\let\state=s
\def\Test{?}
\def\UntilOp{\mathop{\mathcal U\null}}
\def\Utest{\mathop{\$\hspace{0em}}\nolimits}
\def\Var{\ifmmode \mathit{Var}\else $\mathit{Var}$\fi}

\def\sat{\mathrel{\raisebox{\depth}{$\scriptstyle {=}\!{|}$}}}
\def\const{\dashv}
\def\init{\mathit{init}}
\def\thmNL{\vdash_{\NL}}
\def\theoremNL{\thmNL\;}

\def\badVerbChars{\catcode`\&=\other
   \catcode`\^=\other
   \catcode`\_=\other
}
\def\badref{\bgroup\badVerbChars\badrefAux}
\def\badrefAux#1{\ref{#1} %
   (??? {\hyphenpenalty=0 
   \hbadness=10000   
   \textbf{\sffamily #1}} ???)\egroup}
\def\badeqref{\bgroup\badVerbChars\badeqrefAux}
\def\badeqrefAux#1{\eqref{#1} %
   (??? {\hyphenpenalty=0 
   \hbadness=10000   
   \textbf{\sffamily #1}} ???)\egroup}

\begin{document}


\setbox\tempbox=\hbox to 0pt{\ifkcp\else\hss\fi
   \fbox{\relax
    \iffalse
    \large *** A slightly earlier version of this has been submitted for
       publication. ***
    \else
    \parbox{0.9\textwidth}{An earlier version of this appeared in
    \emph{We Will Show Them: Essays in Honour of Dov Gabbay
                  on his 60th Birthday}, Volume 2.
    S. Artemov, H. Barringer, A. S. d'Avila Garcez, L.
    C. Lamb, and J. Woods (eds.),
    pages 371--440,
    College Publications,
    2005,
    URL: \url{http://www.collegepublications.co.uk}.}    
    \fi
    }\hss}
\setbox\tempbox=\vbox to 0pt{\vss\box\tempbox \vskip 6pt}
\ifdistribution
\else
   \setbox\tempbox=\hbox{}
\fi
\ht\tempbox=0pt
\dp\tempbox=0pt

\ifspringer
   \title{
      \mbox{\vbox{\offinterlineskip
          \box\tempbox
          \hbox to 0pt{\hss
             A Hierarchical Analysis of\hss}}} \\
                 Propositional Temporal Logic Based on Intervals}
   \author{Ben Moszkowski\relax
      \thanks{I wish to thank Zohar Manna for introducing me to temporal logic
              and for his guidance and support during my PhD studies.  I would
              also like to thank Nachum Dershowitz for his indefatigable
              efforts in organising the symposium and Festschrift in honour of
              Zohar.}\relax }
   \institute{
       Software Technology Research Laboratory, Hawthorn Building, \\
       De Montfort University,
       The Gateway,
       Leicester LE1 9BH,
       Great Britain \\
       \email{x@y, where x=benm and y=dmu.ac.uk}
   }
   \maketitle
\fi

\ifgenstyle
   \title{\relax
              \mbox{\vbox{\relax
          \box\tempbox
        \hbox to 0pt{\hss A Hierarchical Analysis of\hss}}} \\
        Propositional Temporal Logic Based on Intervals
   }

   \author{{\Large Ben Moszkowski\thanks{\relax
      Part of the research described here has been kindly supported
      by EPSRC research grant GR/K25922.}} \\ [2pt]
   Software Technology Research Laboratory \\
   Gateway House \\
   De Montfort University \\
   The Gateway \\
   Leicester LE1 9BH \\
   Great Britain \\[3pt]
   \ifconfsub
   \begin{tabular}{rl}
   phone: & +44-191-477-6565 \\
   fax: & +44-191-490-1886 \\
   email: & x@y, where x=benm and y=dmu.ac.uk
   \end{tabular}
   \else
      email: \texttt{x@y, where x=benm and y=dmu.ac.uk}
   \fi
   \\[7pt] \ifCoRR \mbox{\ } \else \todayBuiltin \fi
   }

   \let\todayBuiltin=\today
   \def\today{}
   \maketitle
   \ifprintpagenumbers
      \thispagestyle{plain}
   \fi
\fi


\begin{abstract}
  We present a hierarchical framework for analysing propositional linear-time
  temporal logic ($\PTL$) to obtain standard results such as a small model
  property, decision procedures and axiomatic completeness.  Both finite time
  and infinite time are considered and one consequent benefit of the framework
  is the ability to systematically reduce infinite-time reasoning to
  finite-time reasoning.  The treatment of $\PTL$ with both the operator
  $\Until$ and past time naturally reduces to that for $\PTL$ without either
  one.  Our method utilises a low-level normal form for $\PTL$ called a
  \emph{transition configuration}.  In addition, we employ reasoning about
  intervals of time.  Besides being hierarchical and interval-based, the
  approach differs from other analyses of $\PTL$ typically based on sets of
  formulas and sequences of such sets.  Instead we describe models using time
  intervals represented as finite and infinite sequences of states.  The
  analysis relates larger intervals with smaller ones.  Steps involved are
  expressed in Propositional Interval Temporal Logic ($\PITL$) which is better
  suited than $\PTL$ for sequentially combining and decomposing formulas.
  Consequently, we can articulate issues in $\PTL$ model construction of equal
  relevance in more conventional analyses but normally only considered at the
  metalevel.  We also describe a decision procedure based on Binary Decision
  Diagrams.
  
  Beyond the specific issues involving PTL, the research is a significant
  application of ITL and interval-based reasoning and illustrates a general
  approach to formally reasoning about sequential and parallel behaviour in
  discrete linear time.  The work also includes some interesting
  representation theorems.  In addition, it has relevance to hardware
  description and verification since the specification languages PSL/Sugar
  (now IEEE standard 1850) and 'temporal e' (part of IEEE candidate standard
  1647) both contain temporal constructs concerning intervals of time as does
  the related SystemVerilog Assertion language contained in SystemVerilog, an
  extension of the IEEE 1364-2001 Verilog language.
\end{abstract}

\hangindent=\parindent\hangafter=1 %

\section{Introduction}

\label{introduction-sec}

Following the seminal paper by Pnueli~\cite{Pnueli77}, temporal
logic~\cite{MannaPnueli81,Kroeger87,Emerson90} has become one of the main
formalisms used in computer science for reasoning about the dynamic behaviour
of systems.  In particular, propositional linear-time temporal logic ($\PTL$)
and some variants of it have been extensively studied and used.  In a
relatively recent and significant article, Lichtenstein and
Pnueli~\cite{LichtensteinPnueli00} give a detailed analysis of $\PTL$ which is
meant to largely subsume and supercede earlier ones. Indeed, the work appears
to have the rather ambitious goal of coming close to offering the last word on
the subject and is perhaps best described in the authors' own words:
\begin{quote}
  The paper summarizes work of over 20 years and is intended to provide a
  definitive reference to the version of propositional temporal logic used for
  the specification and verification of reactive systems.
\end{quote}
The version of $\PTL$ considered by Lichtenstein and Pnueli has discrete time
and past time.  Both a decision procedure and axiomatic completeness are
investigated and a new simplified axiom system is presented.  The
approach makes use of semantic tableaux and throughout the presentation the
treatment of $\PTL$ with past-time operators runs in parallel with the
future-only version.  The authors choose in particular to use tableaux since
they offer a basis for uniformly showing axiomatic completeness and also
obtaining a practical decision procedure.  The extensive material about past
time is distinctly marked so that one can optionally delete it to obtain an
analysis limited to the future fragment of $\PTL$.

We present a novel framework for investigating $\PTL$ which significantly
differs from the methods of Lichtenstein and Pnueli and earlier treatments
such as~\cite{GabbayPnueli80,Wolper85,Kroeger87,Goldblatt87}.  It is used to
obtain standard results such as a small model property, a practical decision
procedure and axiomatic completeness.  However, instead of relying on semantic
tableaux, filtration and other previous techniques, our method is based on an
interval-oriented analysis of certain kinds of low-level $\PTL$ formulas
called \emph{transition configurations}.  An important feature of this
approach is that it provides a natural hierarchical means of reducing full
$\PTL$ to this subset and also reduces both $\PTL$ with the $\Until$ operator
and past time to versions without them.  Therefore the overwhelming bulk of
the analysis only needs to deal $\PTL$ with neither $\Until$ nor past time.
Moreover, the analysis of $\PTL$ with infinite time naturally reduces to that
for $\PTL$ with just finite time.  The low-level formulas also have associated
practical decision procedures, including a simple symbolic one based on Binary
Decision Diagrams (BDDs)~\cite{Bryant86} which we have implemented.

The basic version of $\PTL$ used here is described in detail in
Sect.~\ref{overview-of-ptl-sec} but we will now briefly summarise some of the
features in order to be able to overview some key aspects of our work.  We
postpone the treatment of $\Until$ and past time in order to later handle them
in a natural hierarchical manner.  Both finite and infinite time are
permitted, whereas most versions of $\PTL$ deal solely with the latter.  One
reason for including finite time is to allow us to naturally capture parts of
our infinite-time analysis within $\PTL$ formulas concerning finite-time
subintervals.  The only two primitive temporal operators initially considered
are $\Next$ (strong next) and $\Diamond$ (eventually) although some others are
definable in terms of them (e.g., $\Box$ (henceforth) and $\SDiamond$ (strict
eventually)).

Our analysis of $\PTL$ extensively employs intervals of time which are
represented as finite and countably infinite sequences of states and described
by formulas in a propositional version of Interval Temporal Logic
($\ITL$)\relax
~\cite{Moszkowski83,Moszkowski83a,HalpernManna83,Moszkowski85,Moszkowski86}
(see also~\cite{ITL}) referred to as $\PITL$.  By using a hierarchical,
interval-oriented framework, the approach differs from that of Lichtenstein
and Pnueli and previous ones which in general utilise sets of formulas and
sequences of such sets (also referred to as \emph{paths}). We instead relate
transition configurations to semantically equivalent formulas in $\PITL$.
Time intervals facilitate an analysis which naturally relates larger intervals
with smaller ones.  The process of doing this can be explicitly expressed in
$\PITL$ in a way not possible within previous frameworks which lack both a
formalisation of intervals and logical operators concerning various kinds of
sequential composition of intervals.

Let us now informally consider as an example a simplified presentation of how
we later establish the existence of periodic models for certain kinds of
low-level formulas involving infinite time.  The analysis for temporal logic
formulas involving infinite time needs to consider formulas of the form
$\Box\SDiamond A$, where $A$ is itself a restricted kind of temporal logic
formula.  Here $\Box\SDiamond A$ is true for an interval, that is, the
interval \emph{satisfies} $\Box\SDiamond A$, iff the interval has infinite
length and $A$ itself is satisfied by an infinite number of the interval's
suffixes.  We want to show that if $\Box\SDiamond A$ is satisfied by some
interval, then there also exists a periodic interval which satisfies
$\Box\SDiamond A$.  We first show a sufficient condition motivated by $A$'s
restricted syntax which ensures that $\Box\SDiamond A$ is semantically
equivalent to the $\PITL$ formula $A\Chopomega$.  This formula is true on an
interval if the interval has infinite length and can be split into an infinite
sequence of finite intervals each satisfying $A$.  We then select one of these
finite intervals and join $\omega$ copies of it together to obtain a periodic
interval satisfying $A\Chopomega$ and hence also the original formula
$\Box\SDiamond A$.  Furthermore, after showing the existence of bounded models
for $A$, we can then establish similar properties for $A\Chopomega$ and hence
also $\Box\SDiamond A$.


We believe that our interval-based analysis complements existing approaches
since it provides a notational way to articulate various issues concerning
$\PTL$ model construction which are equally relevant within a more
conventional analysis but are normally only considered at the metalevel.  It
also illustrates some general techniques for compositional specification and
proof in discrete linear time which are applicable here.  This all fits nicely
with one of the main purposes of a logic which is to provide a notation for
explicitly and formally expressing reasoning processes.  In addition, a number
of the temporal logic formulas encountered can even be used with little or no
change as input to a implementation of a $\PTL$ decision procedure which
supports both finite and infinite time.  The analysis itself is performed
without the need to add any fundamentally new concepts to $\PITL$ but does
require a reader's willingness to acquire some familiarity with $\PITL$ and
various fairly general issues concerning interval-based reasoning.

Another feature of our approach is that it readily generalises to a
finite-time analysis of an important subset of $\PITL$ called \emph{Fusion
  Logic} ($\FL$), which was previously used by us in~\cite{Moszkowski04a} to
hierarchically show the completeness of an axiom system for $\PITL$.  The
analysis of $\FL$ uses a reduction of $\FL$ formulas to $\PTL$ ones.  The
prototype implementation of our $\PTL$ decision procedure also supports $\FL$.
A brief introduction to $\FL$ is given in \S\ref{fusion-logic-subsec} since
$\FL$ is a natural extension of our framework for studying $\PTL$ and
furthermore demonstrates another connection between $\PTL$ and intervals.  We
plan in future work to give a more detailed discussion of the decision
procedure for $\FL$ as well as some other issues concerning $\FL$.

Our preliminary work in~\cite{Moszkowski04} contains an earlier description of
this material but was limited to showing axiomatic completeness for $\PTL$
without past time.  In the mean time, we have significantly extended the
notation, methods and their scope of application.  The structure of
presentation has also been refined.

The use of intervals here seems to go well with a growing general awareness
even in industry of the desirability for temporal logics which go beyond
conventional point-based constructs to also handle behavioural specifications
involving intervals of time.  \index{PSL/Sugar} As evidence for this we
mention the Property Specification Language PSL/Sugar\cite{PSLSugar}.  This is
a modified version of a language Sugar~\cite{BeerBenDavid2001} developed at
IBM/Haifa.  PSL/Sugar has been ratified as IEEE standard 1850 with the purpose
of precisely expressing a hardware system's design properties so that they can
then be tested using simulation and model checking.  It includes a temporal
logic with regular expressions and other operators for sequential composition.
The hardware description language SystemVerilog~\cite{SystemVerilog} is an
extension of the established IEEE 1364-2001 Verilog language and includes
temporal assertions similar to those in PSL/Sugar.  SystemVerilog has itself
been ratified as a standard by Accellera Organization, Inc.\ which also hopes
to obtain ratification from the IEEE.

\index{Verisity Ltd.}
\index{e language@$e$ language}
\index{temporal e@temporal $e$}
In addition, the IEEE Design Automation Standards Committee has recently
approved a project to produce a candidate standard for Verisity
Ltd.'s~\cite{Verisity} \emph{e} language which is intended for testing and
verification\footnote{Verisity has been acquired by Cadence Design
Systems~\cite{Cadence}.}.  A subset of \emph{e} called \emph{temporal e} was
influenced in part by $\ITL$~\cite{Morley99,HollanderMorley2001,Verisity03}.
The IEEE Standards Association has assigned the project the number
1647~\cite{IEEE1647}.

\subsection*{Structure of Presentation}

Let us now summarise the structure of the rest of this paper.
Section~\ref{background-sec} mentions some related work and gives a comparison
with our approach.  Section~\ref{overview-of-ptl-sec} presents the version of
$\PTL$ we use.  Section~\ref{pitl-sec} summarises the propositional version of
$\ITL$ which we use in the analysis.  Section~\ref{t-configurations-sec}
introduces low level $\PTL$ formulas called \emph{transition configurations}
and relates them to some semantically equivalent propositional $\ITL$ formulas
which simplify the subsequent analysis.
Section~\ref{small-models-for-trans-configs-sec} proves the existence of small
models for transition configurations.
Section~\ref{decomposition-of-transition-configurations-sec} shows how to
relate the satisfiability of the two main kinds of transition configurations
with simple interval-oriented tests.  Section~\ref{decision-procedure-sec}
deals with a practical BDD-based decision procedure for transition
configurations.  Section~\ref{axiom-system-for-nl-sec} concerns axiomatic
completeness for an important subset of $\PTL$ in which the only temporal
operator is $\Next$ (next).  Section~\ref{ptl-axiom-system-sec} looks
at a $\PTL$ axiom system and axiomatic completeness for transition
configurations.  Section~\ref{invariants-and-related-formulas-sec} presents
formulas called \emph{invariants} and \emph{invariant configurations} which
together serve as a bridge between the previously mentioned transition
configurations and arbitrary $\PTL$ formulas.
Section~\ref{dealing-with-arbitrary-ptl-formulas-sec} discusses how to
generalise the previous results to work with arbitrary $\PTL$ formulas.
Section~\ref{some-additional-features-sec} hierarchically extends our approach
to deal with both the temporal operators $\Until$ and past time.  It also
briefly looks at a superset of $\PTL$ called \emph{Fusion Logic}.
Section~\ref{discussion-sec} concludes with some brief discussion.

\section{Background}

\label{background-sec}

\index{temporal logic}
Temporal logics have become a popular topic of study in theoretical computer
science and are also being utilised by industry to locate faults in digital
circuit designs, communication protocols and other applications.  Issues such
as small models, proof systems, axiomatic completeness and decision procedures
for $\PTL$ (almost always limited to infinite time) have been extensively
investigated by Gabbay et al.~\cite{GabbayPnueli80}, Wolper~\cite{Wolper85},
Kr{\"o}ger~\cite{Kroeger87}, Goldblatt~\cite{Goldblatt87}, Lichtenstein and
Pnueli~\cite{LichtensteinPnueli00}, Lange and
Stirling~\cite{LangeStirling2001}, Pucella~\cite{Pucella05} (who also
considers $\PTL$ with finite time) and others.  French\cite{French00}
elaborates on the presentation by Gabbay et al.~\cite{GabbayPnueli80}.

Vardi and Wolper~\cite{VardiWolper86} and Bernholtz, Vardi and
Wolper~\cite{BernholtzVardi94} describe decisions procedures for some temporal
logics based on a reduction to $\omega$-automata.  They do not consider
axiomatic completeness.  Wolper~\cite{Wolper01} presents a tutorial on such
decision procedure for $\PTL$ with infinite time.

Ben-Ari et al.~\cite{BenAriManna81,BenAriManna83},
Wolper~\cite{Wolper81,Wolper83} and Banieqbal and
Barringer~\cite{BanieqbalBarringer86} develop closely related proofs of
completeness for logics which include $\PTL$ as a subset or are branching-time
versions of it.  The book by Rescher and Urquhart~\cite{RescherUrquhart71} is
an early source of tableau-based completeness proofs for temporal logics.  The
survey by Emerson~\cite{Emerson90} includes material about axiom systems for
both linear and branching-time temporal logic.

Fisher~\cite{Fisher92,Fisher97} (see also later work by Fisher, Dixon and
Peim~\cite{FisherDixon2001} and Bolotov, Fisher and
Dixon~\cite{BolotovFisher2002}) presents a normal form for $\PTL$ called
\emph{Separated Normal Form} (SNF) which consists of formulas having the
syntax $\Box \bigwedge_i A_i$, where each $A_i$ can be one of the following:
\begin{displaymath}
  \textstyle
  \mathbf{start} \enspace\imp\enspace \bigvee_c l_c
  \qquad
  \Next\bigwedge_a k_a \enspace\imp\enspace \Next\bigvee_d l_d
  \qquad
  \Next\bigwedge_b k_b \enspace\imp\enspace \Diamond l
\enspace.
\end{displaymath}
Here each particular $k_a$, $k_b$, $l$, $l_c$ and $l_d$ is a literal (i.e., a
proposition variable or its negation).  Some versions of SNF permit past-time
constructs or have other relatively minor differences.  Applications include
theorem proving, executable specifications and representing $\omega$-automata.
We mention SNF here since it is a $\PTL$ normal form which somewhat resembles
what we call invariants and formally introduce in
Sect.~\ref{invariants-and-related-formulas-sec}.

\section{Overview of $\protect\PTL$}

\label{overview-of-ptl-sec}

\index{Propositional Temporal Logic|see{PTL}}
\index{PTL}

This section summarises the basic version of $\PTL$ used here.  Later on in
Sect.~\ref{some-additional-features-sec} we augment $\PTL$ with the operator
$\Until$ and past time.

\subsection{Syntax of $\protect\PTL$}

\label{itl-syntax-subsec}

\index{PTL!syntax}

We now describe the syntax of permitted $\PTL$ formulas.  In what follows, $p$
is any propositional variable and both $X$ and $Y$ denote $\PTL$ formulas:
\begin{displaymath}
p \qquad
\True \quad
\Not X \quad
X \Or Y \quad
\Next X \text{ (``strong next'')}\quad
\Diamond X \text{ (``eventually'')}\enspace.
\end{displaymath}
We include $\True$ as a primitive so as to avoid a definition of it which
contains some specific variable.  This is not strictly necessary.  Other
conventional logic operators such as $\False$, $X\And Y$ and $X\imp Y$ ($X$
implies $Y$) are defined in the usual way.  Also, $\Box X$ (``henceforth'') is
defined as $\Not\Diamond\Not X$.

\subsection{Semantics of $\protect\PTL$}

\label{semantics-of-ptl}

\index{PTL!semantics}
\index{interval}
\index{unit of time}
\index{time!unit of}
\index{time!interval}
\index{00@$\sigma$}
The version of $\PTL$ considered here uses discrete, linear time which is
represented by intervals each consisting of a sequence of one or more states.
More precisely, an interval $\sigma$ is any finite or infinite sequence of one
or more states $\sigma_0$, $\sigma_1, \ldotss$.  Each state $\sigma_i$ in
$\sigma$ maps each propositional variable $p$, $q, \ldots$ to one of the
boolean values $\True$ and $\False$.  The value of $p$ in the state $\sigma_i$
is denoted $\sigma_i(p)$.
\index{interval!length}
A finite interval $\sigma$ has an \emph{interval
  length} $\intlen{\sigma}\ge 0$ which equals the number of states minus 1 and
is hence always greater than or equal to 0.  We regard the smallest nonzero
interval length 1 as a \emph{unit} of (abstract) time.  For example, an
interval with 6 states has interval length 5 or equivalently 5 time units.
These units do not correspond to any particular notion of physical time.
The interval length of an infinite interval is taken to be $\omega$.  The term
\index{subinterval}
\emph{subinterval} refers to any interval obtained from some \emph{contiguous}
subsequence of another interval's states.

\index{empty interval}
\index{interval!empty}
\index{unit interval}
\index{interval!unit}
We call a one-state interval (i.e., interval length 0) an \emph{empty
interval}.  A two-state interval (i.e., interval length 1) is called a
\emph{unit interval}.  Both kinds of intervals play an important role in our
analysis.

\index{00@$\vld$} The notation $\sigma \vld X$ denotes that the interval
$\sigma$ \emph{satisfies} the $\PTL$ formula $X$.  We now give a definition of
this using induction on $X$'s syntax:

\begin{itemize}
  
\item Propositional variable: $\sigma \vld p \iff p$ is true in the initial
  state $\sigma_0$ (i.e., $\sigma_0(p)=\True$).

\item True: $\sigma \vld \True$ trivially holds for any $\sigma$.

\item Negation: \( \sigma\vld \Not X  \iff \sigma \not\vld X \).

\item Disjunction: \( \sigma\vld X \Or Y \iff
  \sigma\vld X \text{ or } \sigma\vld Y \).
  
\item Next:
  \( \sigma\vld \Next X \iff \sigma' \vld X, \) \\
where $\sigma$ contains at least two states and $\sigma'$ denotes the suffix
subinterval $\sigma_1 \sigma_2 \ldots$ which starts from second state
$\sigma_1$ in $\sigma$.

\item Eventually: \( \sigma\vld \Diamond X  \iff \sigma' \vld X \), \\
  for some suffix subinterval $\sigma'$ of $\sigma$ (perhaps $\sigma$ itself).

\end{itemize}

Table~\ref{temporal-operators-table} shows a variety of other useful temporal
operators which are definable in $\PTL$.  It includes operators for testing
whether an interval is finite or infinite and whether the interval has exactly
one state or two states.  Most of the operators only become relevant when
finite intervals are permitted.  Therefore, readers who are just familiar with
conventional $\PTL$ and infinite time will have previously encountered only a
few of the operators.
\begin{mytable}
\index{PTL!derived operators}
\def\vgap{\vskip 5pt}%
\def\vdivide{\vgap\hrule\vgap}
\def\theader#1{\multispan3\quad\emph{#1:}\hfil\quad}
\begin{widedisplaymath}
\tempdimen=\fboxsep
\fboxsep=0pt %
\fbox{\vbox{\halign{\strut\ifkcp\else\quad\fi\hskip\tempdimen
   $#$\hfil&$\null\defeqv #$\hfil\quad&#\hfil\hskip\tempdimen
   \ifkcp\else\quad\fi\cr
\noalign{\vgap}
\theader{Standard derived $\PTL$ operators}\cr
\noalign{\vgap}
\Box X &
   \Not \Diamond \Not X & Henceforth\cr
\SDiamond X &
   \Next \Diamond X & Eventually in strict future \cr
\SBox X &
   \Not \SDiamond \Not X & Henceforth in strict future (not used here) \cr
\noalign{\vdivide}
\theader{$\PTL$ operators primarily for finite intervals}\cr
\noalign{\vgap}
\More &
   \Next\True & More than one state \cr
\Empty &
   \Not\More & Only one state (\emph{empty interval})\cr
\WeakNext X &
   \Not\Next\Not X & Weak next (same as $\More \imp \Next X$) \cr
\Skip &
   \Next\Empty & Exactly two states (\emph{unit interval}) \cr
X\Test &
   X \And \Empty & Empty interval with test\cr
\Utest X &
   X \And \Skip & Unit interval with test \cr
\noalign{\vdivide}
\theader{$\PTL$ operators for finite and infinite intervals}\cr
\noalign{\vgap}
\Finite &
   \Diamond \Empty & Finite interval\cr
\Inf &
   \Not\Finite & Infinite interval\cr
\SFin X &
   \Diamond(\Empty \And X) & Strong test of final state\cr
\Fin X &
   \Box(\Empty \imp X)& Weak test of final state\cr
\Dm X &
   \Diamond(\More \And X) & Sometime before the very end\cr
\Bm X &
   \Box(\More \imp X) & Henceforth except perhaps at very end\cr
\noalign{\vgap}}}}
\end{widedisplaymath}
\caption{Some definable $\PTL$ operators}
\label{temporal-operators-table}
\end{mytable}

\emph{Note: Some readers will (quite reasonably) prefer to skim
  Table~\ref{temporal-operators-table} for now and only later consult it in
  more detail when the various operators are actually used.}

Figure~\ref{sample-derived-PTLformulas-fig} assists in the understanding of
Table~\ref{temporal-operators-table} by illustrating a number of the operators
through sample formulas and intervals.  In the figure, the logical values
$\True$ and $\False$ are respectively abbreviated as ``\texttt{t}'' and
``\texttt{f}''.
\begin{myfigure}
  \begin{center}
    \scalebox{1.0}{\input{manna-journal-sample-derived-PTLformulas.half-pstex_t}}
  \end{center}
  \caption{Some examples of formulas with derived $\PTL$ operators}
  \label{sample-derived-PTLformulas-fig}
\end{myfigure}
In what follows, we frequently use $\Bm$ instead of $\Box$ since we need to
test pairs of adjacent states in a interval.  The operator $\Bm$ is better
suited for this since it does not ``run off the end'' when examining finite
intervals.  The fourth example in Figure~\ref{sample-derived-PTLformulas-fig}
serves as an example of this feature.  As a consequence, $\Bm$ is easier to
work with in our interval-based analysis as is later shown in
Theorem~\ref{expressing-transitive-relexive-closure-in-ptl-thm}.

\begin{mydefin}[Satisfiability and Validity]
  \label{satisfiability-and-validity-def}
  \index{satisfiability!PTL}
  \index{validity!PTL}
  \index{PTL!satisfiability}
  \index{PTL!validity}
  \index{00@$\vld$}
  \index{00@$\sat$}
  For any interval $\sigma$ and $\PTL$ formula $X$, if $\sigma$ satisfies $X$
  (i.e., $\sigma\vld X$ holds), then $X$ is said to be \emph{satisfiable},
  denoted as $\sat X$.  A formula $X$ satisfied by all intervals is
  \emph{valid}, denoted as $\vld X$.
\end{mydefin}



We now define an important subset of $\PTL$ involving the operator $\Next$:
\nobreak
\begin{mydefin}[Next Logic]
  \label{next-logic-def}
  \index{Next Logic (NL)}
  The set of $\PTL$ formulas in which the only primitive temporal operator is
  $\Next$ is called \emph{Next-Logic ($\NL$)}.  The subset of $\NL$ in which
  no $\Next$ is nested within another $\Next$ is denoted as $\NLone$.
\end{mydefin}
For example, the $\NL$ formula $p\And \Next q$ is in $\NLone$, whereas the
$\NL$ formula $p\And \Next (q \Or \Next p)$ is not.

The variables $T$, $T'$ and $T''$ denote formulas in $\NLone$.

\begin{mydefin}[Tautologies]
  \label{tautology-def}
  \index{tautology}
  A \emph{tautology} is any formula which is a substitution instance of some
  valid nonmodal propositional formula.
\end{mydefin}
For example, the formula $\Next X \Or \Diamond Y \imp \Diamond Y$ is a
tautology since it is a substitution instance of the valid nonmodal formula
$\vld p \Or q \imp q$.  It is not hard to show that all tautologies are
themselves valid since intuitively a tautology is any valid formula which does
not require modal reasoning to justify its truth.

\paragraph{Convention for variables denoting individual formulas and sets of formulas:}
In what follows, the variable $w$ refers to a \emph{state formula}, that is, a
formula with no temporal operators.  Furthermore, $\PROP$ denotes the set of
all state formulas.  For any finite set of variables $V$, $\PROP_V$ denotes
the set of all state formulas only having variables in $V$.  Likewise, the set
$\PTL_V$ denotes the set of all formulas in $\PTL$ only containing variables
in $V$ and $\NLone_V$ denotes the set of all formulas in $\NLone$ only having
variables in $V$.  For example, the formula $p\And \Diamond q$ is in
$\PTL_{\{p,q\}}$ but not in $\PTL_{\{p\}}$.

\subsection{Example of the Hierarchical Process}

\label{example-of-the-hierarchical-process-subsec}

Our analysis of $\PTL$ reduces arbitrary $\PTL$ formulas to lower level ones
with a much more restricted syntax.  The next $\PTL$ formula serves as a
simple example to motivate some of the notation and conventions later
introduced:
\begin{equation*}
  \Box\Diamond p \And \Box\Diamond\Not p
\enspace.
\end{equation*}
This is reducible to the formula $\Box I \,\And\,w$, where $I$ and $w$ are
given below:
\begin{equation*}
  \begin{array}{l}
    I\colon\enskip
        (r_1\equiv \Diamond p)
        \,\And\, (r_2\equiv \Diamond \Not r_1)
        \,\And\, (r_3\equiv \Diamond \Not p)
        \,\And\, (r_4\equiv \Diamond \Not r_3)
    \\[3pt]
    w\colon\enskip \Not r_2 \,\And\, \Not r_4
\enspace.
  \end{array}
\end{equation*}
The auxiliary variables $r_1$, \ldots, $r_4$ provide a natural way to
eliminate the nesting of temporal operators within other temporal operators in
$I$.  We call the conjunction $I$ an \emph{invariant} and the conjunction
$\Box I \,\And\,w$ an \emph{invariant configuration}.  Both are formally
introduced later in Sect.~\ref{invariants-and-related-formulas-sec}.  It can
be shown that the original formula $\Box\Diamond p \And \Box\Diamond\Not p$ is
satisfiable iff the invariant configuration $\Box I \,\And\,w$ is.

When analysing behaviour in finite time, we further transform the invariant
configuration $\Box I \,\And\,w$ to another special kind of conjunction $\Box
T \And w \And \Finite$, where $T$ and $w$ are as follows:
\begin{equation*}
  \begin{array}{ll}
    T\colon\enskip
      &  (r_1\equiv (p \Or \Next r_1))
        \;\And\; (r_2\equiv (\Not r_1 \Or \Next r_2)) \\
      &  \null \And\; (r_3\equiv (\Not p \Or \Next r_3))
        \;\And\; (r_4\equiv (\Not r_3 \Or \Next r_4))
    \\[3pt]
    w\colon\enskip & \Not r_2 \,\And\, \Not r_4
\enspace.
  \end{array}
\end{equation*}
Here $I$'s first conjunct $r_1\equiv \Diamond p$ is replaced in $T$ by the
$\Diamond$-free formula $r_1\equiv (p \Or \Next r_1)$.  The remaining
conjuncts in $T$ similarly avoid having any $\Diamond$ constructs.  We call
$T$ a \emph{transition formula} and $\Box T \,\And\, w \,\And\, \Finite$ a
\emph{transition configuration} (formally defined in
Section~\ref{t-configurations-sec}).  The formula $T$ is in fact a formula in
the important subset of $\PTL$ called $\NLone$ (previous formally defined in
Definition~\ref{next-logic-def}) in which the only temporal constructs are
$\Next$ operators not nested within other $\Next$ operators.  In addition, in
finite-time intervals the $\PTL$ formulas $\Box I$ and $\Box T$ are
semantically equivalent.  Moreover, it can be shown that the original formula
$\Box\Diamond p \And \Box\Diamond\Not p$ is satisfiable in finite time iff the
transition configuration $\Box T \And w \And \Finite$ is satisfiable.  As is
later shown in Sect.~\ref{t-configurations-sec}, $\NLone$ formulas such as $T$
play a fundamental role in our analysis of transition configurations.

\subsection{Notation for Accessing Parts of Conjunctions}

From the examples just given it can be seen that we often manipulate formulas
which are conjunctions.  The next three definitions provide some helpful
notation for denoting the number of conjuncts of such a formula and for
accessing one or more of them.

\begin{mydefin}[Size of a Conjunction]
  \label{size-of-a-conjunction-def}
  For any conjunction $C$ of zero or more conjuncts, let the notation
  $\size{C}$ denote the number of $C$'s conjuncts.
\end{mydefin}

\begin{mydefin}[Indexing of a Conjunction's Conjuncts]
  \label{indexing-of-a-conjunction's-conjuncts-def}
  For each $k: 1\le k \le \size{C}$, we let $C[k]$ denote the $k$-th conjunct.
\end{mydefin}
Observe that if a conjunction $C$ has length $\size{C}=0$, there are no
conjuncts to be indexed.

\begin{mydefin}[Parts of a Conjunction]
  \label{parts-of-a-conjunction-def}
  Suppose $C$ is a conjunction and $k$ and $l$ are natural numbers such that
  $1\le k\le \size{C}$ and $0 \le l \le \size{C}$.  The notation $C[k:l]$
  denotes the conjunction of consecutive conjuncts in $C$ starting with $C[k]$
  and finishing with $C[l]$, inclusive, i.e., $C[k] \And \cdots \And C[l]$
  (which contains $l-k+1$ conjuncts).
\end{mydefin}  
Note that for any conjunction $C$, the formula $C[1:0]$ denotes $\True$ and
$C[1:\size{C}]$ is identical to $C$.  Also, for any $k: 1\le k \le \size{C}$,
both $C[k]$ and $C[k:k]$ refer to the same conjunct.

\section{Propositional Interval Temporal Logic}

\label{pitl-sec}

\index{Interval Temporal Logic|see{ITL}}
\index{ITL!Propositional|see{PITL}}
\index{Propositional ITL|see{PITL}}
\index{PITL}
We now describe the version of quantifier-free propositional $\ITL$ ($\PITL$)
used here for systematically analysing transition configurations.  More
on
$\ITL$ can be found in \cite{Moszkowski83,Moszkowski83a,HalpernManna83,%
Moszkowski85,Moszkowski86,Moszkowski94,Moszkowski98,%
Moszkowski00a,Moszkowski04a} (see also~\cite{ITL}).  The same discrete-time
intervals are used as in $\PTL$.  In addition, all $\PTL$ constructs are
permitted as well as two other ones.  Hence, any $\PTL$ formula is also a
$\PITL$ formula.

\index{PITL!chop}
\index{PITL!chop-star}
\index{chop!chop-star}
\index{00@$;\,$}
\index{00@$\Chopstar$}
Here is the syntax of $\PITL$'s two extra constructs, where
$A$ and $B$ are themselves $\PITL$ formulas:
\begin{displaymath}
A;B\enskip\mbox{{(\em chop\/)}} \qquad
A\Chopstar\enskip\mbox{{(\em chop-star\/)}}\enspace.
\end{displaymath}
The semantics of the other constructs in $\PITL$ is as in $\PTL$ and is
therefore omitted here.

\index{subinterval} \index{00@$\sigma_{i:j}$} Before defining the semantics of
chop and chop-star, we introduce some notation for describing subintervals of
an interval $\sigma$.  For natural numbers $i$, $j$ with $i\le
j\le\intlen{\sigma}$, let $\sigma_{i:j}$ denotes the subinterval with starting
state $\sigma_i$ and final state $\sigma_j$ and having interval length $j-i$
(i.e., $j-i+1$ states). Furthermore, if $\sigma$ is an infinite interval, let
$\sigma_{i:\omega}$ denote the (infinite) suffix subinterval starting with
state $\sigma_i$.

The formula $A;B$ is true on $\sigma$ (i.e., $\sigma\vld A;B$)
iff one of the following holds:
\begin{itemize}
\item For some natural number $i: 0\le i\le \intlen{\sigma}$, the interval
  $\sigma$ can be divided into two subintervals $\sigma_{0:i}$ and
  $\sigma_{i:\intlen{\sigma}}$ sharing the state $\sigma_i$ such that both
  $\sigma_{0:i}\vld A$ and $\sigma_{i:\intlen{\sigma}}\vld B$ hold.
\item The interval $\sigma$ itself has infinite length and $\sigma\vld A$
  holds.
\end{itemize}

The formula $A\Chopstar$ is true on $\sigma$ (i.e., $\sigma\vld A\Chopstar$)
iff one of the following holds:
\begin{itemize}
\item The interval $\sigma$ has finite length and there exists some natural
  number $n\geq 0$ and finite sequence of natural numbers $l_0\le
  l_1\le\cdots\le l_{n}$ where $l_0=0$ and $l_{n}=\intlen{\sigma}$, such that
  for each $i:0\le i\lt n$, $\sigma_{l_i:l_{i+1}} \vld A$ holds.
  
  The behaviour of chop-star on empty intervals is a frequent source of
  confusion and it is therefore important to note that any formula
  $A\Chopstar$ (including $\False\Chopstar$) is true on a one-state interval.
  This is because in the semantics of chop-star for a one-state interval we
  can always set $n=0$ and therefore ignore the values of variables in the
  interval $\sigma$.
\item The interval $\sigma$ has infinite length and there exists some $n\geq
  0$ and finite sequence of natural numbers $l_0\le l_1\le\cdots\le l_{n}$
  where $l_0=0$, such that for each $i:0\le i\lt n$, $\sigma_{l_i:l_{i+1}}
  \vld A$ holds and also $\sigma_{l_n:\omega} \vld A$ holds.
\item The interval $\sigma$ has infinite length and there exists some
  countably infinite strictly ascending sequence of natural numbers $l_0\lt
  l_1\lt\cdots$ where $l_0=0$, such that for each $i:i\ge 0$,
  $\sigma_{l_i:l_{i+1}} \vld A$ holds.
\end{itemize}

Figure~\ref{informal-ITL-fig} pictorially illustrates the semantics of
\emph{chop\/} and \emph{chop-star} in both finite and infinite time and also
shows some simple $\PITL$ formulas together with intervals which satisfy them.
For some sample formulas we include in parentheses versions using conventional
$\PTL$ logic operators which were previously introduced in
Sect.~\ref{overview-of-ptl-sec}.
\begingroup
\hyphenpenalty=10000 %
\begin{myfigure}
\begin{center}
   \def\xdots{\scalebox{2.0}{\ldots}} %
   \setbox\tempbox=\hbox{\scalebox{0.75}{\relax
            \input{semantics-of-ITL-for-infinite-time-manna-journal-submission.half-pstex_t}}} %
   \subfigure[Informal semantics for finite time]{\vbox to \ht\tempbox{\relax
      \hbox{\scalebox{0.75}{\relax
            \input{semantics-of-ITL-for-finite-time-manna-journal-submission.half-pstex_t}}}\vfil}} %
      \hspace{1cm}
   \subfigure[Informal semantics for infinite time]{\copy\tempbox} %
      \hspace{1cm}
   \subfigure[Some finite-time examples]{\relax
      \scalebox{0.75}{\input{sample-ITL-formulas-for-manna-journal-submission.half-pstex_t}}}
\end{center}
\caption{Informal $\PITL$ semantics and examples}
\label{informal-ITL-fig}
\end{myfigure}
\endgroup

\index{PITL!derived operators|(}
We make use of the following definitions of two straightforward
forms of iteration expressible with chop and chop-star:
\index{00@$\Chopplus$}
\index{chop!chop-plus}
\index{00@$\Chopomega$}
\index{chop!chop-omega}
\begin{equation*}
  A\Chopplus \defeqv A;A\Chopstar \qquad
  A\Chopomega \defeqv (A\And\Finite)\Chopstar \And \Inf
\enspace.
\end{equation*}
In addition, for any $n\ge 0$, we define $A^n$ to be the formula $\Empty$ if
$n=0$ and otherwise to be $A;A^{n-1}$.  The constructs $A^{\le n}$ and $A^{\lt
  n}$ are defined to be the disjunctions $\bigvee_{k\le n} A^k$ and
$\bigvee_{k\lt n} A^k$, respectively.

Other derived operators are also possible.  Table~\ref{di-da-table} shows some
especially useful ones.
\begin{mytable}
  \begin{center}
    \begin{tabular}{L@{$\Defeqv$}Ll}
      \Di A &
        A;\True& $A$ is true in some initial subinterval \\
      \Bi A &
        \Not \Di \Not A & $A$ is true in all initial subintervals \\
      \Da A &
        \Finite;A;\True & $A$ is true in some subinterval \\
      \Ba A &
        \Not \Da \Not A & $A$ is true in all subintervals
    \end{tabular}
    \caption{Some useful derived $\PITL$ operators}
    \label{di-da-table}
    \index{PITL!derived operators|)}
  \end{center}
\end{mytable}

\index{satisfiability!PITL}
\index{validity!PITL}
\index{PITL!satisfiability}
\index{PITL!validity}
The notions of satisfiability and validity already introduced in
Definition~\ref{satisfiability-and-validity-def} for $\PTL$ naturally
generalise to $\PITL$.

Let $\PITLV$ be the set of all $\PITL$ formulas only having variables
in $V$.

The next definition introduces a special kind of state formula which is
indispensable for interval-based reasoning.  It plays the role that sets of
formulas typically do in other analyses of $\PTL$.
\begin{mydefin}[Atoms and $V$-Atoms]
  \label{atoms-def}
  \index{atom} An \emph{atom} is any finite conjunction in which each conjunct
  is some propositional variable or its negation and no two conjuncts share
  the same variable.  The set of all atoms is denoted $\Atoms$.  The Greek
  letters $\alpha$, $\beta$ and $\gamma$ denote individual atoms.  For any
  finite set of propositional variables $V$, let $\Atoms_V$ be some set of
  $2^{\size{V}}$ logically distinct atoms containing exactly the variables in
  $V$.  We refer to such atoms as \emph{$V$-atoms}.
\end{mydefin}
For example, we can let $\Atoms_{\{p,q\}}$ be the set of the four logically
distinct atoms shown below:
\begin{displaymath}
  p \And q
  \qquad
  p \And \Not q
  \qquad
  \Not p \And q
  \qquad
  \Not p \And \Not q
\enspace.
\end{displaymath}
One simple convention is to assume that the propositional variables in an atom
occur from left to right in lexical order.  For any finite set of variables
$V$, this immediately leads to a suitable set of $2^{\size{V}}$ different
$V$-atoms.

\section{Transition Configurations}

\label{t-configurations-sec}

Starting with a finite set of variables $V$, an $\NLone_V$ formula $T$ and a
state formula $\init$ in $\PROP_V$, we consider small models, a decision
procedure and axiomatic completeness for certain low-level formulas referred
to here as \emph{transition configurations}.  These formulas play a central
role in our approach. The analysis of arbitrary $\PTL$ formulas can be
ultimately reduced to that of transition configurations.

Before actually formally defining transition configurations, we need to
introduce the concept of a \emph{conditional liveness formula} which is a
specific kind of conjunction necessary for reasoning about liveness properties
involving infinite time.  The definition therefore makes use of some general
notation already introduced in
Definitions~\ref{size-of-a-conjunction-def}--\ref{parts-of-a-conjunction-def}
for manipulating conjunctions.
\begin{mydefin}[Conditional Liveness Formulas]
  \label{conditional-liveness-formula-def}
  \index{conditional liveness formula}
  A \emph{conditional liveness\kcpbreak formula} $L$ is a conjunction of
  $\size{L}$ implications $L[1]\And\cdots\And L[\size{L}]$.  Each implication
  has the form $w \imp \Dm w'$, where $w$ and $w'$ are two state formulas.
  For convenience, we let $\eta_{L[k]}$ denote the left operand of the $k$-th
  implication in $L$.  Similarly, $\theta_{L[k]}$ denotes the operand of the
  $\Dm$ formula in the $k$-th implication $L[k]$'s right side.  Therefore, for
  each $k: 1\le k\le \size{L}$, the implications $L[k]$ and $\eta_{L[k]} \imp
  \Dm \theta_{L[k]}$ denote the same formula.
  
  For any $V$-atom $\alpha$ and any $k: 1\le k\le \size{L}$, if the formula
  $\alpha\And \eta_{L[k]}$ is satisfiable, we say that $\alpha$ \emph{enables}
  $L$'s $k$-th implication $L[k]$.
\end{mydefin}
Here is a sample conditional liveness formula:
\begin{equation}
  \label{sample-conditional-liveness-formula-eq}
  ((p\Or \Not q) \imp \Dm \Not p) \AND (q \imp \Dm(p \equiv \Not q))
     \AND (\True \imp \Dm(p \imp q))
\enspace.
\end{equation}
Note that $\Dm$ behaves the same as $\Diamond$ on infinite intervals.
However, in finite intervals $\Dm$, like its dual $\Bm$, ignores the final
state.  In principle, either $\Dm$ or $\Diamond$ can be used in conditional
liveness formulas and the choice between them appears to be largely a matter
of taste.  Nevertheless, we choose to use $\Dm$ in part because it facilitates
an interesting generalisation of both conditional liveness formulas and
another kind of formula called an \emph{invariant} which is introduced later
in Sect.~\ref{invariants-and-related-formulas-sec}.  This generalisation will
be mentioned in
\S\ref{generalised-conditional-liveness-formulas-and-invariants-subsec}.  In
addition, the application of $\Dm$ naturally complements our extensive use of
its dual $\Bm$.

Here is the definition of transition configurations:
\begin{mydefin}[Transition Configurations]
  \label{t-configuration-def}
  \index{transition configuration} A \emph{transition configuration} is a
  formula of the form $\Box T \;\And\; X$, where the formula $T$ is in
  $\NLone_V$, and the $\PTL_V$ formula $X$ has one of the four forms shown
  below:
  \begin{center}
    \begin{tabular}{cc}
      Type of transition configuration
        & Syntax of $X$ \\\noalign{\hrule \vskip 2pt}
      Finite-time & $\init \And \Finite$ \\[1pt]
      Infinite-time & $\init \And \Box\SDiamond L$ \\[1pt]
      Final & $w \And \Empty$ \\[1pt]
      Periodic & $\alpha \And L \And \Box\SDiamond (\alpha \And L)$
    \end{tabular}
  \end{center}
  Here $\init$ is a state formula in $\PROP_V$ which corresponds to some
  \emph{initial} condition, $w$ is some state formula in $\PROP_V$, $L$ is a
  conditional liveness formula in $\PTL_V$ and $\alpha$ is a $V$-atom.  If
  $\init$ is the formula $\True$, it can be omitted.  The same applies with
  $w$.
\end{mydefin}
For example, the conjunction $\Box (\More \imp(p \equiv \Next p)) \,\And\, p
\,\And\, \Finite$ is a finite-time transition configuration which is true
exactly for finite intervals in which $p$ is always true.

\paragraph{Note:}
In the course of analysing transition configurations, we will assume that $V$,
$T$, $\init$ and $L$ are fixed.

We will show that finite-time and infinite-time transition configurations are
equivalent to certain $\PITLV$ formulas for which we can more readily
establish such things as the existence of periodic models, small models, a
decision procedure and axiomatic completeness.
Table~\ref{trans-config-to-pitl-table} shows the corresponding $\PITLV$
formula for each kind of transition configuration and where the equivalence of
the two is proved.  Here $\vec V\Tassign \vec V$ denotes that the initial
value of each variable occurring in the set of variables $V$ equals its final
value.  It can be expressed as the $\PTL_V$ formula $\Finite \imp
\bigwedge_{v\in V} (v\equiv \Fin v)$ and is semantically equivalent to the
disjunction $\bigvee_{\alpha\in \Atoms_V} (\alpha \And \Fin\alpha)$.
\begin{mytable}
\centerline{\relax
    \begin{tabular}{@{\ifkcp\else\quad\fi}clc@{\ifkcp\else\quad\fi}}
      Type of transition
        &\multicolumn{1}{c}{$\PITLV$ formula}& Where \\
      configuration
        & &  proved \\\noalign{\hrule \vskip 2pt}
      Finite-time
        & $((\Utest T)\Chopstar \And \init \And \Finite);(T \And \Empty)$
        & Theorem~\ref{reduction-of-finite-time-t-configs-thm}
      \\[3pt]
      Infinite-time
        & $\begin{myarray}
             ((\Utest T)\Chopstar \And \init \And \Finite); \\
             \qquad
                \bigl((\Utest T)\Chopstar \And L
             \And (\vec V\Tassign \vec V) \bigr)\Chopomega
           \end{myarray}$
        & Theorem~\ref{reduction-of-infinite-time-t-configs-thm}
      \\[3pt]
      Final & $T \And w \And \Empty$ & straightforward \\[3pt]
      Periodic & $((\Utest T)\Chopstar \And \alpha \And L)\Chopomega$
        & Theorem~\ref{reduction-of-periodic-t-configs-thm}
    \end{tabular}}
    \caption{Reduction of transition configurations to $\PITLV$ formulas}
    \label{trans-config-to-pitl-table}
\end{mytable}

Theorem~\ref{satisfiable-infinite-time-t-config-thm} will furthermore
establish that the infinite-time transition configuration is satisfiable iff
the next $\PTL$ formula is satisfiable in finite time:
\begin{equation*}
  \Bm T \And \init \And \Diamond (L \And
      \Finite \And \More \And (\vec V\Tassign \vec V))
\enspace.
\end{equation*}

In order to perform interval-based analysis on transition configurations, we
need to relate $\Box T$ to the $\PITL$ formula $(\Utest T)\Chopstar$.  Now the
$\PTL$ formula $\Bm T$, which is very similar to $\Box T$, was previously
defined in Table~\ref{temporal-operators-table} to be true on an interval iff
$T$ is true in all of the interval's nonempty suffix subintervals.  It turns
out that due to $T$ being in $\NLone$, the formula $(\Utest T)\Chopstar$ is
semantically equivalent to $\Bm T$.  Intuitively, this is because an $\NLone$
formula cannot probe past the second state of an interval.  The next lemma
formalises this:
\begin{mylemma}
  \label{nlone-first-two-states-lem}
  Let $\sigma$ and $\sigma'$ be two nonempty intervals which share the same
  first two states (i.e., $\sigma_0=\sigma'_0$ and $\sigma_1=\sigma'_1$).
  Then, for any formula $T$ in $\NLone$, $\sigma$ satisfies $T$ iff $\sigma'$
  satisfies $T$.
\end{mylemma}
\begin{myproof}
  Induction on $T$'s syntax ensures that it cannot distinguish between
  $\sigma$ and $\sigma'$.
\end{myproof}

Consequently, if two nonempty intervals share the same first two states, then
the truth value of $T$ for both intervals is identical.
Figure~\ref{utest-star-eqv-box-m-fig} illustrates this with two instances of
an interval containing 4 states.  The second version uses the concrete
$\NLone$ formula $p\imp\Next\Not p$ and shows specific values for the
proposition variable $p$.
\begin{myfigure}
  \begin{center}
    \bgroup
    \def\T{p{\imp}{\Next}{\Not}p}
    \def\TT{p\imp\Next\Not p}
      \scalebox{1.0}{\input{utest-star-eqv-box-m-with-var-manna-journal-sub-slides.half-pstex_t}}
    \egroup
  \end{center}
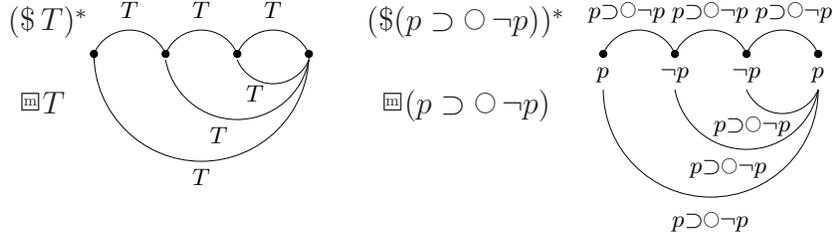
  \caption{Illustration of equivalence of $(\Utest T)\Chopstar$ and $\Bm T$}
  \label{utest-star-eqv-box-m-fig}
\end{myfigure}
Both $(\Utest T)\Chopstar$ and $\Bm T$ test each pair of adjacent states.  The
equivalence consequently permits us to express $(\Utest T)\Chopstar$ in $\PTL$
by means of $\Bm T$.  In addition, it is often useful to express $\Bm T$ as
$(\Utest T)\Chopstar$ because the later turns out to be much more suitable for
interval-based reasoning involving sequential composition and decomposition.

We now formally establish the semantic equivalence of the formulas $(\Utest
T)\Chopstar$ and $\Bm T$:
\begin{mytheorem}
  \label{expressing-transitive-relexive-closure-in-ptl-thm}
  The $\PITL$ formula $(\Utest T)\Chopstar$ and the $\PTL$ formula $\Bm T$ are
  semantically equivalent and hence the equivalence $(\Utest T)\Chopstar
  \equiv \Bm T$ is valid.
\end{mytheorem}
\begin{myproof}
  Given an interval $\sigma$, we can put each two-state (unit) subinterval in
  one-to-one correspondence with the suffix (nonempty) subinterval which
  shares the first two states.  Now $\sigma$ satisfies $(\Utest T)\Chopstar$
  iff $T$ is true on all of $\sigma$'s unit subintervals.  Similarly, $\sigma$
  satisfies $\Bm T$ iff $T$ is true on all of $\sigma$'s nonempty suffix
  subintervals.  By the previous Lemma~\ref{nlone-first-two-states-lem} a
  given unit subinterval satisfies $T$ iff the matching suffix (nonempty)
  subinterval satisfies $T$.  Consequently, the overall interval satisfies
  $(\Utest T)\Chopstar$ iff it satisfies $\Bm T$.
\end{myproof}

It is not hard to check that on a one-state (empty) interval, $\Bm T$ is
trivially true.  On a two-state (unit) interval, it is semantically equivalent
to the formula $T$ itself.

Also note that the $\PTL$ formula $\Box T$ is semantically equivalent to the
$\PTL$ formula $\Bm T \And \Fin T$.  This fact and
Theorem~\ref{expressing-transitive-relexive-closure-in-ptl-thm} together
establish that $\Box T$ is also semantically equivalent to the $\PITL$ formula
$(\Utest T)\Chopstar \And \Fin T$.  Therefore, the $\Box T$ formula in
transition configurations can be readily re-expressed in $\PITL$ as the
conjunction $(\Utest T)\Chopstar \And \Fin T$.  This will assist our
interval-based analysis of transition configurations.

\begin{myremark}
  We have discussed the important semantic equivalence of the formulas
  $(\Utest T)\Chopstar$ and $\Bm T$ with quite a few people who themselves
  have a considerable amount of experience with both $\PTL$ and $\PITL$.
  Originally we thought that this amounted to a straightforward application of
  temporal logic.  However, to our surprise, these people found the
  equivalence and its applications to be nontrivial and interesting.  For this
  reason, we have designated the statement of the equivalence of $(\Utest
  T)\Chopstar$ and $\Bm T$ to be a theorem (i.e., the previous
  Theorem~\ref{expressing-transitive-relexive-closure-in-ptl-thm}), rather
  than merely a lemma.
\end{myremark}

Here is a corollary of
Theorem~\ref{expressing-transitive-relexive-closure-in-ptl-thm} for infinite
time:
\begin{mycorollary}
  \label{expressing-transitive-relexive-closure-in-ptl-for-infinite-time-cor}
  The two formulas $\Box T$ and $(\Utest T)\Chopstar$ are semantically
  equivalent on infinite intervals and hence the implication $\Inf \implies
  \Box T \equiv (\Utest T)\Chopstar$ is valid.
\end{mycorollary}
\begin{myproof}
  This readily follows from
  Theorem~\ref{expressing-transitive-relexive-closure-in-ptl-thm} and the
  semantic equivalence of $\Bm T$ and $\Box T$ on infinite intervals.
\end{myproof}

The next two
Lemmas~\ref{box-t-and-diamond-a-eqv-utest-t-chopstar-and-diamond-box-t-and-a-valid-lem}
and~\ref{box-t-and-diamond-a-eqv-utest-t-chopstar-chop-box-t-and-a-valid-lem}
subsequently provide a basis for relating finite-time transition
configurations to final ones and also for relating infinite-time transition
configurations to periodic ones.
\begin{mylemma}
  \label{box-t-and-diamond-a-eqv-utest-t-chopstar-and-diamond-box-t-and-a-valid-lem}
  For any $\PITL$ formula $A$, the next equivalence is valid:
  \begin{equation*}
    \Valid \Box T \And \Diamond A
      \EQUIV  (\Utest T)\Chopstar \And \Diamond(\Box T \And A)
\enspace.
  \end{equation*}
\end{mylemma}
\begin{myproof}
  We first establish the validity of the $\PTL$ formula $\Box p \EQUIV \Bm p
  \And \Diamond\Box p$ which itself leads to the validity of the formula $\Box
  p \And \Diamond q \EQUIV \Bm p \And \Diamond(\Box p \And q)$. We then
  substitute $T$ into $p$ and $A$ into $q$.  Finally,
  Theorem~\ref{expressing-transitive-relexive-closure-in-ptl-thm} permits us
  to replace $\Bm T$ by $(\Utest T)\Chopstar$.
\end{myproof}

\begin{mylemma}
  \label{box-t-and-diamond-a-eqv-utest-t-chopstar-chop-box-t-and-a-valid-lem}
  For any state formula $w$ and $\PITL$ formula $A$, the next equivalence is
  valid:
  \begin{equation}
    \label{box-t-and-diamond-a-eqv-utest-t-chopstar-and-box-t-and-a-valid-1-eq}
    \Box T \And w \And \Diamond A
    \Equiv ((\Utest T)\Chopstar \And w \And \Finite);(\Box T \And A)
\enspace.
  \end{equation}
\end{mylemma}
\begin{myproof}
  Lemma~\ref{box-t-and-diamond-a-eqv-utest-t-chopstar-and-diamond-box-t-and-a-valid-lem}
  ensures that $\Box T \And \Diamond A$ is semantically equivalent to the
  conjunction $(\Utest T)\Chopstar \And \Diamond(\Box T \And A)$.  This is
  itself semantically equivalent to the next $\PITL$ formula:
  \begin{equation*}
    ((\Utest T)\Chopstar \And \Finite);((\Utest T)\Chopstar \And \Box T \And A)
\enspace.
  \end{equation*}
  Now $\Box T$ trivially implies $\Bm T$ which by
  Theorem~\ref{expressing-transitive-relexive-closure-in-ptl-thm} is
  semantically equivalent to $(\Utest T)\Chopstar$.  This consequently permits
  us to simplify the subformula $(\Utest T)\Chopstar \And \Box T$ into $\Box
  T$ to obtain the next valid equivalence:
  \begin{equation*}
    \Valid \Box T \And \Diamond A
      \Equiv ((\Utest T)\Chopstar \And \Finite);(\Box T \And A)
\enspace.
  \end{equation*}
  Simple temporal reasoning permits us to suitably add the state formula $w$
  to each side to obtain the validity of the
  formula~\eqref{box-t-and-diamond-a-eqv-utest-t-chopstar-and-box-t-and-a-valid-1-eq}.
\end{myproof}

\subsection{Analysis of Finite-Time Behaviour}

The following Lemma~\ref{relation-between-finite-time-and-final-t-configs-lem}
and Theorem~\ref{reduction-of-finite-time-t-configs-thm} concern reducing a
finite-time transition configuration to the associated semantically equivalent
$\PITL$ formula in Table~\ref{trans-config-to-pitl-table} which is easier to
later analyse:
\begin{mylemma}
  \label{relation-between-finite-time-and-final-t-configs-lem}
  The following equivalence is valid for finite-time transition configurations
  and relates them to final configurations:
  \begin{equation}
     \label{relation-between-finite-time-and-final-t-configs-1-eq}
     \Valid \Box T \And \init \And \Finite
       \EQUIV ((\Utest T)\Chopstar \And \init \And \Finite)
                ;(\Box T \And \Empty)
\enspace.
  \end{equation}
\end{mylemma}
\begin{myproof}
  The formula $\Finite$ is defined to be $\Diamond\Empty$.  Therefore
  Lemma~\ref{box-t-and-diamond-a-eqv-utest-t-chopstar-chop-box-t-and-a-valid-lem}
  ensures the validity of the
  equivalence~\eqref{relation-between-finite-time-and-final-t-configs-1-eq}.
\end{myproof}

Theorem~\ref{reduction-of-finite-time-t-configs-thm} builds on
Lemma~\ref{relation-between-finite-time-and-final-t-configs-lem} by reducing a
finite-time transition configuration to a chop formula in $\PITL$ which is
even easier to analysis because its righthand operand is in $\NLone$:
\begin{mytheorem}
  \label{reduction-of-finite-time-t-configs-thm}
  The following equivalence is valid for finite-time transition configurations:
  \begin{equation*}
     \Valid \Box T \And \init \And \Finite
       \EQUIV ((\Utest T)\Chopstar \And \init \And \Finite);(T \And \Empty)
\enspace.
  \end{equation*}
\end{mytheorem}
\begin{myproof}
  This readily follows from
  Lemma~\ref{relation-between-finite-time-and-final-t-configs-lem} and the
  fact that in an empty interval, the formulas $\Box T$ and $T$ are
  equivalent.
\end{myproof}
Note that the $\PITL$ formula $((\Utest T)\Chopstar \And \init \And
\Finite);(T \And \Empty)$ can also be expressed as the semantically equivalent
$\PITL$ formulas $\init\Test;((\Utest T)\Chopstar \And \Finite);T\Test$ and
$\init \And (\Utest T)\Chopstar \And \SFin T$.  Each form has its benefits.
We prefer $T\And \Empty$ over the equivalent $T\Test$ since some readers might
get confused upon seeing the operator $\Test$ with an operand which is a
temporal formula even though this is permitted in $\PITL$.

\subsection{Analysis of Infinite-Time Behaviour}

We now turn to analysing infinite-time transition configurations.  The first
step involves relating them to periodic transition configurations.  The next
Lemma~\ref{relation-between-infinite-time-and-periodic-t-configs-lem} does
this:
\begin{mylemma}
  \label{relation-between-infinite-time-and-periodic-t-configs-lem}
  The following equivalence is valid for infinite-time transition
  configurations:
  \begin{equation}
     \label{relation-between-infinite-time-and-periodic-t-configs-1-eq-lem}
     \begin{myarray}
       \textstyle
       \Box T \And \init \And \Box\SDiamond L \\[3pt]
       \,\EQUIV\, ((\Utest T)\Chopstar \And \init \And \Finite);
          \bigvee_{\alpha\in\Atoms_V}
            (\Box T \And \alpha \And L \And \Box\SDiamond(\alpha \And L))
\enspace.
  \end{myarray}
\end{equation}
\end{mylemma}
\begin{myproof}
  Observe that in an infinite interval if $L$ is always eventually true then
  for at least one of the finite number of $V$-atoms, the conjunction $\alpha
  \And L$ is also always eventually true.  Therefore simple temporal reasoning
  yields that $\Box\SDiamond L$ is semantically equivalent to the disjunction
  $\bigvee_{\alpha\in\Atoms_V} \Box\SDiamond(\alpha \And L)$. The subformula
  $\Box\SDiamond(\alpha \And L)$ can be re-expressed as $\Diamond(\alpha \And
  L \And \Box\SDiamond(\alpha \And L))$ so the next equivalence concerning
  $\Box\SDiamond L$ is valid:
  \begin{equation}
    \label{relation-between-infinite-time-and-periodic-t-configs-2-eq-lem}
    \textstyle
    \Valid \Box\SDiamond L \,\EQUIV\,
      \Diamond\bigvee_{\alpha\in\Atoms_V}
        (\alpha \And L \And \Box\SDiamond(\alpha \And L))
\enspace.
  \end{equation}
  We then use Lemma\relax
  ~\ref{box-t-and-diamond-a-eqv-utest-t-chopstar-chop-box-t-and-a-valid-lem}
  to establish the equivalence below for some arbitrary $V$-atom $\alpha$:
   \begin{multline*}
   \Valid
     \textstyle
      \Box T \And \init
       \And \Diamond
         (\Box T \And \alpha \And L \And \Box\SDiamond(\alpha \And L)) \\[2pt]
     \textstyle
     \quad\EQUIV
       ((\Utest T)\Chopstar \And \init \And \Finite);
             (\Box T \And \alpha \And L \And \Box\SDiamond(\alpha \And L))
\enspace.
  \end{multline*}
  Some simple temporal reasoning involving chop and $\bigvee$ yields the next
  valid equivalence:
  \begin{equation}
    \label{relation-between-infinite-time-and-periodic-t-configs-3-eq-lem}
    {\valid}
    \begin{myarray}
      \textstyle
       \Box T \And \init
        \And \Diamond\bigvee_{\alpha\in\Atoms_V}
          (\Box T \And \alpha \And L \And \Box\SDiamond(\alpha \And L)) \\[2pt]
      \textstyle
      \EQUIV
        ((\Utest T)\Chopstar \And \init \And \Finite);
        \bigvee_{\alpha\in\Atoms_V}
              (\Box T \And \alpha \And L \And \Box\SDiamond(\alpha \And L))
.
    \end{myarray}
  \end{equation}
  The combination of this and the previously mentioned semantic
  equivalence~\eqref{relation-between-infinite-time-and-periodic-t-configs-2-eq-lem}
  establishes the validity of the
  equivalence~\eqref{relation-between-infinite-time-and-periodic-t-configs-1-eq-lem}.
\end{myproof}

\subsubsection{Reduction using Chop-Omega Operator}

Much of the remainder of the analysis consists of showing how to further
reduce a periodic transition configuration $\Box T \And \alpha \And L \And
\Box\SDiamond(\alpha \And L)$ to the semantically equivalent $\PITL$ formula
$((\Utest T)\Chopstar \And \alpha \And L)\Chopomega$.  A general class of
formulas which includes $\alpha \And L$ will now be described.  For any
$\PITL$ formula $A$ in this class, the two formulas $A \And \Box\SDiamond A$
and $A\Chopomega$ will be shown to be semantically equivalent in
Theorem~\ref{a-and-box-sdiamond-a-eqv-a-chopomega-valid-thm}.  We first need
to introduce a derived $\PITL$ operator which turns out to be useful for
analysing periodic behaviour in infinite intervals.

\index{PITL!derived operators}
\begin{mydefin}[The Operator $\Df$]
  \label{df-def}
  For any $\PITL$ formula $A$, let the $\PITL$ formula $\Df A$ is defined to
  be $(A \And \Finite);\True$.  Therefore, $\Df A$ true on an interval iff $A$
  is true on some finite subinterval starting at the beginning of the overall
  interval.
\end{mydefin}
Note that $\Df A$ can also be expressed with the derived operator $\Di$
(itself previously defined in Table~\ref{di-da-table}) as $\Di(A \And
\Finite)$.

\index{PITL!fixpoints of operators}
\index{fixpoints!of PITL operators}
It is worthwhile to define a notion of fixpoints of the operator $\Df$:
\begin{mydefin}[Fixpoints of the Operator $\Df$]
  \label{fixpoints-of-the-operator-df-def}
  A $\PITL$ formula $A$ is a fixpoint of $\Df$ iff the equivalence $A\equiv
  \Df A$ is valid.
\end{mydefin}
Fixpoints of $\Df$ are easier to move out of subintervals than are arbitrary
formulas.  Incidentally, for any $\PITL$ formula $A$, the formula $\Df A$ is a
trivial fixpoint of $\Df$ since $\Df A$ and $\Df\Df A$ are semantically
equivalent.  We will shortly show that all conditional liveness formulas are
$\Df$-fixpoints and later use this in the analysis of infinite intervals.

We extensively investigate fixpoints of various temporal operators and their
application to compositional reasoning in
\cite{Moszkowski94,Moszkowski95a,Moszkowski96,Moszkowski98}.

The next lemma characterises a broad syntactic class of formulas which are
$\Df$-fixpoints and is easy to check:
\begin{mylemma}
  \label{fixpoints-of-the-operator-df-lem}
  Every state formula is a $\Df$-fixpoint.  Furthermore, if the $\PITL$
  formulas $A$ and $B$ are $\Df$-fixpoints, then so are the $\PITL$ formulas
  $A \And B$, $A \Or B$, $\Next A$ and $\Diamond A$.
\end{mylemma}

\begin{mylemma}
  \label{conditional-liveness-formulas-are-df-fixpoints-lem}
  Every conditional liveness formula is a $\Df$-fixpoint.
\end{mylemma}
\begin{myproof}
  A conditional liveness formula is a conjunction of implications each which
  has the form $w\imp \Dm w'$ for some state formulas $w$ and $w'$.  If we
  replace $\imp$ and $\Dm$ by their definitions, then the implication reduces
  to the formula $\Not w \Or \Diamond((\Next\True) \And w')$.
  Lemma~\ref{fixpoints-of-the-operator-df-lem} then ensures that this is a
  $\Df$-fixpoint.  Consequently, the original implication $w\imp \Dm w'$ is
  one as well.  Therefore by Lemma~\ref{fixpoints-of-the-operator-df-lem}, the
  conjunction of such implications which constitutes a conditional liveness
  formula is also a $\Df$-fixpoint.
\end{myproof}
Observe that by Lemmas~\ref{fixpoints-of-the-operator-df-lem}
and~\ref{conditional-liveness-formulas-are-df-fixpoints-lem}, the formula
$\alpha \And L$ is itself a $\Df$-fixpoint because both $\alpha$ and $L$ are
$\Df$-fixpoints.

Now the formula $\alpha \And L \And \Box\SDiamond(\alpha \And L)$ is itself an
instance of the $\PITL$ formula $A \And \Box\SDiamond A$.  We now proof in
Theorem~\ref{a-and-box-sdiamond-a-eqv-a-chopomega-valid-thm} that if $A$ is a
$\Df$-fixpoint, then the formula $A \And \Box\SDiamond A$ can be re-expressed
as the semantically equivalent $\PITL$ formula $A\Chopomega$.  This will let
us re-express $\alpha \And L \And \Box\SDiamond(\alpha \And L)$ as the
semantically equivalent $\PITL$ formula $(\alpha \And L)\Chopomega$.  The
establishment of this equivalence is a key step in the reduction of reasoning
about infinite time behaviour to finite time behaviour and consequently
proving the existence of periodic models for satisfiable periodic transition
configurations.
\begin{mytheorem}
  \label{a-and-box-sdiamond-a-eqv-a-chopomega-valid-thm}
  For any $\PITL$ formula $A$ which is a $\Df$-fixpoint, the next equivalence
  is valid:
  \begin{equation}
    \label{a-and-box-sdiamond--a-eqv-a-chopomega-valid-1-eq}
    \Valid A \And \Box\SDiamond A \Equiv A\Chopomega
\enspace.
  \end{equation}
\end{mytheorem}
\begin{myproof}
  \emph{Left side implies right side:} Suppose that an interval $\sigma$
  satisfies $A \And \Box\SDiamond A$.  Now this conjunction is semantically
  equivalent to the formula $\Df A \And \Box\SDiamond \Df A$ because $A$ is a
  $\Df$-fixpoint.  Therefore $\sigma$ also satisfies the formula $\Df A \And
  \Box\SDiamond \Df A$.  Furthermore, $\sigma$ is clearly an infinite interval
  due to the conjunct containing $\Box\SDiamond$.  Therefore, $\sigma$ has an
  infinite number of finite subintervals which all satisfy $A$ including some
  starting with $\sigma$'s first state.  An infinite sequence of
  nonoverlapping finite-length subintervals all satisfying $A$ can then be
  selected with the first one commencing at $\sigma$'s first state.
  Consequently, $\sigma$ satisfies the $\PITL$ formula $((A \And
  \Finite);\True)\Chopomega$ which is the same as $(\Df A)\Chopomega$.  This
  and the assumption that $A$ is a $\Df$-fixpoint yield that $\sigma$
  satisfies $A\Chopomega$.
  
  \emph{Right side implies left side:} Suppose that an interval $\sigma$
  satisfies $A\Chopomega$.  Therefore $\sigma$ is an infinite interval and has
  an infinite number of finite subintervals all satisfying $A$, including one
  starting with $\sigma$'s initial state.  From this we can readily obtain the
  valid $\PITL$ implication shown below:
  \begin{equation*}
    \Valid A\Chopomega
      \Implies (A\And \Finite);\True
        \,\;\And\,\; \Box\SDiamond ((A\And \Finite);\True)
\enspace.
  \end{equation*}
  This can be re-expressed using $\Df$ as follows:
  \begin{equation*}
    \Valid A\Chopomega \Implies \Df A \;\And\; \Box\SDiamond \Df A
\enspace.
  \end{equation*}
  The assumption that $A$ is a $\Df$-fixpoint then yields the desired validity
  of the semantically equivalent implication $A\Chopomega \implies A \And
  \Box\SDiamond A$.

\end{myproof}

The next Theorem~\ref{reduction-of-periodic-t-configs-thm} relates any
periodic transition configuration with its associated $\PITL$ formula shown
in Table~\ref{trans-config-to-pitl-table}:
\begin{mytheorem}
  \label{reduction-of-periodic-t-configs-thm}
  The next equivalence concerning a periodic transition configuration is valid:
  \begin{equation}
    \label{reduction-of-periodic-t-configs-1-eq}
    \Valid \Box T \And \alpha \And L \And \Box\SDiamond(\alpha \And L)
      \;\EQUIV\; ((\Utest T)\Chopstar \And \alpha \And L)\Chopomega
\enspace.
  \end{equation}
\end{mytheorem}
\begin{myproof}
  Lemmas~\ref{fixpoints-of-the-operator-df-lem}
  and~\ref{conditional-liveness-formulas-are-df-fixpoints-lem} ensure that the
  formula $\alpha \And L$ is itself a $\Df$-fixpoint because both $\alpha$ and
  $L$ are $\Df$-fixpoints.  Therefore
  Theorem~\ref{a-and-box-sdiamond-a-eqv-a-chopomega-valid-thm} yields the
  validity of the equivalence $\alpha \And L \And \Box\SDiamond(\alpha \And L)
  \equiv (\alpha \And L)\Chopomega$.  Now we conjoin $\Box T$ to each side of
  the equivalence.  We then use the fact that $\Box T$ and $(\Utest
  T)\Chopstar$ are semantically equivalent in infinite time (Corollary~\relax
  \ref{expressing-transitive-relexive-closure-in-ptl-for-infinite-time-cor})
  so the equivalence below is valid:
  \begin{equation*}
    \Valid \Box T \And \alpha \And L \And \Box\SDiamond(\alpha \And L)
      \;\EQUIV\; (\Utest T)\Chopstar \And (\alpha \And L)\Chopomega
\enspace.
  \end{equation*}
  Now $(\Utest T)\Chopstar \And (\alpha \And L)\Chopomega$ is an instance of
  the $\PITL$ formula $(\Utest B)\Chopstar \And C\Chopomega$ which itself is
  semantically equivalent to $((\Utest B)\Chopstar \And C)\Chopomega$.  The
  intuition here is that both of them use $\Utest B$ to test exactly all the
  two-state subintervals of the overall interval.  Finally, we use this to
  re-express $(\Utest T)\Chopstar \And (\alpha \And L)\Chopomega$ as $((\Utest
  T)\Chopstar \And \alpha \And L)\Chopomega$, thereby obtaining the validity
  of formula~\eqref{reduction-of-periodic-t-configs-1-eq}.
\end{myproof}

The following Lemma~\ref{reduction-of-bigvee-periodic-t-configs-lem}
concerning a disjunction of periodic transition configurations is needed to
justify our reduction of the satisfiability of a infinite-time transition
configuration to the associated $\PITLV$ formula shown in
Table~\ref{trans-config-to-pitl-table}:
\begin{mylemma}
  \label{reduction-of-bigvee-periodic-t-configs-lem}
  The next equivalence is valid:
  \begin{multline}
    \label{reduction-of-bigvee-periodic-t-configs-1-eq}
      \textstyle
      \Valid \bigvee_{\alpha\in\Atoms_V}(\Box T \And 
          \alpha \And L \And \Box\SDiamond(\alpha \And L))
\ifWideMargins
    \\[3pt]
        \EQUIV
\else
        \Equiv
\fi
         \bigl((\Utest T)\Chopstar \And L
           \And (\vec V\Tassign \vec V) \bigr)\Chopomega
\enspace.
  \end{multline}
\end{mylemma}
\begin{myproof}
  Theorem~\ref{reduction-of-periodic-t-configs-thm} ensures that the
  equivalence given below is valid:
  \begin{equation*}
    \textstyle
    \Valid \Box T \And \alpha \And L \And \Box\SDiamond(\alpha \And L)
      \Equiv
    ((\Utest T)\Chopstar \And \alpha \And L)\Chopomega
\enspace.
  \end{equation*}
  Simple temporal reasoning establishes that the equivalence's righthand
  operand $((\Utest T)\Chopstar \And \alpha \And L)\Chopomega$ can then be
  re-expressed as the formula $\alpha \And ((\Utest T)\Chopstar \And L \And
  (\vec V\Tassign \vec V))\Chopomega$.  Some further simple reasoning
  about the operator $\bigvee$ yields the validity of the
  equivalence~\eqref{reduction-of-bigvee-periodic-t-configs-1-eq}.
\end{myproof}

The equivalence of an infinite-time transition configuration with the
associated $\PITLV$ formula shown in Table~\ref{trans-config-to-pitl-table} is
now established:
\begin{mytheorem}
  \label{reduction-of-infinite-time-t-configs-thm}
  The following equivalence is valid for infinite-time transition
  configurations:
  \begin{multline*}
    \Valid \Box T \And \init \And \Box\SDiamond L
\ifWideMargins
    \\
    \;\EQUIV\;
\else
    \Equiv
\fi
      ((\Utest T)\Chopstar \And \init \And \Finite);
        \bigl((\Utest T)\Chopstar \And L
          \And (\vec V\Tassign \vec V) \bigr)\Chopomega
\enspace.
  \end{multline*}
\end{mytheorem}
\begin{myproof}
  This readily follows from
  Lemma~\ref{relation-between-infinite-time-and-periodic-t-configs-lem} which
  relates infinite-time transition configurations to periodic transition
  configurations and Lemma~\ref{reduction-of-bigvee-periodic-t-configs-lem}
  which re-expresses the disjunction of several periodic transition
  configurations using chop-omega.
\end{myproof}

\subsubsection{Fusion and Canonical Intervals}

Let us consider some general concepts and techniques concerning $\PITL$ and
its notion of intervals.  They will be extensively used later on.

\begin{mydefin}[Fusion]
  \label{fusion-def}
  \index{fusion}
  Let $\sigma$ and $\sigma'$ be two intervals.  The definition of the
  \emph{fusion} of them, denoted $\sigma\circ\sigma'$, has two cases,
  depending on whether $\sigma$ has finite length or not:
  \begin{itemize}
  \item If $\sigma$ has finite length, we require that last state of $\sigma$
    equals the first state of $\sigma'$.  The fusion of the $\sigma$ with
    $\sigma'$ is then the interval obtained by appending the two intervals
    together so as to include only one copy of the shared state.
  \item Otherwise, the fusion is $\sigma$ itself, no matter what $\sigma'$ is.
  \end{itemize}
\end{mydefin}
For example, suppose $s_1$, $s_2$ and $s_3$ are states.
\index{concatenation!versus fusion}
\index{fusion!versus concatenation}
If $\sigma$ is the interval $s_1 s_2$ and $\sigma'$ is the interval $s_2 s_3$,
then their fusion $\sigma\circ\sigma'$ equals the three-state interval $s_1
s_2 s_3$, rather than the four-state interval $s_1 s_2 s_2 s_3$ which
concatenation yields.  Note that when $\sigma$ has finite length and $\sigma$
and $\sigma'$ do not share the relevant state, then their fusion is undefined.
If both $\sigma$ and $\sigma'$ are finite and compatible, then the interval
$\sigma\circ\sigma'$ contains the total sum of states in $\sigma$ and
$\sigma'$ minus one.  Hence the interval length of $\sigma\circ\sigma'$ equals
the sum of the interval lengths of $\sigma$ and $\sigma'$.  Pratt first
defined fusion for describing the semantics of a process logic~\cite{Pratt79}
and called it \emph{fusion product}.

\index{chop!versus fusion}
\index{fusion!versus chop}
\index{interval!chop versus fusion}
It is worth comparing chop and fusion.  Fusion is a general operation
definable for such things as strings (i.e., sequences of letters) or intervals
(i.e., sequences of states). As used here, it starts with two suitable
intervals and joins them together.  In contrast, chop is a logical operator
which starts with an overall interval and then tests for the existence of a
way to split it into two fusible subintervals.  Furthermore, the semantics of
the chop operator can be defined using fusion, whereas fusion is for our
purposes a semantic concept, not a logical construct.

Here is a lemma relating chop with fusion:
\begin{mylemma}
  \label{chop-fusion-lem}
  A $\PITL$ formula $A;B$ is satisfiable iff there exist two intervals
  $\sigma$ and $\sigma'$ such that the fusion of them $\sigma\circ\sigma'$ is
  defined and one of the following is true:
  \begin{itemize}
  \item The interval $\sigma$ has finite length, it satisfies $A$ and
    the interval $\sigma'$ satisfies $B$.
  \item The interval $\sigma$ has infinite length and it satisfies $A$.
  \end{itemize}
\end{mylemma}
This lemma provides a way to reduce the problem of constructing an interval
satisfying $A;B$ to that of constructing intervals satisfying $A$ and $B$.

Before further reducing transition configurations involving infinite time, we
introduce the notion of \emph{canonical intervals} and discuss their use in
relating the satisfiability of chop and chop-omega formulas with
satisfiability of their operands.

The next definition of a notion of canonical states and intervals together
with the subsequent Lemma~\ref{existence-of-canonical-intervals-for-pitl-lem}
will be extensively utilised to facilitate reasoning about intervals.
\begin{mydefin}[Canonical States and Intervals]
  \label{canonical-states-and-intervals-def}
  \index{canonical!state}
  \index{canonical!interval}
  \index{state!canonical}
  \index{interval!canonical}
  For any finite set of variables $V$ and state $s$, we say that $s$ is a
  \emph{$V$-state} if $s$ assigns each variable \emph{not} in $V$ the value
  $\False$.
  
  Similarly, for any finite set of variables $V$ and interval $\sigma$, we say
  that $\sigma$ is a \emph{$V$-interval} if $\sigma$'s states all assign each
  variable \emph{not} in $V$ the value $\False$.
  
  Furthermore, for any set of variables $V$, we can denote a finite $V$-state
  by the unique $V$-atom which the state satisfies.  In addition, a
  $V$-interval can be denoted the unique sequence of $V$-atoms associated with
  its $V$-states.
\end{mydefin}
For example, for any $V$-atoms $\alpha$ and $\beta$, the two-atom sequence
$\alpha\beta$ denotes a finite $V$-interval with $V$-states denoted by
$\alpha$ and $\beta$, respectively.  Hence, $\alpha\beta\vld X$ denotes that
the two-state $V$-interval $\alpha\beta$ satisfies the formula $X$.  If $X$ is
in $\PTL_V$, then $\alpha\beta\vld X$ holds iff the conjunction $\alpha \And
\Next\beta\Test \And X$ is satisfiable.  Furthermore a single $V$-atom can be
regarded as a one-state $V$-interval. For example, $\alpha\vld X$ denotes that
the one-state $V$-interval $\alpha$ satisfies $X$.  For any $X$ in $\PTL_V$,
this is the case iff the conjunction $\alpha \And X \And \Empty$ is
satisfiable.  Similarly, the notation $\alpha\beta\alpha\vld X$ denotes that
the $V$-interval $\alpha\beta\alpha$, which has two identical states,
satisfies the formula $X$.

The next lemma ensures that any satisfiable $\PITLV$ formula is satisfied by
some $V$-interval.
\begin{mylemma}
  \label{existence-of-canonical-intervals-for-pitl-lem}
  An interval $\sigma$ satisfies a $\PITLV$ formula $A$ iff there exists a
  $V$-interval with the same number of states as $\sigma$, agrees with
  $\sigma$ on the values of the variables in $V$ and moreover satisfies $A$.
\end{mylemma}
\begin{myproof}
  Let $\sigma'$ be the $V$-interval obtained from $\sigma$ by setting all
  variables not in the set $V$ to $\False$ in each state.  The semantics in
  $\PITL$ of $A$ ignores such variables.
\end{myproof}

The following lemma employs $V$-atoms and the $\PTL$ construct $\Finite$ to
express a simple sufficient condition which ensures that any two intervals
which respectively satisfy the two parts of a chop formula with a particular
syntax given in the lemma can be fused together into an interval which
satisfies the overall chop formula.
\begin{mylemma}
  \label{chop-coverage-lem}
  For any $V$-atom $\alpha$ and $\PITLV$ formulas $A$ and $B$, the following
  are equivalent:
  \begin{itemize}
  \item[(a)] The formula $(A \And \Finite);(\alpha \And B)$ is satisfiable.
  \item[(b)] The formulas $A \And \SFin\alpha$ and $\alpha \And B$ are
    satisfiable.
  \end{itemize}
\end{mylemma}
\begin{myproof}
  \emph{$(a)\Rightarrow(b)$:} If some interval $\sigma$ satisfies the formula
  $(A \And \Finite);(\alpha \And B)$, then by the semantics of chop there
  exist two subintervals of $\sigma$ denoted here as $\sigma'$ and $\sigma''$
  such that the subinterval $\sigma'$ satisfies $A \And \Finite$ and moreover
  if $\sigma'$ has finite length, then $\sigma''$ satisfies $\alpha \And B$.
  The right subformula $\Finite$ in $A\And \Finite$ ensures that $\sigma'$ is
  indeed finite and therefore $\sigma''$ does satisfies $\alpha \And B$.
  
  \emph{$(b)\Rightarrow(a)$:} If the two formulas $A \And \SFin \alpha$ and
  $\alpha \And B$ are satisfiable, then by
  Lemma~\ref{existence-of-canonical-intervals-for-pitl-lem} some $V$-intervals
  $\sigma$ and $\sigma'$ satisfy them.  Now $\sigma$ is finite due to the
  subformula $\SFin\alpha$.  Also, the last state of $\sigma$ and the first
  state of $\sigma'$ both equal the $V$-state denoted by the $V$-atom
  $\alpha$.  Hence $\sigma$ and $\sigma'$ can be fused and the fusion
  $\sigma\circ\sigma'$ satisfies the formula $(A \And \Finite);(\alpha \And
  B)$.
\end{myproof}

\subsubsection{Periodic Models and Reduction to Finite-Time Behaviour}

The remaining material in this section deals with relating transition
configurations involving infinite time to other formulas involving periodicity
as well as to formulas about finite time.  The connections are interesting in
themselves and also later utilised.

The next Lemmas~\ref{reduce-chopomega-satisfiability-lem}
and~\ref{satisfiable-periodic-t-config-lem} help to establish small models,
decidability and axiomatic completeness for periodic transition
configurations:
\begin{mylemma}
  \label{reduce-chopomega-satisfiability-lem}
  For any $V$-atom $\alpha$ and $\PITLV$ formula $A$, 
the following are equivalent:
  \begin{itemize}
  \item [(a)] The formula $(\alpha \And A)\Chopomega$ is satisfiable.
  \item [(b)] The formula $(\alpha \And A)\Chopomega$ has a periodic model.
  \item [(c)] The formula $\alpha \And A \And \Next\SFin\alpha$ is
    satisfiable (in finite time).
  \end{itemize}
\end{mylemma}
\begin{myproof}
  \emph{$(a)\Rightarrow(c)$:} Suppose the interval $\sigma$ satisfies $(\alpha
  \And A)\Chopomega$.  We can assume each iteration of $\alpha \And A$ occurs
  in a nonempty, finite interval as expressed by the next valid equivalence:
  \begin{equation*}
    \Valid (\alpha \And A)\Chopomega
    \EQUIV (\alpha \And A\And\Finite\And\More)\Chopomega
\enspace.
  \end{equation*}
  Furthermore, each pair of adjacent iterations share a common state
  satisfying $\alpha$ and hence all have $\alpha$ true at the beginning and
  end as is captured by the following valid equivalence:
  \begin{equation*}
    \Valid (\alpha \And  A)\Chopomega
    \EQUIV (\alpha \And A \And \Finite \And \More \And \Fin\alpha)\Chopomega
\enspace.
  \end{equation*}
  Therefore the subformula $\alpha \And A \And
  \Finite \And \More \And \Fin\alpha$ is satisfiable (in finite time) and
  hence the semantically equivalent formula $\alpha \And A \And \Next\SFin
  \alpha$ is also satisfiable.
  
  \emph{$(c)\Rightarrow(b)$:} Suppose the interval $\sigma$ satisfies $\alpha
  \And A \And \Next\SFin\alpha$.  As a consequence of $\alpha$ being a
  $V$-atom and $A$ being a $\PITLV$ formula together with
  Lemma~\ref{existence-of-canonical-intervals-for-pitl-lem}, we can assume
  without loss of generality that $\sigma$ is a $V$-interval.  We then readily
  fuse $\omega$ instances of $\sigma$ together to obtain a periodic interval
  satisfying the formula $(\alpha \And A)\Chopomega$.
  
  \emph{$(b)\Rightarrow(a)$:} Clearly if some periodic interval satisfies
  $(\alpha \And A)\Chopomega$, then this formula is satisfiable.
\end{myproof}

Lemma~\ref{satisfiable-periodic-t-config-lem} shows that any satisfiable
periodic transition configuration has a periodic model.  Subsequently,
Theorem~\ref{satisfiable-infinite-time-t-config-thm} establishes that any
satisfiable infinite-time transition configuration has an ultimately periodic
model (i.e., an interval with a periodic suffix):
\begin{mylemma}
  \label{satisfiable-periodic-t-config-lem}
  For any $V$-atom $\alpha$, the following are equivalent:
  \begin{itemize}
  \item [(a)] The periodic transition configuration $\Box T \And \alpha \And L
    \And \Box\SDiamond (\alpha \And L)$ is satisfiable.
  \item [(b)] The periodic transition configuration $\Box T \And \alpha \And L
    \And \Box\SDiamond (\alpha \And L)$ has a periodic model.
  \item [(c)] The formula $(\Utest T)\Chopstar \And \alpha \And L \And
    \Next\SFin\alpha$ is satisfiable (in finite time).
  \end{itemize}
\end{mylemma}
\begin{myproof}
  Theorem~\ref{reduction-of-periodic-t-configs-thm} reduces the periodic
  transition configuration to the semantically equivalent $\PITLV$ formula
  $((\Utest T)\Chopstar \And \alpha \And L)\Chopomega$.  We then utilise
  Lemma~\ref{reduce-chopomega-satisfiability-lem}.
\end{myproof}

\begin{mylemma}
  \label{reduce-chop-chopomega-satisfiability-lem}
  For any $V$-atom $\alpha$ and $\PITLV$ formulas $A$ and $B$, the following
  are equivalent:
  \begin{itemize}
  \item [(a)] The formula $(A \And \Finite);(\alpha \And B)\Chopomega$ is
    satisfiable.
  \item [(b)] The formula $(A \And \Finite);(\alpha \And B)\Chopomega$ has an
    ultimately periodic model (i.e., an interval with a periodic suffix).
  \item [(c)] The formula $(A \And \Finite);(\alpha \And B \And
    \Next\SFin\alpha)$ is satisfiable (in finite time).
  \end{itemize}
\end{mylemma}
\begin{myproof}
  \emph{$(a)\Rightarrow(c)$:} If the formula $(A \And \Finite);(\alpha \And
  B)\Chopomega$ is satisfiable then the $\PITLV$ formula $(A \And
  \Finite);(\alpha \And B \And \Next\SFin \alpha)\Chopomega$ is also
  satisfiable.  From this readily follows the satisfiability of the formula
  $(A \And \Finite);(\alpha \And B \And \Next\SFin\alpha)$.
  
  \emph{$(c)\Rightarrow(b)$:} If the formula $(A \And \Finite);(\alpha \And B
  \And \Next\SFin\alpha)$ is satisfiable then Lemma~\ref{chop-coverage-lem}
  ensures that the two formulas $A \And \Finite \And \Fin\alpha$ and $\alpha
  \And B \And \Next\SFin\alpha$ are also satisfiable.
  Lemma~\ref{reduce-chopomega-satisfiability-lem} then yields that the formula
  $(\alpha \And B)\Chopomega$ has a periodic model.  Suppose the interval
  $\sigma$ satisfies $A \And \Finite \And \Fin\alpha$ and the interval
  $\sigma'$ is a periodic model of $(\alpha \And B)\Chopomega$.
  Lemma~\ref{existence-of-canonical-intervals-for-pitl-lem} permits us to
  assume that $\sigma$ and $\sigma'$ are $V$-intervals.  We can fuse $\sigma$
  together with $\sigma'$ to obtain an ultimately periodic model for $(A \And
  \Finite);(\alpha \And B)\Chopomega$.
  
  \emph{$(b)\Rightarrow(a)$:} Clearly if some ultimately periodic interval
  satisfies $(A \And \Finite);(\alpha \And B)\Chopomega$, then this formula is
  satisfiable.
\end{myproof}

\begin{mylemma}
  \label{existence-of-ultimately-periodic-interval-lem}
  For any $\PITLV$ formulas $A$ and $B$, the following are equivalent:
  \begin{itemize}
  \item [(a)] The formula $(A \And \Finite);(B \And (\vec V\Tassign \vec
    V))\Chopomega$ is satisfiable.
  \item [(b)] The formula $(A \And \Finite);(B \And (\vec V\Tassign \vec
    V))\Chopomega$ has an ultimately periodic model.
  \item [(c)] The formula $(A \And \Finite);(B \And \More \And \Finite \And
    (\vec V\Tassign \vec V))$ is satisfiable (in finite time).
  \end{itemize}
\end{mylemma}
\begin{myproof}
  This follows from Lemma~\ref{reduce-chop-chopomega-satisfiability-lem} and
  simple temporal reasoning involving chop and the operator $\bigvee$.  We
  also make use of the following valid equivalences concerning $\vec V\Tassign
  \vec V$, the formula $B$ and any $V$-atom $\alpha$:
    \begin{displaymath}
    \begin{myarray}
      \Valid \alpha \And B \And \Next\SFin\alpha
        \Equiv \alpha \And B \And \More \And \Finite
          \And (\vec V\Tassign\vec V) \\[2pt]
      \Valid (\alpha \And B)\Chopomega
        \Equiv \alpha \,\And\, (B \And (\vec V\Tassign\vec V))\Chopomega
\enspace.
    \end{myarray}
  \end{displaymath}
\end{myproof}

\begin{mytheorem}
  \label{satisfiable-infinite-time-t-config-thm}
  The following are equivalent:
  \begin{itemize}
  \item [(a)] The infinite-time transition configuration $\Box T \And \init
    \And \Box\SDiamond L$ is satisfiable.
  \item [(b)] The infinite-time transition configuration $\Box T \And \init
    \And \Box\SDiamond L$ has an ultimately periodic model.
  \item [(c)] The $\PITLV$ formula $\bigl((\Utest T)\Chopstar \And \init \And
    \Finite\bigr);\bigl((\Utest T)\Chopstar \And L \And \More \And \Finite \And
    (\vec V\Tassign \vec V)\bigr)$ is satisfiable (in finite time).
  \item [(d)] The $\PTLV$ formula $\Bm T \And \init \And \Diamond (L \And
    \Finite \And \More \And (\vec V\Tassign \vec V))$ is satisfiable (in
    finite time).
  \end{itemize}
\end{mytheorem}
\begin{myproof}
  We need to obtain formulas which are in a form suitable for
  Lemma~\ref{existence-of-ultimately-periodic-interval-lem}.  First of all,
  Theorem~\ref{reduction-of-infinite-time-t-configs-thm} permits us to
  re-express the infinite-time transition configuration $\Box T \And \init
  \And \Box\SDiamond L$ as the formula $((\Utest T)\Chopstar \And \init \And
  \Finite); \bigl((\Utest T)\Chopstar \And L \And (\vec V\Tassign \vec V)
  \bigr)\Chopomega$.  Recall that
  Theorem~\ref{expressing-transitive-relexive-closure-in-ptl-thm} shows the
  semantic equivalence of the formulas $\Bm T$ and $(\Utest T)\Chopstar$.
  Therefore, simple interval-based temporal reasoning ensures that formulas in
  (c) and (d) are semantically equivalent.  We complete the proof by invoking
  Lemma~\ref{existence-of-ultimately-periodic-interval-lem}.
\end{myproof}

\section{Small Models for Transition Configurations}

\label{small-models-for-trans-configs-sec}

\index{small models!for transition configuration} \index{transition
  configuration!small models} We now turn to giving upper bounds on small
models for satisfiable transition configurations.  This is later used in
Sect.~\ref{decision-procedure-sec} to construct a decision procedure for them.
Table~\ref{summary-of-bounds-for-small-models-for-trans-config-table}
summarises the upper bounds for intervals satisfying the various kinds of
transition configurations and where the results are proved.
\begin{mytable}
\centerline{\relax
    \begin{tabular}{@{}llc@{}}
      Type of transition
        &\multicolumn{1}{c}{Upper bounds}& Where \\
      configuration
        & &  proved \\\noalign{\hrule \vskip 2pt}
      Finite-time
        & Interval length less than $\size{\Atoms_V}$
        & Theorem~\ref{small-model-for-finite-time-t-config-thm}
      \\[3pt]
      Infinite-time
        & Initial part $\lt\size{\Atoms_V}$,
        & Theorem~\ref{small-model-for-infinite-time-t-config-thm}
      \\[3pt]
        & \quad Period $\le(\size{L}+1)\cdot\size{\Atoms_V}$
      \\[3pt]
      Final & Interval length is 0 & straightforward
      \\[3pt]
      Periodic & Period $\le(\size{L}+1)\cdot\size{\Atoms_V}$
        & Lemma~\ref{small-model-for-periodic-t-config-thm}
    \end{tabular}}
  \caption{Summary of upper bounds of intervals for transition configurations}
  \label{summary-of-bounds-for-small-models-for-trans-config-table}
\end{mytable}

It will be necessary to employ the fact (e.g., in
Theorem~\ref{small-model-for-finite-time-t-config-thm} and
Lemma~\ref{bounded-period-for-en-lem}) that the formula $\alpha \And (\Utest
T)\Chopstar \And \SFin \beta$ is satisfiable iff a simple variant of it is
satisfiable in an interval of bounded interval length. The following lemma
deals with this:
\begin{mylemma}
  \label{expressing-bounded-reachability-in-pitl-lem}
  For any $V$-atoms $\alpha$ and $\beta$, the formula $\alpha \And (\Utest
  T)\Chopstar \And \SFin \beta$ is satisfiable iff the formula $\alpha \And
  (\Utest T)^{\lt \size{\Atoms_V}} \And \SFin \beta$ is satisfiable.  Hence,
  the formula $\alpha \And (\Utest T)\Chopstar \And \SFin \beta$ is
  satisfiable iff it is satisfiable in an interval having interval length less
  than $\size{\Atoms_V}$.
\end{mylemma}
\begin{myproof}
  Any interval satisfying $\alpha \And (\Utest T)^{\lt \size{\Atoms_V}} \And
  \SFin \beta$ can be readily seen to also satisfy $\alpha \And (\Utest
  T)\Chopstar \And \SFin \beta$.  Let us now establish the converse by doing a
  proof by contradiction.  Suppose $\alpha \And (\Utest T)\Chopstar \And \SFin
  \beta$ is satisfiable but $\alpha \And (\Utest T)^{\lt \size{\Atoms_V}} \And
  \SFin \beta$ is not.  Let $\sigma$ be any interval which has the smallest
  length of those which satisfy $\alpha \And (\Utest T)\Chopstar \And \SFin
  \beta$.  Lemma~\ref{existence-of-canonical-intervals-for-pitl-lem} permits
  us to assume that $\sigma$ is a $V$-interval.  Now $\sigma$'s length is
  greater than or equal to $\size{\Atoms_V}$ and therefore contains at least
  $\size{\Atoms_V}+1$ states.  Consequently, some $V$-state occurs at least
  twice in $\sigma$.  Let the $V$-atom $\gamma$ denote this state.  It follows
  that $\sigma$ satisfies the following $\PITLV$ formula:
  \begin{equation*}
    \alpha \,\And\, \bigl((\Utest T)\Chopstar;\gamma\Test;(\Utest T)\Chopplus;
      \gamma\Test;(\Utest T)\Chopstar\bigr) \,\And\, \SFin \beta
\enspace.
  \end{equation*}
  Therefore $\sigma$ contains two proper subintervals $\sigma'$ and $\sigma''$
  which respectively satisfy the $\PITLV$ formulas $\alpha \And (\Utest
  T)\Chopstar \And \SFin\gamma$ and $\gamma \And (\Utest T)\Chopstar \And
  \SFin\beta$.  In addition, the last state of $\sigma'$ is the same as the
  first one of $\sigma''$ so $\sigma'$ and $\sigma''$ can be fused together.
  The fusion $\sigma'\circ\sigma''$ has length strictly less than that of
  $\sigma$ and furthermore, like $\sigma$, satisfies the formula $\alpha \And
  (\Utest T)\Chopstar \And \SFin \beta$.  But this violates the assumption
  that $\sigma$ was amongst the shortest such intervals and yields a
  contradiction.
\end{myproof}

\begin{mytheorem}
  \label{small-model-for-finite-time-t-config-thm}
  If a finite-time transition configuration $\Box T \And \init \And \Finite$
  is satisfiable, then it is satisfied by some finite interval of length less
  than $\size{\Atoms_V}$.
\end{mytheorem}
\begin{myproof}
  Theorem~\ref{reduction-of-finite-time-t-configs-thm} ensures that the
  finite-time transition configuration $\Box T \And \init \And \Finite$ is
  semantically equivalent to the formula $((\Utest T)\Chopstar \And \init \And
  \Finite);(T \And \Empty)$.  This is satisfiable iff for some $V$-atom
  $\alpha$, the formula $((\Utest T)\Chopstar \And \init \And \Finite);(\alpha
  \And T \And \Empty)$ is satisfiable.  Now Lemma~\ref{chop-coverage-lem}
  ensures that this itself is satisfiable iff the formulas $(\Utest
  T)\Chopstar \And \init \And \SFin\alpha$ and $\alpha \And T \And \Empty$ are
  both satisfiable.  By
  Lemma~\ref{expressing-bounded-reachability-in-pitl-lem}, the first of these
  is satisfiable iff the formula $(\Utest T)^{\lt \size{\Atoms_V}} \And \init
  \And \SFin \alpha$ is satisfiable.
  Lemma~\ref{existence-of-canonical-intervals-for-pitl-lem} permits us to
  assume without loss of generality that the intervals satisfying the formulas
  $(\Utest T)^{\lt \size{\Atoms_V}} \And \init \And \SFin \alpha$ and $\alpha
  \And T \And \Empty$ are $V$-intervals.  We then fuse the intervals together
  to obtain one of interval length less than $\size{\Atoms_V}$ which satisfies
  $((\Utest T)\Chopstar \And \init \And \Finite);(T \And \Empty)$ and hence
  also satisfies the semantically equivalent finite-time transition
  configuration.
\end{myproof}

The next definition is required for analysing infinite-time
configurations and makes use
of the earlier
Definitions~\ref{size-of-a-conjunction-def}--\ref{parts-of-a-conjunction-def}
concerning conjunctions and Definition~\ref{conditional-liveness-formula-def}
concerning conditional liveness formulas
\begin{mydefin}[Enabled Liveness Formula]
  \label{enabled-liveness-formula-def}
  An \emph{enabled liveness formula} $\En$ is a conjunction of $\size{\En}$
  formulas in which for each $k: 1\le k\le \size{\En}$, the subformula
  $\En[k]$ is of the form $\Dm w$, for some state formula $w$.  The state
  formulas $\theta_{\En[1]}$, \ldots, $\theta_{\En[\size{\En}]}$ denote the
  $\size{\En}$ liveness tests in $\En$ so that $\En[k]$ and $\Dm
  \theta_{\En[k]}$ refer to the same formula.
  
  For any $V$-atom $\alpha$ and conditional liveness formula $L$, we will also
  define $\En_{L,\alpha}$ to be the enabled liveness formula containing the
  $L$'s liveness tests which are enabled by $\alpha$ (recall
  Definition~\ref{conditional-liveness-formula-def}).  Let $S$ be the set of
  indices of $L$'s implications which are enabled by $\alpha$.  Then
  $\En_{L,\alpha}$ is the conjunction $\bigwedge_{j\in S} \Dm \theta_{L[j]}$.
\end{mydefin}
For example, suppose $V$ is the set $\{p,q\}$, $\alpha$ is the $V$-atom $\Not
p\And q$ and $L$ is the conditional liveness formula $((p\Or \Not q) \imp \Dm
\Not p) \And (q \imp \Dm(p \equiv \Not q)) \And (\True \imp \Dm(p \imp q))$
mentioned earlier as formula~\eqref{sample-conditional-liveness-formula-eq}.
Then $\En_{L,\alpha}$ is the conjunction $\Dm(p \equiv \Not q) \And \Dm(p \imp
q)$.

\begin{mylemma}
  \label{alpha-and-l-eqv-alpha-and-en-l-alpha-eqv-lem}
  For any $V$-atom $\alpha$ and conditional liveness formula $L$ in $\PTL_V$,
  the conjunctions $\alpha \And L$ and $\alpha \And \En_{L,\alpha}$ are
  semantically equivalent
\end{mylemma}

Not surprisingly, the hardest part of the proof of existence of small models
for infinite-time transition configurations involves finding small models for
periodic transition configurations.  Recall that
Lemma~\ref{satisfiable-periodic-t-config-lem} relates the satisfiability of
the periodic transition configuration $\Box T \And \alpha \And L \And
\Box\SDiamond (\alpha \And L)$ to that of the $\PITLV$ formula $(\Utest
T)\Chopstar \And \alpha \And L \And \Next\SFin\alpha$.  We will use the
equivalence of $\alpha \And L$ and $\alpha \And \En_{L,\alpha}$ to assist in
the analysis of bounded models of $(\Utest T)\Chopstar \And \alpha \And L \And
\Next\SFin\alpha$.  These can then be used to obtain a bounded periodic model
for the original periodic transition configuration.
\begin{mylemma}
  \label{alpha-and-l-and-sfin-alpha-eqv-alpha-and-en-and-sfin-alpha-valid-lem}
  For any $V$-atom $\alpha$ and conditional liveness formula $L$ in $\PTL_V$,
  the following equivalence is valid:
  \begin{equation*}
    \Valid (\Utest T)\Chopstar \And \alpha \And L \And \Next\SFin\alpha
      \;\EQUIV\;
        (\Utest T)\Chopstar \And \alpha \And \En_{L,\alpha}
          \And \Next\SFin\alpha
\enspace.
  \end{equation*}
\end{mylemma}
\begin{myproof}
  This readily follows from the earlier
  Lemma~\ref{alpha-and-l-eqv-alpha-and-en-l-alpha-eqv-lem} concerning the
  semantic equivalence of the formulas $\alpha \And L$ and $\alpha \And
  \En_{L,\alpha}$.
\end{myproof}

The next Lemma~\ref{bounded-period-for-en-lem} shortens the nonempty,
finite model expressed by the formula $(\Utest T)\Chopstar \And \alpha \And
\En \And \Next\SFin\alpha$ to one having a bounded length by adapting the
technique presented earlier in
Lemma~\ref{expressing-bounded-reachability-in-pitl-lem} concerning a bounded
model for the formula $(\Utest T)\Chopstar \,\And\, \alpha \And \SFin \beta$.
\begin{mylemma}
  \label{bounded-period-for-en-lem}
  For any $V$-atom $\alpha$ and enabled liveness formula $\En$ in $\PTL_V$, if
  the formula $(\Utest T)\Chopstar \And \alpha \And \En \And \Next\SFin\alpha$
  is satisfiable, then it is satisfied by a interval having interval length at
  most $(\size{\En}+1)\,\size{\Atoms_V}$.
\end{mylemma}
\begin{myproof}
  If the formula $(\Utest T)\Chopstar \And \alpha \And \En \And
  \Next\SFin\alpha$ is satisfiable, then by
  Lemma~\ref{existence-of-canonical-intervals-for-pitl-lem} there exists some
  satisfying $V$-interval.  We can fuse $\size{\En}+1$ copies of this interval
  together to obtain a $V$-interval $\sigma$ which satisfies the formula
  $\bigl((\Utest T)\Chopstar \And \alpha \And \En \And
  \Finite\bigr)^{\size{\En}+1} \And \Next\SFin\alpha$.  It is not hard to
  check than $\sigma$ itself satisfies the original formula $(\Utest
  T)\Chopstar \And \alpha \And \En \And \Next\SFin\alpha$ since each liveness
  test in $\En$ is satisfied somewhere in $\sigma$ prior to the last state.
  Furthermore, there exist a sequence of $\size{\En}$ $V$-atoms $\gamma_1,
  \ldots, \gamma_{\size{\En}}$ such that for each $j:1 \le j\le \size{\En}$,
  the state formula $\gamma_j \And \theta_{\En[j]}$ is satisfied by some state
  prior to the last one and the $V$-interval $\sigma$ satisfies the next
  formula:
  \begin{equation*}
    \alpha \;\And\; \bigl((\Utest T)\Chopstar;\gamma_1\Test;
      \ldots;
      (\Utest T)\Chopstar;
      \gamma_{\size{\En}}\Test;(\Utest T)\Chopplus\bigr)
      \;\And\; \Next\SFin\alpha
\enspace.
  \end{equation*}
  If a gap between two of the $\size{\En}$ selected states satisfying their
  respective liveness tests has interval length of at least $\size{\Atoms_V}$,
  then within the gap, some state occurs twice.  Such a gap can then be
  shortened in the manner of
  Lemma~\ref{expressing-bounded-reachability-in-pitl-lem}.  By means of this
  we obtain from the $V$-interval $\sigma$ another $V$-interval having bounded
  length and satisfying the formula below:
  \begin{equation*}
    \begin{myarray}
    \alpha \;\And\; \bigl((\Utest T)^{\lt\size{\Atoms_V}};\gamma_1\Test;
      \ldots;
      (\Utest T)^{\lt\size{\Atoms_V}};
      \gamma_{\size{\En}}\Test;
      (\Utest T)^{\le\size{\Atoms_V}}\bigr)
   \\[3pt]
     \hphantom{\alpha}
      \;\And\; \Next\SFin\alpha
\enspace.
  \end{myarray}
\end{equation*}
  The resulting new interval is nonempty and has interval length not exceeding
  $(\size{\En}+1)\,\size{\Atoms_V}$.  Moreover it still
  satisfies $(\Utest T)\Chopstar \And \alpha \And \En \And \Next\SFin\alpha$.
\end{myproof}

\begin{mylemma}
  \label{bounded-period-for-l-with-alpha-lem}
  If the formula $(\Utest T)\Chopstar \And \alpha \And L \And
  \Next\SFin\alpha$ is satisfiable, then it is satisfiable on a finite,
  nonempty interval with interval length at most
  $(\size{L}+1)\,\size{\Atoms_V}$.
\end{mylemma}
\begin{myproof}
  From Lemma~\ref{bounded-period-for-en-lem} we have that if the formula
  $(\Utest T)\Chopstar \And \alpha \And \En_{L,\alpha} \And \Next\SFin\alpha$
  is satisfiable, then it is satisfiable on a finite, nonempty interval having
  interval length at most $(\size{\En_{L,\alpha}}+1)\,\size{\Atoms_V}$.
  Lemma~\ref{alpha-and-l-eqv-alpha-and-en-l-alpha-eqv-lem} ensures that the
  conjunctions $\alpha \And L$ and $\alpha \And \En_{L,\alpha}$ are
  semantically equivalent.  In addition, we have $\size{\En_{L,\alpha}}\le
  \size{L}$.  Therefore, if the formula $(\Utest T)\Chopstar \And \alpha \And
  L \And \Next\SFin\alpha$ is satisfiable, then it is satisfiable on a finite,
  nonempty interval with interval length at most
  $(\size{L}+1)\,\size{\Atoms_V}$.
\end{myproof}


\begin{mylemma}
  \label{small-model-for-periodic-t-config-thm}
  If the periodic transition configuration $\Box T \And \alpha \And L \And
  \Box\SDiamond (\alpha \And L)$ is satisfiable, then it is satisfied by a
  periodic interval with period of interval length at most
  $(\size{L}+1)\,\size{\Atoms_V}$.
\end{mylemma}
\begin{myproof}
  Lemma~\ref{satisfiable-periodic-t-config-lem} ensures that if the periodic
  transition configuration is satisfiable, then the formula $(\Utest
  T)\Chopstar \And \alpha \And L \And \Next\SFin\alpha$ is satisfiable.  By
  Lemma~\ref{bounded-period-for-l-with-alpha-lem}, if this is satisfiable,
  then it has a satisfying interval having interval length at most
  $(\size{L}+1)\,\size{\Atoms_V}$.
  Lemma~\ref{existence-of-canonical-intervals-for-pitl-lem} permits us to
  assume without loss of generality that the interval is a $V$-interval.  We
  can then fuse $\omega$ copies of it together to obtain a periodic interval
  which has a period with interval length at most
  $(\size{L}+1)\,\size{\Atoms_V}$ and also satisfies the formula $((\Utest
  T)\Chopstar \And \alpha \And L)\Chopomega$.
  Theorem~\ref{reduction-of-periodic-t-configs-thm} establishes that this
  formula is equivalent to the original periodic transition configuration.
\end{myproof}

\begin{mytheorem}
  \label{small-model-for-infinite-time-t-config-thm}
  If the infinite-time transition configuration $\Box T \And \init \And
  \Box\SDiamond L$ is satisfiable, then it is satisfied by an ultimately
  periodic interval consisting of an initial segment having interval length
  less than $\size{\Atoms_V}$ fused with a periodic interval having a period
  with interval length of at most $(\size{L}+1)\,\size{\Atoms_V}$.
\end{mytheorem}
\begin{myproof}
  If some interval satisfies the formula $\Box T \And \init \And \Box\SDiamond
  L$, then
  Lemma~\ref{relation-between-infinite-time-and-periodic-t-configs-lem}
  ensures that the interval also satisfies the next semantically equivalent
  formula:
  \begin{equation}
    \label{bounded-ultimately-periodic-config-1-eq}
    \textstyle
    ((\Utest T)\Chopstar \And \init \And \Finite);
      \bigvee_{\alpha\in\Atoms_V}
        (\Box T \And \alpha \And L \And \Box\SDiamond(\alpha \And L))
\enspace.
  \end{equation}
  Lemma~\ref{chop-coverage-lem} and simple temporal reasoning establish that
  for some $V$-atom $\alpha$ the two formulas $(\Utest T)\Chopstar \And \init
  \And \SFin\alpha$ and $\Box T \And \alpha \And L \And \Box\SDiamond(\alpha
  \And L)$ are satisfiable.  By
  Lemma~\ref{expressing-bounded-reachability-in-pitl-lem}, the first formula
  is satisfiable in some interval $\sigma$ having interval length less than
  $\size{\Atoms_V}$.  Lemma~\ref{small-model-for-periodic-t-config-thm} yields
  some periodic interval $\sigma'$ which satisfies the second formula and
  possesses a period with interval length of at most
  $(\size{L}+1)\,\size{\Atoms_V}$.
  Lemma~\ref{existence-of-canonical-intervals-for-pitl-lem} permits us to
  assume that $\sigma$ and $\sigma'$ are $V$-intervals.  Therefore the last
  state of $\sigma$ is the same as the first one of $\sigma'$ since both
  states satisfy $\alpha$.  The fusion $\sigma\circ\sigma'$ is itself
  ultimately periodic and satisfies the
  formula~\eqref{bounded-ultimately-periodic-config-1-eq}.  Hence it also
  satisfies the semantically equivalent original infinite-time transition
  configuration $\Box T \And \init \And \Box\SDiamond L$ as well.  In
  addition, the interval $\sigma\circ\sigma'$ has an initial segment having
  interval length less than $\size{\Atoms_V}$ fused with a periodic interval
  with period of interval length at most $(\size{L}+1)\,\size{\Atoms_V}$.
\end{myproof}

\section{Decomposition of Transition Configurations}

\label{decomposition-of-transition-configurations-sec}

We now prove the two Theorems~\ref{finite-time-config-decompose-thm}
and~\ref{infinite-time-config-decompose-thm} which respectively relate the
satisfiability of finite-time and infinite-time transition configurations with
simple interval-oriented tests involving finite time.  These theorems are
later used in Sect.~\ref{decision-procedure-sec} as part of the justification
of the our $\PTL$ decision procedure and in Sect.~\ref{ptl-axiom-system-sec}
as part of the completeness proof of an axiom system for $\PTL$.

\begin{mytheorem}[Decomposing Finite-Time Transition Configurations]
  \label{finite-time-config-decompose-thm}
  The following are equivalent:
  \begin{itemize}
  \item[(a)] The finite-time configuration $\Box T \And \init \And \Finite$ is
    satisfiable.
  \item[(b)] For some $V$-atoms $\alpha$ and $\beta$, the three formulas below
    are satisfiable:
    \begin{displaymath}
      \alpha \And \init
        \qquad (\Utest T)\Chopstar \And \alpha \And \SFin\beta
        \qquad T \And \beta \And \Empty
\enspace.
    \end{displaymath}
  \end{itemize}
\end{mytheorem}
\begin{myproof}
  Theorem~\ref{reduction-of-finite-time-t-configs-thm} ensures
  that the finite-time configuration is semantically equivalent to the next
  $\PITLV$ formula:
  \begin{equation*}
    ((\Utest T)\Chopstar \And \init \And \Finite);(T \And \Empty)
\enspace.
  \end{equation*}
  Now simple interval-based reasoning guarantees that this is satisfiable iff
  for some $V$-atoms $\alpha$ and $\beta$, the next formula is satisfiable:
  \begin{equation*}
    ((\Utest T)\Chopstar \And \alpha \And \init \And \Finite);
      (T \And \beta \And \Empty)
\enspace.
  \end{equation*}
  Lemma~\ref{chop-coverage-lem} ensures that this is itself satisfiable iff
  the next two formulas are:
  \begin{equation*}
    (\Utest T)\Chopstar \And \alpha \And \init \And \SFin\beta
    \qquad T \And \beta \And \Empty
\enspace.
  \end{equation*}
  Finally, simple temporal reasoning ensures that the first of these is itself
  is satisfiable iff the following two formulas are satisfiable:
  \begin{equation*}
    \alpha \And \init
    \qquad (\Utest T)\Chopstar \And \alpha \And \SFin\beta
\enspace.
  \end{equation*}
\end{myproof}

We now turn to decomposing an infinite-time transition configuration:
\begin{mylemma}
  \label{infinite-time-config-reduction-with-atoms-lem}
  The infinite-time transition configuration $\Box T \And \init \And
  \Box\SDiamond L$ is satisfiable iff for some $V$-atoms $\alpha$ and $\beta$,
  the following formulas are satisfiable:
  \begin{equation}
    \label{infinite-time-config-reduction-with-atoms-1-eq}
      (\Utest T)\Chopstar \And \alpha \And \init \And \SFin\beta
      \qquad (\Utest T)\Chopstar \And \beta \And \En_{L,\beta}
        \And \Next\SFin\beta
\enspace.
  \end{equation}
\end{mylemma}
\begin{myproof}
  Theorem~\ref{satisfiable-infinite-time-t-config-thm} ensures that the
  infinite-time configuration is satisfiable iff the next $\PITLV$ formula is
  satisfiable:
  \begin{equation*}
    \bigl((\Utest T)\Chopstar \And \init \And \Finite\bigr);
      \bigl((\Utest T)\Chopstar \And L \And \More \And \Finite \And
        (\vec V\Tassign \vec V)\bigr)
\enspace.
  \end{equation*}
  Simple interval-based temporal reasoning ensures that this itself is
  satisfiable iff for some $V$-atoms $\alpha$ and $\beta$, next formula is
  satisfiable:
  \begin{equation}
    \label{infinite-time-config-reduction-with-atoms-2-eq}
    \bigl((\Utest T)\Chopstar \And \alpha \And \init \And \Finite\bigr);
    \bigl((\Utest T)\Chopstar \And \beta \And L \And \Next\SFin\beta\bigr)
\enspace.
  \end{equation}
  Now Lemma~\ref{alpha-and-l-eqv-alpha-and-en-l-alpha-eqv-lem} guarantees the
  semantic equivalence of the conjunctions $\beta \And L$ and $\beta \And
  \En_{L,\beta}$. We therefore can replace $L$ by $\En_{L,\beta}$ in
  formula~\eqref{infinite-time-config-reduction-with-atoms-2-eq}.  Finally,
  Lemma~\ref{chop-coverage-lem} yields that the resulting formula is itself
  satisfiable iff the two formulas
  in~\eqref{infinite-time-config-reduction-with-atoms-1-eq} are satisfiable.
\end{myproof}

The next lemma concerning enabled liveness formulas is shortly used in
Theorem~\ref{infinite-time-config-decompose-thm} to analyse the satisfiability
of infinite-time configurations:
\begin{mylemma}
  \label{enabled-liveness-formula-decompose-lem}
  For any $V$-atom $\alpha$ and enabled liveness formula $\En$, the following
  are equivalent:
  \begin{itemize}
  \item[(a)] The formula $(\Utest T)\Chopstar \And \alpha \And \En \And
    \Next\SFin\alpha$ is satisfiable.
  \item[(b)] For some $\size{\En}$ $V$-atoms $\gamma_1$, \ldots,
    $\gamma_{\size{\En}}$ (not necessarily distinct), the following are all
    satisfiable:
    \begin{displaymath}
      \begin{myarray}
        (\Utest T)\Chopstar \And \alpha \And \Next\SFin\alpha \\[3pt]
        \text{for each } \gamma_i\colon
          \enskip (\Utest T)\Chopstar \And \alpha \And \SFin\gamma_i \quad
          \gamma_i \And \theta_{\En_{L,\alpha}[i]} \quad
          (\Utest T)\Chopstar \And \gamma_i \And \SFin\alpha
\enspace.
      \end{myarray}
    \end{displaymath}
  \end{itemize}
\end{mylemma}
\begin{myproof}
  Induction on the length of $\En$ and simple interval-based reasoning can be
  used to demonstrate that the formula $(\Utest T)\Chopstar \And \alpha \And
  \En \And \Next\SFin\alpha$ is satisfiable iff the formula $(\Utest
  T)\Chopstar \And \alpha \And \Next\SFin\alpha$ is satisfiable and
  also for some $V$-atoms $\gamma_1$, \ldots, $\gamma_{\size{\En}}$,
  for each $\gamma_i$ the following formula is satisfiable:
  \begin{equation}
  \label{enabled-liveness-formula-decompose-1-eq}
    (\Utest T)\Chopstar \And \alpha
      \And \Diamond(\gamma_i \And \theta_{\En[i]})
      \And \SFin\alpha
\enspace.
  \end{equation}
  This guarantees that for each liveness test $\theta_{\En[i]}$ in $\En$, the
  $V$-atom $\alpha$ can reach some $V$-atom $\gamma_i$ which satisfies
  $\theta_{\En[i]}$ and this $V$-atom $\gamma_i$ itself can reach back to
  $\alpha$.  We can re-express~\eqref{enabled-liveness-formula-decompose-1-eq}
  as the semantically equivalent formula below:
  \begin{equation*}
    \bigl((\Utest T)\Chopstar \And \alpha \And \Finite\bigr);
      \bigl((\Utest T)\Chopstar \And \gamma_i \And \theta_{\En[i]}
        \And \SFin\alpha\bigr)
\enspace.
  \end{equation*}
  Lemma~\ref{chop-coverage-lem} ensures that this is satisfiable iff the next
  two formulas are:
  \begin{equation*}
    (\Utest T)\Chopstar \And \alpha \And \SFin\gamma_i
    \qquad (\Utest T)\Chopstar \And \gamma_i \And \theta_{\En[i]}
      \And \SFin\alpha
\enspace.
  \end{equation*}
  The second one is satisfiable iff the two formulas shown below are
  satisfiable:
  \begin{displaymath}
     \gamma_i \And \theta_{\En[i]}
     \qquad (\Utest T)\Chopstar \And \gamma_i \And \SFin\alpha
\enspace.
  \end{displaymath}
\end{myproof}

\begin{mytheorem}[Decomposing Infinite-Time Transition Configurations]
  \label{infinite-time-config-decompose-thm}
  The following are equivalent:
  \begin{itemize}
  \item[(a)] The infinite-time configuration $\Box T \And \init \And
    \Box\SDiamond L$ is satisfiable.
  \item[(b)] For some $V$-atoms $\alpha$, $\beta$ and $\gamma_1$, \ldots,
    $\gamma_{\size{\En_{L,\beta}}}$ (not necessarily distinct), the following
    are all satisfiable:
    \begin{displaymath}
      \begin{myarray}
        \alpha \And \init
        \qquad
        (\Utest T)\Chopstar \And \alpha \And \SFin\beta
        \qquad
        (\Utest T)\Chopstar \And \beta \And \Next\SFin\beta \\[3pt]
        \text{for each } \gamma_i\colon
          \quad (\Utest T)\Chopstar \And \beta \And \SFin\gamma_i \quad
          \gamma_i \And \theta_{\En_{L,\beta}[i]} \quad
          (\Utest T)\Chopstar \And \gamma_i \And \SFin\beta
\enspace.
      \end{myarray}
    \end{displaymath}
  \end{itemize}
\end{mytheorem}
\begin{myproof}
  Lemma~\ref{infinite-time-config-reduction-with-atoms-lem} establishes that
  the infinite-time configuration $\Box T \And \init \And \Box\SDiamond L$ is
  satisfiable iff there exist some $V$-atoms $\alpha$ and $\beta$ for which
  the next two formulas are satisfiable:
  \begin{equation}
    \label{infinite-time-config-decompose-2-eq}
      (\Utest T)\Chopstar \And \alpha \And \init \And \SFin\beta
      \qquad (\Utest T)\Chopstar \And \beta \And \En_{L,\beta}
        \And \Next\SFin\beta
\enspace.
  \end{equation}
  Now simple temporal reasoning ensures that the first of these is itself is
  satisfiable iff the following two formulas are satisfiable:
  \begin{equation*}
    \alpha \And \init
    \qquad (\Utest T)\Chopstar \And \alpha \And \SFin\beta
\enspace.
  \end{equation*}
  Furthermore, Lemma~\ref{enabled-liveness-formula-decompose-lem} guarantees
  that the second formula in~\eqref{infinite-time-config-decompose-2-eq} is
  satisfiable iff the formula $(\Utest T)\Chopstar \And \beta \And
  \Next\SFin\beta$ is satisfiable and furthermore for some $V$-atoms
  $\gamma_1$, \ldots, $\gamma_{\size{\En_{L,\beta}}}$ (not necessarily
  distinct), the following are all satisfiable for each $\gamma_i$:
  \begin{displaymath}
    (\Utest T)\Chopstar \And \beta \And \SFin\gamma_i
     \qquad \gamma_i \And \theta_{\En_{L,\beta}[i]}
     \qquad (\Utest T)\Chopstar \And \gamma_i \And \SFin\beta
\enspace.
  \end{displaymath}
\end{myproof}

\section{A Decision Procedure}

\label{decision-procedure-sec}

\index{PTL!decision procedure for transition configuration|(}
\index{decision procedure|(}
\index{transition configuration!decision procedure|(}
We now describe a decision procedure for finite-time and infinite-time
transition configurations based on \index{BDDs} \index{Binary Decision
Diagrams} Binary Decision Diagrams (BDDs) \cite{Bryant86,Bryant92} which
provide an efficient basis for performing many computational tasks involving
reductions to reasoning about formulas in propositional logic.  We had little
difficultly implementing the decision procedure using the popular Colorado
University Decision Diagram Package (CUDD)\ifkcp\else~\fi\cite{CUDD}
developed by Somenzi.  Our prototype tool consists of a front-end coded in the
CLISP~\cite{CLISP} implementation of Common Lisp~\cite{ANSICommonLisp} as well
as a back-end coded in Perl~\cite{Perl}.  The back-end employs a Perl-oriented
interface to CUDD written by Somenzi and called PerlDD~\cite{PerlDD}.  The
front-end accepts arbitrary $\PTL$ formulas and converts them to transition
configurations using methods later described in
Sections~\ref{invariants-and-related-formulas-sec}
and~\ref{dealing-with-arbitrary-ptl-formulas-sec}.  The transition
configurations are then passed to the back-end which analyses them using BDDs.
In this section we describe the basis for performing this analysis.

The remainder of this section assumes that the reader already has some
familiarity with BDDs.

Our algorithm for finite-time transition configurations adapts methods for
\index{symbolic state space traversal} \emph{symbolic state space traversal}
described by Coudert, Berthet and Madre
\cite{CoudertBerthet89,CoudertBerthet89a,CoudertMadre90} (see also
Kropf~\cite{Kropf99,ClarkeGrumberg00}) for use with BDD-based representations
of formulas in propositional logic.  It simultaneously greatly benefits from
closely related methods first employed by McMillan in symbolic model
checking~\cite{McMillan93,BurchClarke92,ClarkeGrumberg00} which also include
the automatic generation of counterexamples for unsatisfiable formulas and,
similarly, witnesses for satisfiable ones.  Recall that
Theorem~\ref{finite-time-config-decompose-thm} shows that the finite-time
transition configuration $\Box T \And \init \And \Finite$ is satisfiable iff
for some $V$-atoms $\alpha$ and $\beta$, the next three formulas are
satisfiable:
\begin{center}
  $\alpha \And \init$
  \qquad
  $(\Utest T)\Chopstar \And \alpha \And \SFin\beta$
  \qquad
  $T \And \beta \And \Empty$\relax
\enspace.
\end{center}
We can readily search for suitable $V$-atoms using BDDs.  Three BDDs
$\Gamma_1$, $\Gamma_2$ and $\Gamma_3$ are initially constructed.  In what
follows, please recall the notion $\sat X$ introduced in
Definition~\ref{satisfiability-and-validity-def} to denote that the formula
$X$ is satisfiable.  We first describe the roles of the BDDs $\Gamma_1$,
$\Gamma_2$ and $\Gamma_3$ before actually constructing them:
\begin{itemize}
\item The BDD $\Gamma_1$ represents the state formula $\init$ and hence the
  set of $V$-atoms satisfying $\init$ (i.e., the set $\{\alpha \in
  \Atoms_V\colon \alpha \vld \init \}$).  This is the same as the set
  $\{\alpha \in \Atoms_V\colon \sat \alpha\And \init \}$.
\item The second BDD $\Gamma_2$ captures all pairs of $V$-atoms corresponding
  to unit (i.e., two-state) intervals satisfying $T$.  In other words, it
  corresponds to the set $\{\langle \alpha,\beta \rangle \in \Atoms_V^2\colon
  \alpha\beta\vld T \}$.  This is the same as the set $\{ \langle \alpha,\beta
  \rangle \in \Atoms_V^2\colon \sat T \And \alpha \And \Skip \And \SFin\beta
  \}$.
\item The third BDD $\Gamma_3$ captures the behaviour of $T$ in an empty
  interval.  Therefore $\Gamma_3$ represents the set of all $V$-atoms
  satisfying the formula $T \And \Empty$ (i.e., the set $\{\alpha \in
  \Atoms_V\colon \alpha \vld T \}$).  This is the same as the set $\{\alpha
  \in \Atoms_V\colon \sat T \And \alpha \And \Empty \}$
\end{itemize}

In the course of manipulating the BDDs we make use of two finite sets of
propositional variables.  They include the original ones (e.g., $p$, $r_1$,
\ldots, $r_4$) as well as primed versions (e.g., $p'$, $r'_1$, \ldots,
$r'_4$).  For convenience, we often do not distinguish between a BDD and the
propositional logic formula it represents.

Let $V$ and $V'$ respectively denote the two sets of variables.  We
now construct the BDDs $\Gamma_1$, $\Gamma_3$ and $\Gamma_2$ as follows:
\begin{itemize}
\item Let $\Gamma_1$ be the formula $\init$.
\item Obtain $\Gamma_2$ from the formula $T$ by replacing all variables in the
  scope of any $\Next$ constructs by corresponding ones in $V'$ and then
  deleting all $\Next$ operators (but not the associated operands) to obtain a
  formula in conventional propositional logic.  We refer to this process of
  constructing $\Gamma_2$ from $T$ by the term \emph{flattening}.
\item Obtain $\Gamma_3$ from the formula $T$ by replacing each $\Next$
  construct by $\False$.
\end{itemize}
The BDDs $\Gamma_1$ and $\Gamma_3$ both only can contain variables in $V$
whereas $\Gamma_2$ can contain variables in $V$ and $V'$.

Suppose $T$ and $\init$ are the following formulas mentioned earlier in
\S\ref{example-of-the-hierarchical-process-subsec}:
\begin{equation*}
    T\colon\enskip
      \begin{myarray}
        (r_1\equiv (p \Or \Next r_1))
        \And (r_2\equiv (\Not r_1 \Or \Next r_2)) \\
        \null \And (r_3\equiv (\Not p\Or \Next r_3))
        \And (r_4\equiv (\Not r_3 \Or \Next r_4))
      \end{myarray}
    \qquad
    \init\colon\enskip \Not r_2 \And \Not r_4
\enspace.
\end{equation*}

Here are the associated $\Gamma_1$, $\Gamma_2$ and $\Gamma_3$ for these $T$
and $\init$:
\begin{equation*}
  \begin{array}{ll}
    \Gamma_1\colon\enskip & \Not r_2 \And \Not r_4 \\[5pt]
    \Gamma_2\colon\enskip
      &  (r_1\equiv (p \Or r'_1))
        \And (r_2\equiv (\Not r_1 \Or r'_2)) \\
      &  \null \And (r_3\equiv (\Not p \Or r'_3))
        \And (r_4\equiv (\Not r_3 \Or r'_4))
    \\[5pt]
    \Gamma_3\colon\enskip &
        (r_1\equiv (p \Or \False))
        \And (r_2\equiv (\Not r_1 \Or \False)) \\
      &  \null \And (r_3\equiv (\Not p \Or \False))
        \And (r_4\equiv (\Not r_3 \Or \False))
\enspace.
  \end{array}
\end{equation*}

The connection between the BDDs for $\Gamma_1$ and $\Gamma_3$ and the
previously mentioned sets of $V$-atoms they are meant to capture is
straightforward.  In order to justify the less intuitive relationship between
the construction for $\Gamma_2$ and the earlier associated set of pairs of
$V$-atoms, we shortly present
Lemma~\ref{reduction-of-sat-t-and-alpha-and-skip-and-sfin-beta-lem} relating
$\Gamma_2$ with $T$.  However, the following lemma concerning $\NLone$
formulas is first given since it is used in the proof of
Lemma~\ref{reduction-of-sat-t-and-alpha-and-skip-and-sfin-beta-lem}.
\begin{mylemma}
  \label{sat-nlone-iff-nlone-and-skip-lem}
  The following are equivalent for any $\NLone$ formula $T$:
  \begin{itemize}
  \item[(a)] The formula $T$ is satisfiable in some nonempty interval.
  \item[(b)] The formula $\Skip \And T$ is satisfiable.
\end{itemize}
\end{mylemma}
\begin{myproof}
  \emph{$(a)\Rightarrow(b)$:} Suppose some nonempty interval $\sigma$
  satisfies the formula $T$.  Now $\sigma$ contains at least two states.  Let
  $\sigma'$ denote the subinterval consisting the first two states in
  $\sigma$. Now $\sigma'$ satisfies the formula $\Skip$. Furthermore, the
  formula $T$ is in $\NLone$.  Lemma~\ref{nlone-first-two-states-lem}
  consequently ensures that the interval $\sigma'$, like $\sigma$, satisfies
  the formula $T$ because both two intervals share the same first two states.
  Therefore $\sigma'$ satisfies the formula $\Skip \And T$.
  
  \emph{$(b)\Rightarrow(a)$:} If some interval $\sigma$ satisfies the $\PTL$
  formula $\Skip \And T$, then $\sigma$ is clearly nonempty and also satisfies
  $T$.
\end{myproof}

\begin{mylemma}
  \label{reduction-of-sat-t-and-alpha-and-skip-and-sfin-beta-lem}
  For any $V$-atoms $\alpha$ and $\beta$, the following are
  equivalent:
  \begin{itemize}
  \item[(a)] The formula $T \And \alpha \And \Skip \And \SFin\beta$ is
    satisfiable (i.e., $\alpha\beta\vld T$).
  \item[(b)] The propositional logic formula $\Gamma_2 \And \alpha \And
    \beta_V^{V'}$ is satisfiable.
\end{itemize}
\end{mylemma}
\begin{myproof}
  \emph{$(a)\Rightarrow(b)$:} Suppose the formula $T \And \alpha \And \Skip
  \And \SFin\beta$ is satisfiable. Then the flattening of $T$ into $\Gamma_2$
  readily yields that the formula $\Gamma_2 \And \alpha \And \beta_V^{V'}$
  is satisfiable.
    
  \emph{$(b)\Rightarrow(a)$:} If the propositional logic formula $\Gamma_2
  \And \alpha \And \beta_V^{V'}$ is satisfiable, then the flattening of
  $\Next$ constructs in $\Gamma_2$ readily yields that the $\NLone$ formula $T
  \And \alpha \And \Next \beta$ is satisfiable.  Clearly any interval
  satisfying it has at least two states.  Hence by the previous
  Lemma~\ref{sat-nlone-iff-nlone-and-skip-lem} the formula $\Skip \And T \And
  \alpha \And \Next \beta$ is satisfiable.  Simple temporal reasoning then
  ensures that the semantically equivalent formula $T \And \alpha \And \Skip
  \And \SFin\beta$ is also satisfiable.
\end{myproof}

We use $\Gamma_2$ together with the first BDD $\Gamma_1$ to iteratively
calculate a sequence of BDDs $\Delta_0$, \ldots, $\Delta_k$, \ldots so that
for any $k$, $\Delta_k$ describes all $V$-atoms which can be reached from one
which satisfies $\init$ in exactly $k$ steps.  In other words, $\Delta_k$
represents the following set:
\begin{equation*}
  \{\beta\in \Atoms_V\colon \text{for some } \alpha\in\Atoms_V,
    \sat (\Utest T)^k \And \alpha \And \init \And \SFin \beta \}
\enspace.
\end{equation*}
We set $\Delta_0$ to be $\Gamma_1$.  Therefore, every variable in
$\Delta_0$ is in $V$.  Each $\Delta_{k+1}$ is calculated to be semantically
equivalent to the next quantified propositional logic formula in which
renaming ensures that all free variables are in $V$:
\begin{equation}
  \label{bdd-exists-and-eq}
  \bigl(\Exists{V}(\Delta_k \And \Gamma_2)\bigr)_{V'}^V
\enspace.
\end{equation}
Due to the final renaming, the sole variables left in the BDD $\Delta_{k+1}$
itself are elements of $V$.  The only BDD operations required to calculate
$\Delta_{k+1}$ from~\eqref{bdd-exists-and-eq} are logical-and, existential
quantification (which actually yields a BDD representing a semantically
equivalent quantifier-free formula) and renaming which are all standard ones.

\begin{myremark}
  Within the CUDD system, the entire calculation for obtaining
  $\Exists{V}(\Delta_k \And \Gamma_2)$ can even be done by a single CUDD
  operation tailored to handle this specific kind of common BDD manipulation.
  Furthermore, the renaming of variables in $V'$ to those in $V$ is actually
  achieved by taking the BDD obtained for $\Exists{V}(\Delta_k \And \Gamma_2)$
  and then performing a single CUDD operation which yields another BDD in
  which the variables in $V$ are swapped with the corresponding ones in $V'$.
\end{myremark}

For any given $\Delta_k$ which has been calculated, we next determine the
logical-and of $\Gamma_3$ and $\Delta_k$ and then proceed as follows:
\begin{enumerate}
\item If the logical-and is not false, then there is some $V$-atom $\beta$
  satisfying $T \And \Empty$ which can be reached in $k$ steps from a $V$-atom
  $\alpha$ satisfying $\init$.  Therefore the next three formulas are all
  satisfiable:
  \begin{displaymath}
    \alpha \And \init
    \qquad (\Utest T)^k \And \alpha \And \SFin\beta
    \qquad T \And \beta \And \Empty
\enspace.
  \end{displaymath}
  Now the second formula ensures the satisfiability of the formula $(\Utest
  T)\Chopstar \And \alpha \And \SFin\beta$.  Therefore
  Theorem~\ref{finite-time-config-decompose-thm} can be invoked to obtain the
  satisfiability of the original finite-time transition configuration $\Box T
  \And \init \And \Finite$.  We therefore do not need to calculate any further
  $\Delta_k$'s.
\item Otherwise, the logical-and is false so we must continue to iterate.
\end{enumerate}
During the iteration process, we maintain a BDD representing the set of all
$V$-atoms so far reachable from one satisfying $\init$.  This BDD corresponds
to the formula $\bigvee_{0\le i\le k} \Delta_i$ which equals the next set:
\begin{equation*}
  \begin{myarray}
    \{\beta\in \Atoms_V\colon \text{for some }\alpha\in\Atoms_V,
      \sat (\Utest T)^{\le k} \And \alpha \And \init \And \SFin\beta \}
\enspace.
  \end{myarray}
\end{equation*}
If no such $\beta$ exists which also satisfies $T \And \Empty$, the BDD
eventually converges to a value corresponding to the set of all $V$-atoms
reachable from $V$-atoms which satisfy $\init$.  The following set denotes
this:
\begin{equation*}
  \{ \beta\in \Atoms_V: \text{ for some }\alpha\in\Atoms_V,\ 
   \sat (\Utest T)\Chopstar \And \alpha \And \init \And \SFin\beta \}
\enspace.
\end{equation*}
We then terminate the algorithm with a report that the original transition
configuration $\Box T \And \init \And \Finite$ is unsatisfiable.  Even though
Theorem~\ref{small-model-for-finite-time-t-config-thm} bounds the number of
iterations, in some cases convergence takes too long.  This necessitates a
preset iteration limit or a facility for manual intervention in order to force
premature termination of the loop.

\begin{kcpomit}
\subsection{Construction of Model}
\end{kcpomit}

If for some $n$, the algorithm succeeds after $n$ iterations and determines
that the transition configuration is satisfiable, then a sample $V$-interval
having $n+1$ states and which satisfies the formula can be calculated.  This
involves standard BDD methods for constructing such examples and is done by
working backward through the BDDs $\Delta_n$, $\Delta_{n-1}$, \ldots
$\Delta_0$ to find a suitable sequence of $n+1$ $V$-atoms to serve as a
$V$-interval satisfying the transition configuration.  The algorithm can be
also readily adapted to only determine values for a subset of the variables in
$V$.
\begin{kcpomit}
However, we omit the details of this useful variant.

We start by picking some $V$-atom $\gamma_n$ in $\Delta_n\And\Gamma_3$ so that
the relation $\gamma_n\sat \Delta_n\And\Gamma_3$ holds.  The procedure for
doing this involves sequentially constructing a $V$-atom in
$\Delta_n\And\Gamma_3$ by regarding the propositional variables in $V$ as
having some ordering and then iteratively testing for the possible values of
each of them, assuming that the previous ones have already been fixed.
Suppose the elements of $V$ are $p_1$, \ldots, $p_m$.  For any $V$-atom
$\alpha$, our notation for manipulation conjunctions (recall
Definitions~\ref{size-of-a-conjunction-def}--\ref{parts-of-a-conjunction-def})
ensures that $\alpha[j]$ denote the value of the $j$-th propositional variable
in $\alpha$.  This is used in the procedure below for calculating the $V$-atom
$\gamma_n$.  The procedure involves $m$ iterations during which it performs
destructive assignments on a BDD variable referred to here as $\Pi$.
Furthermore, another variable $J$ is incrementally assigned all natural number
between $0$ and $m$, inclusively. Here is the algorithm for determining
$\gamma_n$:
\begin{enumerate}
\item Initialise the variable $J$ to be 1 and the BDD variable $\Pi$ to equal
  the BDD $\Delta_n\And\Gamma_3$.
\item If $J\gt m$, exit.
\item Otherwise, proceed as follows: If the BDD $\Pi \And p_J$
  is not false then set the value of the $J$-th propositional variable $p_J$
  in the $V$-atom $\gamma_n$ to be true (i.e., $\vld \gamma_n[J]\equiv p_J$)
  and destructively assign the BDD variable $\Pi$ the current value
of $\Pi \And p_J$.
  Otherwise, set the value of the $J$-th propositional variable $p_J$ in
  $\gamma_n$ to be false (i.e., $\vld \gamma_n[J]\equiv \Not p_J$) and
  destructively set the BDD variable $\Pi$ to be $\Pi \And \Not p_J$.
\item Increment $J$ by 1 and loop back to step 2.
\end{enumerate}
Each time step 2 is encountered, the invariant consisting of the next two
properties holds:
\begin{displaymath}
  \sat \Pi
  \qquad \valid \Pi \equiv (\Delta_n\And\Gamma_3 \And \gamma_n[1:J-1])
\enspace.
\end{displaymath}
Consequently, at the end, the BDD $\Delta_n\And\Gamma_3 \And \gamma_n[1:m]$ is
satisfiable. This BDD is equivalent to the BDD $\Delta_n\And\Gamma_3 \And
\gamma_n$ because $m$ equals the number of conjuncts in $\gamma_n$ (i.e.,
$m=\size{\gamma_n}$).  Therefore by simple propositional reasoning our
immediate goal, namely, the relation $\gamma_n\vld \Delta_n\And\Gamma_3$,
holds.

The fact that $\Delta_n\And\Gamma_3$ is not false ensures that at least one
suitable $V$-atom exists for use as $\gamma_n$.  The chosen $V$-atom serves as
the final state of the constructed $V$-interval. When viewed as a one-state
$V$-interval, $\gamma_n$ satisfies the formula $T \And \Empty$.  Next we
proceed \emph{in reverse} to calculate some $V$-atoms $\gamma_{n-1}$, \ldots,
$\gamma_0$ as is now described.  For each $V$-atom $\gamma_k$, we obtain a BDD
representing the following nonempty set of all $V$-atoms in $\Delta_{k-1}$
which can reach $\gamma_k$ in one step using $T$:
\begin{equation*}
  \{ \alpha\in \Atoms_V \colon
      \alpha \vld \Delta_{k-1} \text{ and } \alpha\gamma_k \vld T \}
\enspace.
\end{equation*}
We now re-express this set as the nonempty set shown below:
\begin{equation}
  \label{decision-procedure-1-eq}
  \{ \alpha\in \Atoms_V \colon
      \sat \alpha \And \Delta_{k-1}
         \text{ and } \sat T \And \Skip \And \alpha \And \Next\gamma_k \}
\enspace.
\end{equation}
The BDD representing the nonempty set~\eqref{decision-procedure-1-eq} is itself
determined using the following formula:
\begin{equation}
  \label{decision-procedure-2-eq}
  \Exists{V'}\bigl(\Delta_{k-1} \And \Gamma_2 \And (\gamma_k)_V^{V'}\bigr)
\enspace.
\end{equation}
The proof involves a minor generalisation of the earlier
Lemma~\ref{reduction-of-sat-t-and-alpha-and-skip-and-sfin-beta-lem}.  Due to
the set~\eqref{decision-procedure-1-eq} being nonempty, the BDD calculated
from~\eqref{decision-procedure-2-eq} is guaranteed to contain at least one
$V$-atom.  Some $V$-atom in this BDD is then selected to be $\gamma_{k-1}$.
This is done in the manner described earlier for the BDD
$\Delta_n\And\Gamma_3$ except that $\Pi$ is now initialised to the BDD
calculated using the formula~\eqref{decision-procedure-2-eq} instead of
$\Delta_n\And\Gamma_3$.  The construction ensures that $\gamma_{k-1}$, when
viewed as a $V$-state, satisfies $\Delta_{k-1}$ (i.e.,
$\gamma_{k-1}\vld\Delta_{k-1}$).  Moreover, the construction's encoding of the
$\NLone$ formula $T$ as the BDD $\Gamma_2$ ensures that the adjacent pair of
$V$-atoms $\gamma_{k-1}\gamma_k$, when viewed as a two-state $V$-interval,
satisfies the formula $T$ (i.e., $\gamma_{k-1}\gamma_k\vld T$).  This process
continues until some $V$-atom $\gamma_0$ is determined.  Now $\gamma_0$ is in
$\Delta_0$ and the definition of $\Delta_0$ ensures that all of its elements
satisfy the initial condition $\init$.  Therefore the following satisfiability
relations hold:
\begin{displaymath}
  \gamma_0 \vld \init
  \qquad \gamma_0\ldots\gamma_n \vld (\Utest T)\Chopstar
  \qquad \gamma_n \vld T \And \Empty
\enspace.
\end{displaymath}
Consequently the finite $V$-interval $\gamma_0\ldots\gamma_n$ satisfies the
$\PITLV$ formula $((\Utest T)\Chopstar \And \init \And \Finite);(T \And
\Empty)$.  Theorem~\ref{reduction-of-finite-time-t-configs-thm} then yields
that the $V$-interval $\gamma_0\ldots\gamma_n$ satisfies the original
semantically equivalent finite-time transition configuration $\Box T \And
\init \And \Finite$.
\end{kcpomit}

\subsection{Dealing with Infinite Time}

For testing an infinite-time transition configuration $\Box T \And \init \And
\Box\SDiamond L$, we can make use of
Theorem~\ref{satisfiable-infinite-time-t-config-thm} which guarantees that
this formula is satisfiable iff the next $\PTLV$ formula is satisfiable:
\begin{equation*}
  \Bm T \And \init
    \And \Diamond (L \And \Finite \And \More \And (\vec V\Tassign \vec V))
\enspace.
\end{equation*}
The previously described satisfiability algorithm for finite-time can
therefore be utilised.  However, we must first transform this second formula
to some suitable finite-time transition configuration using techniques later
described in Sect.~\ref{dealing-with-arbitrary-ptl-formulas-sec} for reducing
arbitrary $\PTL$ formulas to finite-time transition configurations.
Alternatively, more sophisticated algorithms using
Theorem~\ref{infinite-time-config-decompose-thm} can be employed to
directly analyse the infinite-time transition configuration using BDD-based
techniques.  Space does not permit more details here.

\index{PTL!decision procedure for transition configuration|)}
\index{decision procedure|)}
\index{transition configuration!decision procedure|)}

\section{Axiom System for $\protect\NL$}

\label{axiom-system-for-nl-sec}

In preparation for the proof of axiomatic completeness for $\PTL$, we now
consider an axiom system for $\NL$.  The axiomatic completeness of $\NL$ later
plays a major role in the completeness proof for $\PTL$.

\emph{Within this section, the variables $X$, $X'$, $X_0$ and $X'_0$ denote
  $\NL$ formulas.}

\index{Next Logic (NL)!axiom system}
\index{axiom system!Next Logic (NL)}
Table~\ref{nl-axiom-sys-table} contains a complete axiom system for $\NL$
adapted from the modal logic $\mathrm{K{+}D_c}$.  Here $\WeakNext$ (``weak
next''), previously defined in Table~\ref{temporal-operators-table} to be a
derived operator, is instead regarded as a primitive construct.  We can
consider $\Next X$ to be an abbreviation for $\Not\WeakNext\Not X$.  Hughes
and Cresswell \cite[Problem 6.8 on p.\ 123 with solution on p.\
379]{HughesCresswell96} briefly discuss how to show deductive completeness of
the logic $\mathrm{K{+}D_c}$.
\begin{mytable}
\centerline{\begin{math}
\begin{array}[t]{@{}l>{\enskip}l}
\multicolumn{2}{@{}l}{\mbox{Axioms:}} \\[3pt]
\mathrm{N1\ (K).} &
   \thm \WeakNext(X\imp X') \implies \WeakNext X \imp \WeakNext X' \\
\mathrm{N2\ (D_c).} &
   \thm \Next X \implies \WeakNext X \\[7pt]
\multicolumn{2}{@{}l}{\mbox{Inference rules:}} \\[3pt]
\mathrm{NR1.} & \mbox{If $X$ is a tautology, then $\thm X$} \\
\mathrm{NR2\ (MP).} & \mbox{If $\thm X\imp X'$ and $\thm X$, then $\thm X'$} \\
\mathrm{NR3\ (RN).} & \mbox{If $\thm X$, then $\thm \WeakNext X$}
\end{array}
\end{math}}
\caption{Complete axiom system for $\NL$ (Modal system $\mathrm{K{+}D_c}$)}
\label{nl-axiom-sys-table}
\end{mytable}

\index{Next Logic (NL)!alternative axiom system}
Table~\ref{nl-axiom-alt-sys-table} contains a complete axiom system for $\NL$
in which $\Next$, rather than $\WeakNext$, is the primitive operator.
Consequently, $\WeakNext$ is derived in the manner already shown in
Table~\ref{temporal-operators-table}.  The axiom system is essentially one of
several $M$-based axiomatisations of normal systems of modal logic covered by
Chellas~\cite{Chellas80} with the addition of the axiom $\mathrm{D_c}$.  This
second axiom system appears preferable for our purposes since our definition
of $\PTL$ also takes $\Next$ to be primitive.
\begin{mytable}
\centerline{\begin{math}
\begin{array}[t]{@{}l>{\hspace{2pt}}l>{\enskip}l>{\hspace{2pt}}l}
\multicolumn{2}{@{}l}{\mbox{Axioms:}} \\[3pt]
\mathrm{N1'\ (N\!\Diamond).} &
   \thm \Not\Next\False \\
\mathrm{N2'\ (C\!\Diamond).} &
   \thm \Next(X \Or X') \implies \Next X \,\Or\, \Next X' \\
\mathrm{N3'\ (\mathrm{D_c}).} &
   \thm \Next X \implies \WeakNext X \\[7pt]
\multicolumn{2}{@{}l}{\mbox{Inference rules:}} \\[3pt]
\mathrm{NR1'.} & \mbox{If $X$ is a tautology, then $\thm X$} \\
\mathrm{NR2'\ (MP).}& \mbox{If $\thm X\imp X'$ and $\thm X$, then $\thm X'$} \\
\mathrm{NR3'\ (RM\!\Diamond).}
   & \mbox{If $\thm X\imp X'$, then $\thm \Next X \imp \Next X'$}
\end{array}
\end{math}}
\caption{Alternative complete axiom system for $\NL$ based on $\Next$}
\label{nl-axiom-alt-sys-table}
\end{mytable}
We therefore use this axiom system here although the methods employed can
be easily adapted to the first $\NL$ axiom system.

\begin{mydefin}[Theoremhood and Consistency for $\NL$]
  \index{Next Logic (NL)!theorem} If some $\NL$ formula $X$ is deducible from
  the axiom system, we call it an \emph{$\NL$ theorem} and denote this
  theoremhood as $\thm_{\NL} X$.  \index{consistent formula} \index{Next Logic
    (NL)!consistent formula} We define $X$ to be \emph{$\NL$-consistent} if
  $\Not X$ is \emph{not} an $\NL$ theorem, i.e., $\not\thm_{\NL} \Not X$.
\end{mydefin}

Below are some representative lemmas about satisfiability and consistency of
$\NL$ formulas.  They are subsequently used in the completeness proof for the
$\NL$ axiom system in Table~\ref{nl-axiom-alt-sys-table}.
\begin{mylemma}
  \label{nl-if-sat-w-then-sat-w-and-not-next-t-lem}
  For any state formula $w$ and $\NL$ formula $X$, if $w$ is satisfiable, then
  the $\NL$ conjunction $w \And \Not\Next X$ is satisfied by some
  one-state interval.
\end{mylemma}
\begin{kcpomit}
\begin{myproof}
  Suppose the state formula $w$ is satisfied by some interval $\sigma$. Let
  the empty interval $\sigma'$ be obtained from the first state of $\sigma$.
  Now $\sigma'$ also satisfies $w$ because $w$ is a state formula.  Also, a
  negated $\Next$ formula is trivially true in any one-state interval.  Hence
  the interval $\sigma'$ satisfies $\Not\Next X$. Consequently, it also
  satisfies the conjunction $w \And \Not\Next X$.
\end{myproof}
\end{kcpomit}

\begin{mylemma}
  \label{nl-if-sat-w-and-sat-t-then-sat-w-and-next-t-lem}
  For any state formula $w$ and $\NL$ formula $X$, if both $w$ and $X$ are
  satisfiable, then so is the formula $w
  \And \Next X$. \\
  In such as case, if $X$ itself is satisfied by an interval having at most
  $n$ states, then $w \And \Next X$ is satisfied by an interval having at most
  $n+1$ states,
\end{mylemma}

\begin{mylemma}
  \label{nl-if-const-next-t-then-const-t-lem}
  For any $\NL$ formula $X$, if $\Next X$ is $\NL$-consistent, then so $X$.
\end{mylemma}

For any $\NL$ formulas $X$ and $X'$, the following are deducible as $\NL$
theorems and shortly used to simplify formulas:
\begin{align}
  \label{nl-thm-next-t-and-t'-eqv-next-t-and-next-t'-eq}
  & \theoremNL \Next(X \And X') \EQUIV \Next X \,\And\, \Next X' \\
  \label{nl-thm-next-t-and-not-t'-eqv-next-t-and-not-next-t'-eq}
  & \theoremNL \Next(X \And \Not X') \EQUIV \Next X \,\And\, \Not\Next X' \\
  \label{nl-thm-not-next-t-or-t'-eqv-not-next-t-and-not-next-t'-eq}
  & \theoremNL \Not\Next(X \Or X') \EQUIV \Not\Next X \,\And\, \Not\Next X'
\enskip.
\end{align}

Axiomatic completeness is usually defined to mean that every valid formula is
deducible as a theorem.  However, we will make use of the following variant
way of expressing completeness:
\begin{mylemma}[Alternative Notion of Completeness]
  \index{axiomatic completeness!alternative notion}
  \label{alternative-completeness-lem}
  A logic's axiom system is complete iff each consistent formula is
  satisfiable.
\end{mylemma}

\begin{kcpomit}
We now demonstrate axiomatic completeness for $\NL$:
\end{kcpomit}
\begin{mytheorem}[Completeness of Alternative $\NL$ Axiom System]
 \label{nl-completeness-thm}
 The $\NL$\kcpbreak axiom system in Table~\ref{nl-axiom-alt-sys-table} is
 complete.
\end{mytheorem}
\begin{myproof}
  The proof involves the kind of consistency-based reasoning found later in
  the paper.  Using Lemma~\ref{alternative-completeness-lem}, we show that any
  $\NL$ formula $X_0$ which is $\NL$-consistent (i.e., $\not\thmNL \Not X_0$)
  has a satisfying finite interval.  Let $n$ be the next-height of $X_0$,
  i.e., the maximum nesting of $\Next$s in $X_0$.  We do induction on $n$ to
  show that $X_0$ is satisfied by some interval with at most $n+1$ states.
\ifkcpinclude
  For $n\gt 0$, we regard the temporal constructs in $X_0$ which are not
  nested in other temporal constructs as being primitive.  Then conventional
  propositional reasoning yields a deducibly equivalent formula $X'_0$ in
  disjunctive normal form.  As least one disjunct is consistent.
  Lemmas~\ref{nl-if-sat-w-then-sat-w-and-not-next-t-lem}--\relax
  \ref{nl-if-const-next-t-then-const-t-lem} and
  equivalences~\eqref{nl-thm-next-t-and-t'-eqv-next-t-and-next-t'-eq}--\relax
  \eqref{nl-thm-not-next-t-or-t'-eqv-not-next-t-and-not-next-t'-eq} are
  invoked to obtain a satisfying interval.
\fi
  \begin{kcpomit}
  \par
  \ProofStep{Base case for $n=0$:} The $\NL$ formula $X_0$ contains no
  temporal operators.  Therefore, its assumed $\NL$-consistency ensures that
  the formula $\Not X_0$ is not a propositional tautology. Therefore, $X_0$ is
  immediately satisfiable by a interval with one state.
  
  \ProofStep{Inductive case for $n\gt 0$:} We regard all of the temporal
  constructs in $X_0$ which are not nested in other temporal constructs as
  being primitive.  Now use conventional propositional reasoning to obtain a
  formula $X'_0$ in disjunctive normal form which is both deducibly equivalent
  and semantically equivalent to $X_0$.  The logical equivalence $X_0 \equiv
  X'_0$ is in fact a tautology by the inference rule $\mathrm{NR1'}$ and
  therefore does not require any temporal reasoning.  For example, if $X_0$ is
  the $\NL$ formula $\bigl(p \Or (\Next r \And \Next(p\Or \Next \Not q))\bigr)
  \And (r \imp \Not\Next\Next r)$, then the following deducibly equivalent
  $\NL$ disjunction can be used for $X'_0$:
  \begin{mycomment}
(progn

  (setq t1
        '(and (or (var p)
                  (and (next (var r))
                       (next (or (var p) (next (not (var q)))))))
              (implies (var r)
                       (not (next (next (var r)))))))

  (setq t2
        '(or (and (var p)
                  (or (not (var r))
                      (not (next (next (var r))))))
             (and (and (next (var r))
                       (or (next (or (var p) (next (not (var q)))))))
                  (or (not (var r))  (not (next (next (var r))))))))

  (dd-vld `(equiv ,t1 ,t2))

  (setq t3
        '(or (and (var p) (not (var r)))
             (and (var p)
                  (not (next (next (var r)))))
             (and (next (var r))
                  (next (or (var p) (next (not (var q)))))
                  (not (var r)))
             (and (next (var r))
                  (next (or (var p) (next (not (var q)))))
                  (not (next (next (var r)))))))

  (dd-vld `(and (equiv ,t1 ,t2) (equiv ,t2 ,t3))))
  ***Testing for validity with finite time.
  ***Starting iteration 0.
  Cudd current number of live nodes=108.
  1 iteration(s) performed.
  Valid with finite time.

  (dd-sat t1)
  ***# of dependencies=4, # of independent variables=3.
  ***Testing for satisfiability with finite time.
  0 iteration(s) performed.
  Satisfiable with finite time.
  ***Now start to find a model.
  $#SuccWList=-1
  $j=0
  Here is a model with 1 state:
  ***State 1:  P=1  Q=1  R=1.

  (dd-sat `(and (not (var p)) ,t3))
  ***# of dependencies=4, # of independent variables=3.
  ***Testing for satisfiability with finite time.
  ***Starting iteration 0.
  Cudd current number of live nodes=79.
  1 iteration(s) performed.
  Satisfiable with finite time.
  ***Now start to find a model.
  $#SuccWList=0
  $j=1
  $j=0
  Here is a model with 2 states:
  ***State 1:  P=0  Q=1  R=1.
  ***State 2:  P=1  Q=1  R=1.
  \end{mycomment}
  \begin{equation}
    \label{nl-completeness-1-eq}
    (p \And \Not r)
      \;\Or\; (p \And \Not\Next\Next r)
      \;\Or\; \bigl(\Next r
                \And \Next(p\Or \Next \Not q)
                \And \Not r \bigr)
      \;\Or\; \bigl(\Next r
                \And \Next(p\Or \Next \Not q)
                \And \Not\Next\Next r\bigr)
\enspace.
  \end{equation}
  Like $X_0$, the formula $X'_0$ is $\NL$-consistent.  Hence by propositional
  reasoning $X'_0$ must have at least one $\NL$-consistent disjunct.  This
  disjunct is itself a conjunction of one or more formulas.  Each is either a
  propositional variable, the primitive formula $\True$, a $\Next$-formula or
  the negation of one of them.  We now assume without loss of generality that
  there is at least one nontemporal formula (e.g., $\True$) and consolidate
  all nonnegated and negated $\Next$-formulas using $\NL$
  theorems~\eqref{nl-thm-next-t-and-t'-eqv-next-t-and-next-t'-eq}--\relax
  \eqref{nl-thm-not-next-t-or-t'-eqv-not-next-t-and-not-next-t'-eq}.  Let us
  illustrate this process by showing consolidated versions of the four
  disjuncts for the sample formula~\eqref{nl-completeness-1-eq}:
  \begin{displaymath}
    \begin{array}{l}
      p \And \Not r
      \qquad p \And \Not\Next\Next r
      \qquad \Not r \And \Next(r \And (p\Or \Next \Not q))
      \\[2pt]
      \True \And \Next\bigl(r \And (p\Or \Next \Not q)
               \And \Not\Next r\bigr)
    \end{array}
\enspace.
  \end{displaymath}
  There are now two cases to consider, depending on whether or not the
  resulting disjunct contains a positive (i.e., nonnegated) instance of
  $\Next$:
  \begin{itemize}
  \item \emph{The consolidated version of the disjunct does not contain a
      positive instance of $\Next$:} If the result contains no positive
    instance of $\Next$ (as in the first two sample consolidated disjuncts),
    then the remaining nontemporal part is itself $\NL$-consistent and
    immediately satisfiable in a single state as in the base case.  This
    serves as a single-state interval.  In such as case, even if there is some
    negated $\Next$ formula, it can be safely ignored since it is trivially
    true in an empty interval (see
    Lemma~\ref{nl-if-sat-w-then-sat-w-and-not-next-t-lem}).  Hence interval
    satisfies the consolidated version of the disjunct and hence the
    disjunction $X'_0$ itself.  Therefore the interval also satisfies the
    semantically equivalent formula $X_0$.
  \item \emph{The consolidated version of the disjunct contains a positive
      instance of $\Next$:} In this case (as in the last two sample
    consolidated disjuncts), the only temporal construct in the consolidated
    version of the disjunct is one nonnegated $\Next$-construct.  Its single
    operand is $\NL$-consistent by
    Lemma~\ref{nl-if-const-next-t-then-const-t-lem} and has next-height of at
    most $n-1$.  By induction, this operand has some satisfying interval
    $\sigma$ with at most $n$ states.  We also use the disjunct's remaining
    nontemporal part to obtain an $\NL$-consistent formula which has a
    satisfying interval with a single state.
    Lemma~\ref{nl-if-sat-w-and-sat-t-then-sat-w-and-next-t-lem} is used to add
    this state to the front of the interval $\sigma$ to obtain a longer
    interval $\sigma'$ with at most $n+1$ states satisfying the disjunct and
    hence also the disjunction $X'_0$.  Therefore $\sigma'$ also satisfies the
    semantically equivalent original formula $X_0$.
  \end{itemize}
  \end{kcpomit}
\end{myproof}

\section{Axiomatic Completeness for Transition Configurations}

\label{ptl-axiom-system-sec}

\index{axiomatic completeness!for set of formulas}
We now turn to describing a $\PTL$ axiom system with which axiomatic
completeness can be shown for transition configurations.

\index{PTL!axiom system}
\index{axiom system!PTL}
The $\PTL$ axiom system used here is shown in
Table~\ref{modified-pnueli-dx-ptl-axiom-sys-table} and is adapted from another
similar $\PTL$ axiom system $\mathrm{DX}$ proposed by Pnueli~\cite{Pnueli77}.
Gabbay et al.~\cite{GabbayPnueli80} showed that $\mathrm{DX}$ is complete.
Pnueli's original system uses strong versions of $\Diamond$ and $\Box$ (which
we denote as $\SDiamond$ and $\SBox$, respectively) which do not examine the
current state.  In addition, Pnueli's system only deals with infinite time.
However, Gabbay et al.~\cite{GabbayPnueli80} also include a variant system
called $\mathrm{D^0X}$ based on the conventional $\Diamond$ and $\Box$
operators which examine the current state.  The version presented here does
this as well and furthermore permits both finite and infinite time.
\begin{mytable}
\centerline{\begin{math}
\begin{array}[t]{@{}ll}
\multicolumn{2}{@{}l@{}}{\mbox{Axioms:}} \\[3pt]
\text{T1.} & \thm \Box(X\imp Y) \implies \Box X \imp \Box Y \\
\text{T2.} & \thm \Next X \implies \WeakNext X \\
\text{T3.} & \thm \Next(X\imp Y) \implies \Next X \imp \Next Y \\
\text{T4.} & \thm \Box X \implies X \And \WeakNext\Box X \\
\text{T5.} & \thm \Box(X \imp \WeakNext X) \implies X \imp \Box X
\end{array}
\quad
\begin{array}[t]{@{}ll}
\multicolumn{2}{@{}l@{}}{\mbox{Inference rules:}} \\[3pt]
\text{R1.} & \mbox{If $X$ is a tautology, then $\thm X$} \\
\text{R2.} & \mbox{If $\thm X\imp Y$ and $\thm X$, then $\thm Y$} \\
\text{R3.} & \mbox{If $\thm X$, then $\thm \Box X$}
\end{array}
\end{math}}
\caption{Modified version of Pnueli's complete $\PTL$ axiom system
   $\mathrm{DX}$}
\label{modified-pnueli-dx-ptl-axiom-sys-table}
\end{mytable}

\begin{mydefin}[Theoremhood and Consistency for $\PTL$]
  \index{PTL!theorem}
  If the $\PTL$ formula $X$ is deducible from the axiom system, we call it a
  \emph{$\PTL$ theorem} and denote this theoremhood as $\thm X$.
\index{consistent formula}
\index{PTL!consistent formula}
  We define $X$ to be \emph{consistent} if $\Not X$ is \emph{not} a theorem,
  i.e., $\not\thm \Not X$.
\end{mydefin}

In the course of proving completeness for $\PTL$ we make use of a definition
of completeness for sets of formulas such as sets of transitions
configurations:
\begin{mydefin}[Completeness for a Set of Formulas]
  \index{axiomatic completeness!for set of formulas}
  An axiom system is said to be complete for a set of formulas
  $\{X_1,\ldots,X_n\}$ if the consistency of any $X_i$ implies that $X_i$ is
  also satisfiable.
\end{mydefin}
Now the Alternative Notion of Completeness
(Lemma~\ref{alternative-completeness-lem}) can also be readily adapted to sets
of formulas.  Indeed, our goal in the rest of this section is to show that any
consistent transition configuration is also satisfiable.

The next lemma permits us to utilise within $\PTL$ the axiomatic completeness
of the $\NL$ proof system:
\begin{mytheorem}[Completeness for $\NL$ in $\PTL$]
  \label{completeness-for-nl-in-ptl-thm}
  \index{axiomatic completeness!for NL formulas in PTL}
  The $\PTL$ axiom system is complete for the set of $\NL$ formulas.
\end{mytheorem}
\begin{myproof}
  Theorem~\ref{nl-completeness-thm} establishes the completeness of the
  alternative $\NL$ axiom system in Table~\ref{nl-axiom-alt-sys-table}.  We
  then show that any $\NL$ theorem is also a $\PTL$ theorem.  This can be done
  by demonstrating that all axioms and inferences rules in the $\NL$ axiom
  system are derivable from $\PTL$ ones.
\end{myproof}

\subsection{Some Basic Lemmas for Completeness}

\label{some-basic-lemma-for-completeness-subsec}

In this subsection, we deal with another part of the completeness proof.  We
utilise ways to go from certain specific kinds of consistent formulas
involving reachability to intervals in order to later construct models for
consistent transition configurations in
\S\ref{completeness-for-t-configurations-subsec}.
Table~\ref{consistency-to-satisfiablity-for-trans-rels-table} summarises the
basic lemmas proved here.  Within the table, we use the
notation $\sat X$ already introduced in
Definition~\ref{satisfiability-and-validity-def} to denote that the formula
$X$ is satisfiable and $\const X$ to denote that $X$ is consistent.
\begin{mytable}
  \centerline{\relax
    \begin{tabular}{@{}c@{\enspace}l@{}}
      Lemma
        & \multicolumn{1}{c}{Summary}
        \\\noalign{\hrule \vskip 2pt}
      \ref{consistency-to-nextrel-for-atoms-lem}
        & If $\const\Bm T \And \alpha \And \Next \beta$, then
          $\sat T \And \alpha \And \Skip \And \SFin\beta$
      \\[1pt]
      \ref{consistency-to-transrel-for-atoms-lem}
        & If $\const\Bm T \And \alpha \And \Diamond\beta$, then
          $\sat (\Utest T)\Chopstar \And \alpha \And \SFin\beta$
      \\[1pt]
      \ref{consistency-to-stransrel-for-atoms-lem}
        & If $\const\Bm T \And \alpha \And \SDiamond\beta$, then
          $\sat (\Utest T)\Chopstar \And \alpha \And \Next\SFin\beta$
    \end{tabular}}
  \caption{Summary of some basic lemmas for consistency and satisfiability}
  \label{consistency-to-satisfiablity-for-trans-rels-table}
\end{mytable}

\begin{mylemma}
  \label{consistency-to-nextrel-for-atoms-lem}
  For any $V$-atoms $\alpha$ and $\beta$, if the formula $\Bm T \And \alpha
  \And \Next\beta$ is consistent, then the formula $T \And \alpha \And \Skip
  \And \SFin\beta$ is satisfiable.
\end{mylemma}
\begin{myproof}
  From the consistency of the formula $\Bm T \And \alpha \And \Next\beta$ and
  simple temporal reasoning, we obtain the consistency of the $\NLone_V$
  formula $T \And \alpha \And \Next\beta$.
  Theorem~\ref{completeness-for-nl-in-ptl-thm} concerning axiomatic
  completeness for $\NL$ formulas in the $\PTL$ axiom system then ensures that
  this is satisfiable.  Clearly any interval satisfying it has at least two
  states.  Hence by the earlier Lemma~\ref{sat-nlone-iff-nlone-and-skip-lem}
  the formula $\Skip \And T \And \alpha \And \Next\beta$ is also
  satisfiable.  Consequently, simple temporal reasoning yields that the
  semantically equivalent formula $T \And \alpha \And \Skip \And \SFin\beta$
  is satisfiable as well.
\end{myproof}

For any $V$-atom $\alpha$, within the next two lemmas we let $S_\alpha$ denote
the subset of $\Atoms_V$ containing exactly every $V$-atom $\gamma$ for which
the following formula, which concerns reachability from $\alpha$, is
satisfiable:
\begin{equation*}
   (\Utest T)\Chopstar \And \alpha \And \SFin\gamma
\enspace.
\end{equation*}
Here is a more formal definition of $S_\alpha$:
\begin{equation*}
   S_\alpha \Defeq \{\, \gamma \in\Atoms_V:\null
     \sat (\Utest T)\Chopstar \And \alpha \And \SFin\gamma \,\}
\enspace.
\end{equation*}

\begin{mylemma}
  \label{|-[m]z-&-alpha-gamma-box-bigvee-gamma-lem}
  For any $V$-atom $\alpha$, the following formula is a $\PTL$ theorem:
  \begin{equation}
    \label{|-[m]z-&-alpha-gamma-box-bigvee-gamma-eq}
    \Theorem \Bm T \And \alpha
       \implies \Box\bigvee_{\gamma \in S_\alpha}\gamma \enspace.
  \end{equation}
\end{mylemma}
\begin{myproof}
  The following formulas are valid and in $\NLone$.  Hence, they are theorems
  by the completeness of the $\PTL$ axiom system for $\NLone$ formulas
  (Theorem~\ref{completeness-for-nl-in-ptl-thm}):
   \begin{displaymath}
     \Theorem \alpha \implies \bigvee_{\gamma \in S_\alpha}\gamma
     \qquad\qquad
     \Theorem \More \And T \And \bigvee_{\gamma \in S_\alpha}\gamma
       \Implies \Next\bigvee_{\gamma \in S_\alpha}\gamma
\enspace.
  \end{displaymath}
  From these and simple temporal reasoning we can readily deduce our
  goal~\eqref{|-[m]z-&-alpha-gamma-box-bigvee-gamma-eq}.
\end{myproof}

\begin{mylemma}
  \label{consistency-to-transrel-for-atoms-lem}
  For any $V$-atoms $\alpha$ and $\beta$, if the formula $\Bm T \And \alpha
  \And \Diamond \beta$ is consistent, then the formula $(\Utest T)\Chopstar
  \And \alpha \And \SFin\beta$ is satisfiable.
\end{mylemma}
\begin{myproof}
  Suppose on the contrary that $(\Utest T)\Chopstar \And \alpha \And
  \SFin\beta$ is unsatisfiable.  Now $\alpha$ is in the set $S_\alpha$,
  whereas $\beta$ is not.  Hence, the following formula concerning
  $\beta$ not being in $S_\alpha$ is valid and thus a propositional
  tautology:
  \begin{equation}
    \label{|-alpha-imp-bigvee-not-beta-eq}
    \Theorem \bigvee_{\gamma \in S_\alpha}\!\!\gamma \Implies \Not\beta
\enspace.
  \end{equation}
  Furthermore, the previous
  Lemma~\ref{|-[m]z-&-alpha-gamma-box-bigvee-gamma-lem} ensures that the next
  implication is a $\PTL$ theorem:
  \begin{equation}
    \label{|-bm-t-and-alpha-imp-box-bigvee-...-eq}
    \Theorem \Bm T \And \alpha
       \implies \Box\bigvee_{\gamma \in S_\alpha}\!\!\gamma \enspace.
  \end{equation}
  The two implications~\eqref{|-alpha-imp-bigvee-not-beta-eq}
  and~\eqref{|-bm-t-and-alpha-imp-box-bigvee-...-eq} together with some simple
  temporal reasoning let us deduce that $\alpha$ can never reach $\beta$:
  \begin{equation*}
     \Theorem \Bm T \And \alpha
        \implies \Box \Not\beta
\enspace.
  \end{equation*}
  From this and the general equivalence
  $\thm \Box\Not\beta\equiv\Not\Diamond\beta$
  we can deduce the following $\PTL$ theorem:
  \begin{equation*}
     \Theorem \Bm T \And \alpha
        \implies \Not\Diamond\beta
\enspace.
  \end{equation*}
  Therefore, the formula $\Bm T \And \alpha \And \Diamond \beta$ is
  inconsistent.  This contradicts the lemma's assumption.
\end{myproof}

\begin{mylemma}
  \label{consistency-to-stransrel-for-atoms-lem}
  For any $V$-atoms $\alpha$ and $\beta$, if the formula $\Bm T \And \alpha
  \And \SDiamond \beta$ is consistent, then the formula $(\Utest T)\Chopstar
  \And \alpha \And \Next\SFin \beta$ is satisfiable.
\end{mylemma}
\begin{myproof}
  From the consistency of the formula $\Bm T \And \alpha \And \SDiamond
  \beta$, we readily deduce for some $V$-atom $\gamma$ the consistency of the
  two $\PTL_V$ formulas below:
  \begin{displaymath}
    \Bm T \And \alpha \And \Diamond\gamma
    \qquad \Bm T \And \gamma \And \Next \beta
\enspace.
  \end{displaymath}
  The consistency of the first formula $\Bm T \And \alpha \And \Diamond\gamma$
  and Lemma~\ref{consistency-to-stransrel-for-atoms-lem} yield that the
  formula $(\Utest T)\Chopstar \And \alpha \And \SFin\gamma$ is satisfiable.
  Lemma~\ref{consistency-to-nextrel-for-atoms-lem} and the second formula $\Bm
  T \And \gamma \And \Next \beta$ then guarantee that the formula $T \And
  \gamma \And \Skip \And \SFin\beta$ is satisfiable.
  Lemma~\ref{chop-coverage-lem} then yields that the next formula is
  satisfiable:
  \begin{equation*}
    \bigl((\Utest T)\Chopstar \And \alpha \And \Finite\bigr);
      (T \And \gamma \And \Skip \And \SFin\beta)
\enspace.
  \end{equation*}
  From this and some further simple interval-based reasoning we can establish
  our goal, namely, that the formula $(\Utest T)\Chopstar \And \alpha \And
  \Next\SFin \beta$ is satisfiable.
\end{myproof}

\subsection{Completeness for Transition Configurations}

\label{completeness-for-t-configurations-subsec}

\index{axiomatic completeness!for transition configurations|(}
\index{transition configuration!axiomatic completeness|(}
We now apply the material presented in the previous
\S\ref{some-basic-lemma-for-completeness-subsec} to ultimately establish
completeness for finite- and infinite-time transition configurations.  Here is
a summary of the completeness theorems for them:
\begin{center}
  \begin{tabular}{cc}
    Type of transition
      & Where proved
   \\\noalign{\hrule \vskip 2pt}
    Finite-time
      & Theorem~\ref{completeness-for-finite-time-t-configs-thm}
    \\[1pt]
    Infinite-time
      & Theorem~\ref{completeness-for-infinite-time-t-configs-thm}
  \end{tabular}
\end{center}
The remaining two kinds of transition configurations are subordinate to these.
For the sake of brevity, we do not consider them here.

\begin{mytheorem}
  \label{completeness-for-finite-time-t-configs-thm}
  Completeness holds for any finite-time transition configuration $\Box T \And
  \init \And \Finite$.
\end{mytheorem}
\begin{myproof}
  From the consistency of the finite-time transition configuration $\Box T
  \And \init \And \Finite$ and simple temporal reasoning we can demonstrate
  that for some $V$-atoms $\alpha$ and $\beta$, the next formula is
  consistent:
  \begin{equation*}
    \Bm T \And \alpha \And \init \And \SFin(T \And \beta)
\enspace.
  \end{equation*}
  From this and further simple temporal reasoning it is readily follows that
  the following formulas are all consistent:
  \begin{displaymath}
    \alpha \And \init
    \qquad
    \Bm T \And \alpha \And \Diamond\beta
    \qquad
    T \And \beta \And \Empty
\enspace.
  \end{displaymath}
  The first of these is itself satisfiable since any consistent formula in
  $\PROP$ is satisfiable.  The second one and
  Lemma~\ref{consistency-to-transrel-for-atoms-lem} yields that the $\PITL$
  formula $(\Utest T)\Chopstar \And \alpha \And \SFin\beta$ is satisfiable.
  The third formula $T \And \beta \And \Empty$ is in $\NLone$ and hence by
  Theorem~\ref{completeness-for-nl-in-ptl-thm} satisfiable.
  Hence
  the following formulas are all satisfiable:
  \begin{displaymath}
    \alpha \And \init
    \qquad
    (\Utest T)\Chopstar \And \alpha \And \SFin\beta
    \qquad
    T \And \beta \And \Empty
\enspace.
  \end{displaymath}
  This and Theorem~\ref{finite-time-config-decompose-thm} then yield the
  satisfiability of the finite-time transition configuration $\Box T \And
  \init \And \Finite$.
\end{myproof}

\begin{mytheorem}
  \label{completeness-for-infinite-time-t-configs-thm}
  Completeness holds for any infinite-time transition configuration $\Box T
  \And \init \And \Box\SDiamond L$.
\end{mytheorem}
\begin{myproof}
  From the consistency of the infinite-time transition configuration $\Box T
  \And \init \And \Box\SDiamond L$ and simple temporal reasoning we can
  demonstrate that for some $V$-atoms $\alpha$ and $\beta$, the next formula
  is consistent:
  \begin{equation}
    \label{completeness-for-infinite-time-t-configs-1-eq}
    \Bm T \And \alpha \And \init \And \Box\SDiamond(\beta \And L)
\enspace.
  \end{equation}
  Lemma~\ref{alpha-and-l-eqv-alpha-and-en-l-alpha-eqv-lem} ensures that the
  formulas $\beta \And L$ and $\beta \And \En_{L,\beta}$ are semantically
  equivalent.  The proof of this only requires simple propositional reasoning
  not involving the temporal operators in $L$.  Hence the next equivalence is
  readily deducible as a $\PTL$ theorem using substitution into a
  propositional tautology (see Definition~\ref{tautology-def} and $\PTL$
  inference rule $\text{R1}$ in
  Table~\ref{modified-pnueli-dx-ptl-axiom-sys-table}):
  \begin{equation}
    \label{completeness-for-infinite-time-t-configs-2-eq}
    \Theorem \beta \And L  \EQUIV  \beta \And \En_{L,\beta}
\enspace.
  \end{equation}
  From the consistency of
  formula~\eqref{completeness-for-infinite-time-t-configs-1-eq} and the
  deducibility of
  formula~\eqref{completeness-for-infinite-time-t-configs-2-eq}, we can show
  the consistency of the next formula:
  \begin{equation*}
    \Bm T \And \alpha \And \init \And \Box\SDiamond(\beta \And \En_{L,\beta})
\enspace.
  \end{equation*}
  This and simple temporal reasoning then together yield the consistency of
  the following formulas involving some additional $V$-atoms $\gamma_1$,
  \ldots, $\gamma_{\size{\En_{L,\beta}}}$ (not necessarily distinct):
  \begin{displaymath}
    \begin{myarray}
      \alpha \And \init
      \qquad
      \Bm T \And \alpha \And \Diamond\beta
      \qquad
      \Bm T \And \beta \And \SDiamond\beta \\[3pt]
      \text{for each } \gamma_i\colon
        \quad \Bm T \And \beta \And \Diamond\gamma_i \quad
        \gamma_i \And \theta_{\En_{L,\beta}[i]} \quad
        \Bm T \And \gamma_i \And \Diamond\beta
\enspace.
    \end{myarray}
  \end{displaymath}
  The consistency of the propositional formulas $\alpha \And \init$ and
  $\gamma_i \And \theta_{\En_{L,\beta}[i]}$ for each $V$-atom $\gamma_i$
  ensures they are satisfiable.
  Lemma~\ref{consistency-to-transrel-for-atoms-lem} is then applied to the
  remaining consistent formulas, except for $\Bm T \And \beta \And
  \SDiamond\beta$ which requires
  Lemma~\ref{consistency-to-stransrel-for-atoms-lem}.  The combined result is
  that the following formulas are all satisfiable:
  \begin{displaymath}
    \begin{myarray}
      \alpha \And \init
      \qquad
      (\Utest T)\Chopstar \And \alpha \And \SFin\beta
      \qquad
      (\Utest T)\Chopstar \And \beta \And \Next\SFin\beta \\[3pt]
      \text{for each } \gamma_i\colon
        \quad (\Utest T)\Chopstar \And \beta \And \SFin\gamma_i \quad
        \gamma_i \And \theta_{\En_{L,\beta}[i]} \quad
        (\Utest T)\Chopstar \And \gamma_i \And \SFin\beta
\enspace.
    \end{myarray}
  \end{displaymath}
  Hence by Theorem~\ref{infinite-time-config-decompose-thm}, the original
  consistent infinite-time transition configuration is indeed satisfiable.
\end{myproof}

\index{axiomatic completeness!for transition configurations|)}
\index{transition configuration!axiomatic completeness|)}

\section{Invariants and Related Formulas}

\label{invariants-and-related-formulas-sec}

We will shortly introduce the concepts of invariants and invariant
configurations which together act as a natural middle level between transition
configurations and full $\PTL$ and involve the use of auxiliary variables.
These variables provide a way to reduce the nesting of temporal operators
within other temporal operators and thereby simplify further analysis.
Satisfiability, existence of small models, decidability and axiomatic
completeness for invariant configurations can be readily related to the
analysis of transition configurations.  Furthermore, it is not hard to reduce
arbitrary $\PTL$ formulas to invariant configurations by utilising such
auxiliary variables.

The analysis of invariant configurations and arbitrary $\PTL$ formulas does
not require any further interval-based reasoning or
$\PITL$.

\begin{mydefin}[Invariant]
  \label{invariant-def}
  \index{invariant}
  An \emph{invariant} is any finite conjunction of zero or more equivalences
  in which each equivalence's left side is a distinct propositional
  variable and each equivalence's right side is one of the following:
  \begin{itemize}
  \item Some $\PTL$ formula of the form $\Diamond w$, for some state formula
    $w$.
  \item Some $\NLone$ formula.
  \end{itemize}
\end{mydefin}
\index{dependent variable}
\index{invariant!dependent variables of}
\index{independent variable}
\index{invariant!independent variables of}
\index{dependency}
\index{invariant!dependency in}
The variables occurring on the left sides of equivalences are called
\emph{dependent variables} and any other variables are called
\emph{independent variables}.
The right sides are called \emph{dependent formulas} and each equivalence is
itself called a \emph{dependency}.  Hence for a given invariant $I$, it
follows that $\size{I}$ denotes the number of dependencies in $I$.  Also, for
any $k: 1\le k\le \size{I}$, $I[k]$ denote the $k$-th dependency in $I$.  Each
dependency containing $\Diamond$ is referred to as a
\index{invariant!00@$\Diamond$-dependency in}
\emph{$\Diamond$-dependency}. Observe that a dependent variable can be
referenced in any dependent formula including the one associated with it.

Below is a sample invariant referred to as $I_1$:
\begin{align*}
  I_1\colon\enspace&
    (r_1 \equiv \Diamond (p \And \Not q))
    \;\And\; (r_2 \equiv (r_1 \And \Next r_2))
\enspace.
\end{align*}
Here $\size{I_1}$ equals 2, the first dependency $I[1]$ is the equivalence
$r_1 \equiv \Diamond (p \And \Not q)$ and the second dependency $I[2]$ is the
equivalence $r_2 \equiv (r_1 \And \Next r_2)$.

Note that an invariant is not necessarily satisfiable as in $r_1\equiv \Not
r_1$.  Also note that dependencies of the two forms $r\equiv w$ and $r\equiv
\Next w$, for some propositional variable $r$ and state formula $w$, are both
subsumed by the second case in Definition~\ref{invariant-def}.  If desired, a
more restrictive definition of invariants limited to dependencies of the form
$w$, $\Next w$ and $\Diamond w$ is possible.

We can view an invariant $I$ as being any conjunction having the form
$\bigwedge_{k:1\le k\le \size{I}} (u_k \equiv \phi_k)$ so that $u_k$ is the
$k$-th dependent propositional variable and $\phi_k$ is the $k$-th dependent
formula in $I$.  Observe that for any $k: 1\le k\le \size{I}$, the conjunct
$I[k]$ has the form $u_k \equiv \phi_k$ and $I$ itself can be expressed as
$\bigwedge_{k:1\le k\le \size{I}} I[k]$.

Starting with an invariant $I$, we analyse certain low-level formulas referred
to here as \emph{invariant configurations}.
\begin{mydefin}[Invariant Configurations]
  \label{i-configuration-def}
  \index{invariant configuration}
  An \emph{invariant configuration} is a formula of the form $\Box I \,\And\,
  X$ where the $\PTL$ formula $X$ is in one of three categories shown below:
  \begin{center}
    \begin{tabular}{cc}
      Type of invariant configuration
        & Syntax of $X$ \\\noalign{\hrule \vskip 2pt}
      Basic & $w$ \\
      Finite-time & $w \And \Finite$ \\
      Infinite-time & $w \And \Inf$
    \end{tabular}
  \end{center}
  Here $w$ is a state formula.
\end{mydefin}
For example, the conjunction $\Box I_1 \,\And\, r_2$ is a basic invariant
configuration which is true for intervals which are infinite, have $r_1$ and
$r_2$ always true and $p$ and $\Not q$ both always eventually true.

The next definition helps to simplify the notation used in the reduction
of invariant configurations to transition configurations:
\begin{mydefin}[Ordered Invariant]
  \label{ordered-invariant}
  An invariant is said to be \emph{ordered} if all of its
  $\Diamond$-dependencies precede any others.
\end{mydefin}
It is not hard to rearrange an arbitrary invariant's dependencies to obtain a
semantically equivalent ordered invariant.  In the rest of this section, we
will without loss of generality limit our attention to ordered invariants and
invariant configurations based on them.

We now associate with an ordered invariant $I$ a transition formula $T_I$ and
a conditional liveness formula $L_I$.  They serve to expeditiously reduce
invariant configurations to transition configurations previously analysed in
earlier sections.  Definition~\ref{transition-formula-def} below describes
$T_I$.  The subsequent
Definition~\ref{conditional-liveness-formula-of-an-invariant-def} describes
the form of $L_I$.
\begin{mydefin}[Transition Formula for an Ordered Invariant]
  \label{transition-formula-def}
  \index{invariant!transition formula for}
  For an ordered invariant $I$, the associated transition formula $T_I$ is an
  $\NLone$ formula which captures $I$'s transitional behaviour between pairs
  of adjacent states. It is obtained from $I$ by replacing each
  $\Diamond$-dependency with another dependency not containing $\Diamond$ and
  leaving the remaining $\Diamond$-free dependencies unchanged.  More
  precisely, each dependency in $I$ of the form $r\equiv \Diamond w$, for some
  propositional variable $r$ and state formula $w$, is replaced by the
  $\Diamond$-free equivalence $r\equiv (w \Or \Next r)$.
\end{mydefin}
Observe that the transition formula $T_I$ is in $\NLone$ and is also a
well-formed invariant. Also, for any $k: 1\le k \le \size{I}$, if the
dependency $I[k]$ does not contain $\Diamond$, then it and $T_I$'s
corresponding dependency $T_I[k]$ are identical.

Here is the transition formula $T_{I_1}$ associated with $I_1$:
\begin{equation*}
  T_{I_1}\colon\enspace
    \bigl(r_1 \equiv ((p \And \Not q) \Or \Next r_1)\bigr)
    \;\And\; (r_2 \equiv (r_1 \And \Next r_2))
\enspace.
\end{equation*}

Let us now introduce some simple notation needed for reasoning about liveness
and $\Diamond$-dependencies.  This will be used in the definition of an ordered
invariant's associated conditional liveness formula.
\begin{mydefin}[Liveness Tests of an Ordered Invariant]
  \label{liveness-tests-of-invariant-def}
  \index{liveness tests}
  \index{invariant!liveness tests of}
  For any ordered invariant $I$ having $n$ $\Diamond$-depen\-den\-cies, define
  $n$ different \emph{liveness tests} $\theta_{I[1]}$, \ldots, $\theta_{I[n]}$
  so that for each $k: 1\le k\le n$, the $k$-th dependency in $I$ is
  expressible as $u_k \equiv \Diamond \theta_{I[k]}$.
\end{mydefin}
For instance, the sample invariant $I_1$ has a single liveness test
$\theta_{I[1]}$ which denotes the formula $p \And \Not q$.  Note that each
$\theta_{I[k]}$ is always a state formula.  If an invariant $I$ has $n$
$\Diamond$-dependencies, then for each $k: 1\le k\le n$, $T_I$'s dependency
$T_I[k]$ identical to the equivalence $u_k\equiv (\theta_{I[k]} \Or\Next
u_k)$.

Given an ordered invariant $I$, we now associate a specific conditional
liveness formula $L_I$ with it:
\begin{mydefin}[Conditional Liveness Formula of an Ordered Invariant]
  \label{conditional-liveness-formula-of-an-invariant-def}
  \index{conditional liveness formula!of invariant}
  \index{invariant!conditional liveness formula of}
  The conditional liveness formula $L_I$ of an ordered invariant $I$ which has
  $n$ $\Diamond$-dependencies is itself a conjunction of $n$ implications.
  For each $k: 1\le k\le n$, the $k$-th implication is obtained by simply
  replacing the outermost equivalence operator in $I$'s $k$-th
  $\Diamond$-dependency by the implication operator and using $\Dm$ instead of
  $\Diamond$.  Therefore, for each $k: 1\le k\le n$, the dependency $I[k]$ has
  the form $u_k\equiv \Diamond \theta_{I[k]}$ and the implication $L_I[k]$ has
  the form $u_k\imp \Dm\theta_{I[k]}$.
\end{mydefin}
The definition of $I$'s conditional liveness formula $L_I$ intentionally
ignores any $\NLone$ dependencies in $I$ since $T_I$ already adequately deals
with them.  As a result, $L_I$ can contain fewer conjuncts than $I$ and $T_I$.
Below is the conditional liveness formula $L_{I_1}$ associated with ordered
invariant $I_1$:
\begin{equation*}
  L_{I_1}\colon\enspace
    (r_1 \imp \Dm (p \And \Not q))
\enspace.
\end{equation*}
It is not hard to see that, unlike $I$'s transition formula, the conditional
liveness formula associated with $I$ is not a well-formed invariant.

\subsection{Reduction of Basic Invariant Configurations}

\label{reduction-of-basic-configurations-subsec}

Starting with an ordered invariant $I$, let us now consider the relationship
between its basic invariant configuration and the associated finite-time and
infinite-time invariant configurations.  This permits us to focus the
remaining analysis on the two later kinds of invariant configurations.

\begin{mylemma}
  \label{sat-basic-i-config-iff-sat-finite-or-infinite-time-one-lem}
  A basic invariant configuration $\Box I \And w$ is satisfiable iff at least
  one of its associated finite-time and infinite-time invariant configurations
  is satisfiable.
\end{mylemma}
\begin{myproof}
  This follows from the validity of the formula $\Finite \Or \Inf$ and simple
  propositional reasoning.
\end{myproof}

The finite-time and infinite-time invariant configurations for the ordered
invariant $I$ each have a corresponding semantically equivalent transition
configuration of the same kind as is now shown:
\begin{center}
  \begin{tabular}{@{}lcccc@{}}
    & Invariant
      & Transition
      & Where
    \\
    & configuration
      & configuration
      & proved
    \\\noalign{\hrule \vskip 2pt}
    Finite time & $\Box I \And w \And \Finite$
      & $\Box T_I \And w \And \Finite$
      & Theorem~\ref{semantic-equivalent-of-finite-time-i-and-t-configs-thm}
    \\[3pt]
    Infinite time & $\Box I \And w \And \Inf$
      & $\Box T_I \And w \And \Box\SDiamond L_I$
      & Theorem~\ref{semantic-equivalent-of-infinite-time-i-and-t-configs-thm}
  \end{tabular}
\end{center}
Observe that the reductions from the two types of the invariant configurations
to the corresponding transition configurations do not introduce any extra
variables.  In what follows we prove that a finite-time invariant
configuration is semantically equivalent to its associated finite-time
transition configuration and similarly a infinite-time invariant configuration
is semantically equivalent to its associated infinite-time transition
configuration.

In what follows we will often abstract the behaviour of a
$\Diamond$-dependency by using two propositional variables $p$ and $q$ and
representing the dependency as the $\PTL$ equivalence $p\equiv \Diamond q$.
This technique is used to establish the next lemma:
\begin{mylemma}
  \label{box-i-eqv-box-t-valid-lem}
  The formulas $\Box I$ and $\Box T_I$ are semantically equivalent on finite
  intervals. In other words, the following implication is valid:
  \begin{equation*}
    \Valid \Finite \Implies \Box I \EQUIV \Box T_I
\enspace.
  \end{equation*}
\end{mylemma}
\begin{myproof}
  We can represent $\Box I$ as the conjunction $\bigwedge_{k: 1\le k \le
    \size{I}} \Box I[k]$ and similarly represent $\Box T_I$ as the conjunction
  $\bigwedge_{k: 1\le k \le \size{I}} \Box T_I[k]$. For any $k: 1\le k \le
  \size{I}$, if $I[k]$ is in $\NLone$ then $T_I[k]$ is identical to it and
  hence $\Box I[k]$ and $\Box T_I[k]$ are identical.  Otherwise, $\Box I[k]$
  can be seen as a substitution instance of the $\PTL$ formula $\Box (p \equiv
  \Diamond q)$ containing the two propositional variables $p$ and $q$.  Now
  $\Box T_I[k]$ therefore corresponds to the formula $\Box (p \equiv (q \Or
  \Next p))$.  Simple temporal reasoning can then be used to show that each of
  these implies the other in any finite interval.
\end{myproof}
Let us note that the validity for finite time of the relevant equivalence
$\Box (p \equiv \Diamond q) \equiv \Box (p \equiv (q \Or \Next p))$ can even
be readily checked by a computer implementation of a decision procedure for
$\PTL$ with finite time.

\begin{mytheorem}
  \label{semantic-equivalent-of-finite-time-i-and-t-configs-thm}
  The finite-time invariant configuration for $I$ is semantically equivalent
  to the associated finite-time transition configuration.
\end{mytheorem}
\begin{myproof}
  This readily follows from Lemma~\ref{box-i-eqv-box-t-valid-lem} and
  propositional reasoning.
\end{myproof}

Unfortunately, the equivalence $\Box I \equiv \Box T_I$ can fail to be valid
for infinite time if $I$ contains $\Diamond$-dependencies because $T_I$ does
not fully capture the liveness requirements of such dependencies.
Lemma~\ref{box-i-eqv-box-t-and-box-sdiamond-l-valid-lem} later on corrects for
this problem by showing that in infinite time the two formulas $\Box I$ and
$\Box T_I \And \Box\SDiamond L_I$ are semantically equivalent.  The reason
that $\Box I \equiv \Box T_I$ is not necessarily valid is because when we
consider an individual $\Diamond$-dependency, the formulas $\Box (p \equiv
\Diamond q)$ and $\Box (p \equiv (q \Or \Next p))$ are not semantically
equivalent on infinite-time intervals since on such an interval, the first
formula can be false and the second one true.  An example of this occurs in
any infinite interval where $p$ is always true and $q$ is always false.
Therefore, if $I$ contains $\Diamond$-dependencies, then $\Box I$ can be false
on an infinite-time interval even though $\Box T_I$ is true on the interval.
However, the next lemma holds even for infinite time:
\begin{mylemma}
  \label{box-i-imp-box-t-valid-lem}
  The $\PTL$ implication $\Box I \implies \Box T_I$ is valid.
\end{mylemma}
\begin{myproof}
  The $\NLone$-dependencies in $I$ and $T_I$ are identical.  Furthermore, for
  the $\Diamond$-dependencies we make use of the valid $\PTL$ formula $\Box
  (p\equiv \Diamond q) \implies \Box (p \equiv (q \Or \Next p))$.
\end{myproof}

We see from Lemma~\ref{box-i-imp-box-t-valid-lem} that the formula $\Box I
\imp \Box T_I$ is valid for both finite and infinite time.  However if $I$
contains $\Diamond$-dependencies, then the converse implication $\Box T_I\imp
\Box I$ is not necessarily valid for infinite time because the implication
$\Box (p \equiv (q \Or \Next p)) \implies \Box (p \equiv \Diamond q)$ fails to
be valid.  We now discuss the principles which successfully correct for this.
First of all, the following weakened implication concerning an individual
$\Diamond$-dependency is valid:
\begin{equation*}
   \Valid \Box (p\equiv (q \Or \Next p)) \Implies \Box (\Diamond q \implies p)
\enspace.
\end{equation*}
Here we use the formula $\Diamond q \imp p$ instead of the stronger
equivalence $p \equiv \Diamond q$.  The following equivalence then strengthens
the effect of $\Box (p \equiv (q \Or \Next p))$ by adding the formula $\Box(p
\implies \Diamond q)$:
\begin{equation*}
  \Valid \Box (p\equiv \Diamond q)
    \Equiv \Box (p\equiv (q \Or \Next p))
        \,\And\, \Box (p \implies \Diamond q)
\enspace.
\end{equation*}
In fact, we can even replace the conjunct $\Box (p \implies \Diamond q)$ by
the weaker formula $\Box\Diamond(p \implies \Diamond q)$ which adds a
$\Diamond$:
\begin{equation*}
  \Valid \Box (p\equiv \Diamond q)
    \Equiv \Box (p\equiv (q \Or \Next p))
        \,\And\, \Box\Diamond (p \implies \Diamond q)
\enspace.
\end{equation*}
All three valid formulas only contain the propositional variables $p$ and $q$
and can consequently be readily checked for infinite-time validity by any
computer implementation of a decision procedure for $\PTL$ with infinite time.

Now suppose the ordered invariant $I$ has $m$ $\Diamond$-dependencies and
hence $m=\size{L_I}$.  If we have $m$ pairs of propositional variables $p_1$,
$q_1$, \ldots, $p_m$, $q_m$ (corresponding to $I$'s $\Diamond$-dependencies)
then the following generalisation of the previous valid equivalence is itself
valid:
\begin{multline*}
  \Valid \Box\bigwedge_{1\le k\le m} (p_k\equiv \Diamond q_k)
  \\
    \EQUIV\quad \Box\bigwedge_{1\le k\le m} (p_k\equiv (q_k \Or \Next p_k))
        \;\And\;\Box\Diamond\bigwedge_{1\le k\le m} (p_k \implies \Diamond q_k)
\enspace.
\end{multline*}
The left side of the equivalence corresponds to the invariant $I[1:m]$.
Similarly, the first conjunct on the right side corresponds to $T_I[1:m]$ and
the second one to $L_I$, except for the use of $\Diamond$ instead of $\Dm$.

\begin{mycomment}
It readily follows from $L_I$'s definition together with the preceding
discussion about the relationship between $\Box I$ and $\Box T_I$ in infinite
time that the equivalence $\Box I \equiv (\Box T_I \And \Box L_I)$ is valid
for infinite intervals as is the semantically equivalent, more concise one
$\Box I \equiv \Box(T_I \And L_I)$.  Consequently, we have the following
lemma:
\begin{mylemma}
  \label{box-i-eqv-box-(z-and-l)-valid-lem}
  The formula $\Inf \implies \bigl(\Box I \equiv \Box(T_I \And L_I)\bigr)$ is
  valid.
\end{mylemma}
\begin{myproof}
  Only the $\Diamond$-dependencies in $I$ and $T_I$ differ.  Hence, the proof
  mainly involves checking that the following similar equivalence holds for
  each $\Diamond$-dependency $I[k]$ for $k: 1\le k\le \size{L_I}$:
  \begin{equation*}
    \Valid \Inf \implies \bigl(\Box I[k] \equiv \Box(T_I[k] \And L_I[k])\bigr)
\enspace.  
  \end{equation*}
  This is a substitution of the following valid $\PTL$ formula:
  \begin{equation*}
    \Valid \Inf \implies \Bigl(\Box (p\equiv \Diamond q)
           \,\equiv\,
             \Box\bigl((p\equiv (q \Or\Next p))
                         \,\And\, (p\imp \Diamond q)\bigr)
      \Bigr)
\enspace.  
  \end{equation*}
  The proof of validity uses the valid
  implication~\eqref{box-p-eqv-q-or-next-p-imp-box-diamond-q-imp-p-eq}.
\end{myproof}

The next few lemmas lead up to
Theorem~\ref{semantic-equivalent-of-infinite-time-i-and-t-configs-thm} which
establishes the semantic equivalence of the ordered invariant $I$'s
infinite-time invariant configuration and its associated infinite-time
transition configuration:
\begin{mylemma}
  \label{t-and-next-l-imp-l-valid-lem}
  The implication $T_I \And \Next L_I \imp L_I$ is valid.
\end{mylemma}
\begin{myproof}
  Ensuring validity reduces to showing that for each $k: 1\le k\le
  \size{L_I}$, the implication $T_I[k] \And \Next L_I[k] \imp L_I[k]$ is
  valid.  Now $I[k]$ is a $\Diamond$-dependency. Therefore $T_I[k]$ denotes
  $u_k \equiv \Diamond\theta_{I[k]}$ and $L_I[k]$ denotes $u_k \imp
  \Dm\theta_{I[k]}$.  We make use of the valid $\PTL$ formula now given which
  captures the essence of the implication $T_I[k] \And \Next L_I[k] \imp
  L_I[k]$:
  \begin{equation*}
    \Valid (p \equiv (q \Or \Next p)) \;\And\;\Next(p\imp \Dm q)
      \Implies (p\imp \Dm q)
\enspace.
  \end{equation*}
  By substituting $u_k$ into $p$ and $\theta_{I[k]}$ into $q$, we obtain the
  validity of the formula $T_I[k] \And \Next L_I[k] \imp L_I[k]$.
\end{myproof}

\begin{mylemma}
  \label{box-t-and-sdiamond-l-imp-box-l-valid-lem}
  The implication $\Box (T_I \And \SDiamond L_I) \imp \Box L_I$ is valid.
\end{mylemma}
\begin{myproof}
  We use simple temporal reasoning to ensure the validity of the following
  $\PTL$ implication:
  \begin{equation*}
    \Valid \Box((p \And \Next q) \implies q)
       \Implies \bigl(\Box(p \And \SDiamond q) \implies \Box q\bigr)
\enspace.
  \end{equation*}
  Substitution of $T_I$ and $L_I$ into $p$ and $q$, respectively, together
  with Lemma~\ref{t-and-next-l-imp-l-valid-lem} (to obtain $\vld \Box(T_I \And
  \Next L_I \imp L_I)$) and modus ponens yields our goal.
\end{myproof}

\begin{mylemma}
  \label{box-t-and-l-eqv-box-t-and-sdiamond-l-valid-lem}
  The equivalence $\Box (T_I \And L_I) \equiv \Box (T_I \And \SDiamond L_I)$
  is valid.
\end{mylemma}
\begin{myproof}
  We use simple temporal reasoning to ensure the validity of the following
  $\PTL$ implication:
  \begin{equation*}
    \Valid \bigl(\Box (p \And \SDiamond q) \imp \Box q\bigr)
       \Implies \bigl(\Box p
                    \implies (\Box\SDiamond q \equiv \Box q) \bigr)
\enspace.
  \end{equation*}
  The substitution of $T_I$ and $L_I$ into $p$ and $q$, respectively, together
  with Lemma~\ref{box-t-and-sdiamond-l-imp-box-l-valid-lem} and modus ponens
  yields the validity of the next formula:
  \begin{equation*}
    \Valid \Box T_I \Implies (\Box\SDiamond L_I \equiv \Box L_I)
\enspace.
  \end{equation*}
  This and some simple temporal reasoning ensures our goal, namely, the
  validity of the equivalence $\Box (T_I \And L_I) \equiv \Box (T_I \And
  \SDiamond L_I)$.
\end{myproof}

\begin{mylemma}
  \label{box-i-eqv-box-t-and-sdiamond-l-valid-lem}
  The formula $\Inf \implies \bigl(\Box I \equiv \Box (T_I \And \SDiamond
  L_I)\bigr)$ is valid.
\end{mylemma}
\begin{myproof}
  We use Lemmas~\ref{box-i-eqv-box-(z-and-l)-valid-lem}
  and~\ref{box-t-and-l-eqv-box-t-and-sdiamond-l-valid-lem} and
  propositional reasoning.
\end{myproof}
\end{mycomment}

Now within infinite time, $\Box\Diamond$ and $\Box\SDiamond$ have the same
behaviour and in addition $\Diamond$ and $\Dm$ act identically.  We use this
to obtain the next lemma which expresses $I$ in terms of $T_I$ and $L_I$:
\begin{mylemma}
  \label{box-i-eqv-box-t-and-box-sdiamond-l-valid-lem}
  The formula $\Inf \implies \bigl(\Box I \equiv (\Box T_I \And \Box\SDiamond
  L_I)\bigr)$ is valid.
\end{mylemma}

\begin{mytheorem}
  \label{semantic-equivalent-of-infinite-time-i-and-t-configs-thm}
  An infinite-time invariant configuration $\Box I \And w \And \Inf$ for the
  ordered invariant $I$ is semantically equivalent to the associated
  infinite-time transition configuration $\Box T_I \And w \And \Box\SDiamond
  L_I$.
\end{mytheorem}
\begin{myproof}
  This readily follows from
  Lemma~\ref{box-i-eqv-box-t-and-box-sdiamond-l-valid-lem} and simple temporal
  reasoning.
\end{myproof}

The soundness of the reductions to the associated transition configurations
ensures that we can use the decision procedure described in
Sect.~\ref{decision-procedure-sec}.

\subsection{Bounded Models for Basic Invariant Configurations}

The theorem given below gives the small model property for basic invariant
configurations:
\begin{mytheorem}
  \label{small-model-for-invariant-configs-thm}
  \index{invariant configuration!small models}
  Suppose $V$ is a finite set of variables and the variables in the ordered
  invariant $I$ and the state formula $w$ are all elements of $V$.  Then the
  basic invariant configuration $\Box I \And w$ is satisfiable iff it is
  satisfied by some some finite interval with interval length less than
  $\size{\Atoms_V}$ or by an infinite, ultimately periodic one consisting of
  an initial segment with interval length at most $\size{\Atoms_V}$ fused with
  a remaining infinite periodic part with a period having interval length at
  most $(\size{L_I}+1)\,\size{\Atoms_V}$.
\end{mytheorem}
\begin{myproof}
  Suppose $\Box I \And w$ is satisfiable. We will consider the two cases of
  finite and infinite intervals separately:
  
  \emph{Case for finite intervals:}
  Theorem~\ref{semantic-equivalent-of-finite-time-i-and-t-configs-thm} ensures
  that the finite-time invariant configuration $\Box I \And w \And \Finite$
  and its associated finite-time transition configuration $\Box T_I \And w
  \And \Finite$ are semantically equivalent.  The construction of $T_I$
  ensures that any variable occurring in it is a member of the set $V$.
  Lemma~\ref{small-model-for-finite-time-t-config-thm} therefore establishes
  that if the conjunction $\Box T_I \And w \And \Finite$ is satisfiable, then
  a satisfying interval exists having less interval length than
  $\size{\Atoms_V}$. This interval consequently also satisfies the basic
  invariant configuration $\Box I \And w$.
  
  \emph{Case for infinite intervals:}
  Theorem~\ref{semantic-equivalent-of-infinite-time-i-and-t-configs-thm}
  ensures that the infinite-time invariant configuration $\Box I \And w \And
  \Inf$ and its associated infinite-time transition configuration $\Box T_I
  \And w \And \Box\SDiamond L_I$ are semantically equivalent.  From
  Lemma~\ref{small-model-for-infinite-time-t-config-thm} we have that this
  second formula is satisfied by an infinite interval consisting of an initial
  segment having interval length less than $\size{\Atoms_V}$ fused with a
  periodic interval with period having interval length at most
  $(\size{L_I}+1)\,\size{\Atoms_V}$.  The overall ultimately periodic interval
  therefore also satisfies the formula $\Box I \And w$.
\end{myproof}

\subsection{Axiomatic Completeness for Invariant Configurations}

\label{axiomatic-completeness-for-i-configs-subsec}

\begin{mytheorem}
  \label{completeness-for-finite-and-infinite-time-i-configs-thm}
  Completeness holds for finite- and infinite-time invariant configurations.
\end{mytheorem}
\begin{myproof}
  Suppose we have some invariant $I$.  Assume without loss of generality that
  $I$ is ordered since otherwise we can trivially rearrange its dependencies
  to obtain an ordered invariant which is both semantically and deducibly
  equivalent to $I$.
  Subsection~\ref{reduction-of-basic-configurations-subsec} already described
  how to construct a semantically equivalent transition configuration from any
  finite-time or infinite-time invariant configuration associated with $I$.
  The various valid formulas mentioned there can be deduced as $\PTL$ theorems
  to establish that each such finite-time and infinite-time invariant
  configuration is also deducibly equivalent to the associated transition
  configuration.  This and the previously shown axiomatic completeness for
  finite-time and infinite-time transition configurations respectively proved
  in Theorems~\ref{completeness-for-finite-time-t-configs-thm}
  and~\ref{completeness-for-infinite-time-t-configs-thm} ensure that any
  consistent finite-time or infinite-time invariant configuration associated
  with $I$ is satisfiable.  Hence, we establish our immediate goal of
  completeness for finite- and infinite-time invariant configurations.
\end{myproof}

\begin{mytheorem}
  \label{completeness-for-basic-i-configs-thm}
  \index{invariant configuration!axiomatic completeness for}
  Completeness holds for basic invariant configurations.
\end{mytheorem}
\begin{myproof}
  Suppose we have some consistent basic invariant configuration $\Box I
  \,\And\, w$.  Now the disjunction $\Finite\Or \Inf$ is easily deduced as a
  propositional tautology since $\Inf$ is defined to be $\Not\Finite$ (see
  Table~\ref{temporal-operators-table}).  It is then straightforward to show
  using purely propositional reasoning that $\Box I \,\And\, w$ is deducibly
  equivalent to the disjunction of its associated finite-time or infinite-time
  invariant configurations:
  \begin{equation*}
    \Theorem \Box I \,\And\, w
      \Equiv (\Box I \,\And\, w \,\And\, \Finite)
         \;\Or\; (\Box I \,\And\, w \,\And\, \Inf)
\enspace.
  \end{equation*}
  Hence at least one of the latter is also consistent.  The previous
  Theorem~\ref{completeness-for-finite-and-infinite-time-i-configs-thm}
  ensures that any such consistent finite- or infinite-time invariant
  configuration is satisfiable as well.  An interval which satisfies it can
  also serve as a model for the basic invariant configuration.  This
  demonstrates the desired axiomatic completeness for all basic invariant
  configurations.
\end{myproof}

\section{Dealing with Arbitrary $\protect\PTL$ Formulas}

\label{dealing-with-arbitrary-ptl-formulas-sec}

So far we have only looked at bounded models and axiomatic completeness for
certain kinds of $\PTL$ formulas.  For an arbitrary $\PTL$ formula $X$, it is
straightforward to construct an invariant $I$ linearly bounded by the size of
$X$ and containing a finite number of dependent variables $u_1$, $u_2$,
\ldots, $u_{\size{I}}$ not themselves occurring in $X$ so as to mimic the
semantics of $X$ in the sense that $X$ is satisfiable iff $\Box I \And
u_{\size{I}}$ is satisfiable and in addition the implication $\Box I \imp
(u_{\size{I}}\equiv X)$ is valid.

One possible translation will be detailed shortly.  Before describing it, we
need to discuss a convention for systematically renaming an invariant's
dependent variables.  Normally, the first dependent variable in an invariant
$I$ constructed here from an $\PTL$ formula is $r_1$ and the last is
$r_{\size{I}}$.  However, we inductively construct the invariants by combining
smaller invariants into larger ones and often must alter the indices of the
dependent variables to avoid clashes.  A operator on formulas to
suitably do this is now defined:
\begin{mydefin}[Shifting of Subscripts in Invariants]
  \label{shifting-of-subscripts-in-invariants-def}
  \index{invariant!shifting of subscripts}
  \index{00@$\shift$}
  For any invariant $I$, the operation $I \shift k$ is defined to be the
  invariant obtained by replacing $u_1$,\ldots, $u_{\size{I}}$ by
  $r_{1+k}$,\ldots, $r_{\size{I}+k}$, i.e.,
  $I_{u_1,\ldots,u_{\size{I}}}^{r_{1+k},\ldots,r_{\size{I}+k}}$.
\end{mydefin}
It is not hard to see that if $I$'s dependent variables are themselves the
distinct variables $r_1, \ldots, r_{\size{I}}$, then $I \shift k$ shifts the
subscripts of them so that each $r_j$ becomes $r_{j+k}$.  Therefore, the first
dependent variable becomes $r_{1+k}$ instead of $r_1$, the second becomes
$r_{2+k}$ and so forth.  In other words, $I \shift k$ denotes the same formula
as the conjunction $\bigwedge_{1\le j\le \size{I}} (r_{j+k}\equiv
(\phi_j)_{u_1,\ldots,u_{\size{I}}}^{r_{1+k},\ldots,r_{\size{I}+k}})$.

Without loss of generality, let $X$ be a $\PTL$ formula which does not contain
any of the variables $r_1, r_2, \ldots$.
Table~\ref{ptl-invariant-associated-with-an-ptl-formula-table} contains the
definition of a function $\H(X)$ which translates $X$ into an invariant
containing some of the variables $r_1$, $r_2$, \ldots\ as dependent variables.
\begin{mytable}
  \centerline{\begin{math}
    \begin{array}{cl}
      X & \qquad \H(X) \\\noalign{\hrule \vskip 2pt}
      T & r_1\equiv T, \text{ for any $\NLone$ formula $T$.} \\[3pt]
      \Not Y & \H(Y)
        \;\And\; (r_{\size{\H(Y)}+1}\equiv \Not r_{\size{\H(Y)}})
      \\[3pt]
      Y\Or Y' & \H(Y) \;\And\; \H(Y') \shift m
          \;\And\; (r_{m+n+1} \equiv r_m\Or r_{m+n})
\enspace, \\[3pt]
      \multicolumn{1}{l}{} &
        \mbox{where $m=\size{\H(Y)}$ and $n=\size{\H(Y')}$.} \\[3pt]
      \Diamond Y & \H(Y)
        \;\And\; (r_{\size{\H(Y)}+1}\equiv \Diamond r_{\size{\H(Y)}})
\end{array}
\end{math}}
  \caption{Definition of $\H(X)$}
  \label{ptl-invariant-associated-with-an-ptl-formula-table}
\end{mytable}
In order to reduce the number of dependent variables, the first case is used
whenever the formula is in $\NLone$ even if one of the next two cases
for negation and logical-or is applicable.

Table~\ref{example-of-invariant-obtained-by-applying-h-to-a-ptl-formula-table}
contains a sample $\PTL$ formula $X_0$, an equivalent formula $X'_0$ having no
logical-ands, implications or $\Box$ constructs, and the invariant $\H(X'_0)$
and the initial condition $r_{\size{\H(X'_0)}}$.
We also include a version of
  $X'_0$ which shows how the dependencies correspond to the subformulas in
  $X'_0$.
\begin{mycomment}
\begin{verbatim}
(dd-sat '(box (and (implies (var p) (next (diamond (var q))))
                        (diamond (or (not (var p)) (next (var q)))))))


(dd-sat '(not (diamond (not (not (or (not (or (not (var p))
                                                   (next (diamond (var q)))))
                                     (not (diamond (or (not (var p)) (next (var q)))))))))))
Here is a model with 1 state:
***State 1:  P=0  Q=1.
\end{verbatim}
\end{mycomment}
\begin{mytable}
  \begin{equation*}
    \def\myunder#1#2{\underbrace{\strut #2}_{r_{#1}}}
    \def\myover#1#2{\overbrace{\strut #2}^{r_{#1}}}
    \fbox{$\begin{array}{l@{\ifkcp\else\quad\fi}l}
      \noalign{\vskip 5pt}
      X_0\colon\enskip
        &  \Box\bigl((p\implies \Next\Diamond q)
             \,\And\, \Diamond (\Not p \Or \Next q)\bigr)
      \\[3pt]
      X'_0\colon\enskip
        &  \Not\Diamond\Not\Bigl(\Not \bigl(\Not(\Not p\Or\Next\Diamond q)
             \,\Or\, \Not\Diamond (\Not p \Or \Next q)\bigr)\Bigr)
      \\
      \rlap{$\begin{array}[b]{@{}l@{}}
        X'_0
        \text{ with} \\
        \text{dependent} \\
        \text{variables} \\
        \text{shown}\colon\enskip
      \end{array}$}
        &  \scalebox{\ifkcp\else 1.0\fi}
           {$\myover{14}{\Not\;\;\myunder{13}{\!
             \Diamond\;\;\myover{12}{\Not\Bigl(\;\;
               \myunder{11}{\Not\;\;\myover{10}{\bigl(\;\;
                 \myunder6{\Not
                   \myover5{(\;\myunder1{\Not p}
                       \;\Or\;\myunder4{\!\Next\,\myover3{\Diamond
                            \myunder2{q}}}\;)}}\,
             \;\Or\; \myunder9{
                       \Not\;\myover8{
                               \Diamond \;\myunder7{\!
                                           (\Not p \Or \Next q)}\;}}
          \;\bigr)}\;}\;\;\Bigr)}}}$}
      \\
  \noalign{\vspace{10pt}}
      \H(X'_0)\colon\enskip & 
             (r_1 \equiv \Not p)
             \,\And\, (r_2 \equiv q)
             \,\And\, (r_3 \equiv \Diamond r_2)
             \,\And\, (r_4 \equiv \Next r_3)
        \\
        &
             \quad\,\And\, (r_5 \equiv (r_1 \Or r_4))
             \,\And\, (r_6 \equiv \Not r_5)
             \,\And\, (r_7 \equiv (\Not p \Or \Next q))
        \\
        &
             \quad\,\And\, (r_8 \equiv \Diamond r_7)
             \,\And\, (r_9 \equiv \Not r_8)
             \,\And\, (r_{10} \equiv (r_6 \Or r_9))
        \\
        &
             \quad\,\And\, (r_{11} \equiv \Not r_{10})
             \,\And\, (r_{12} \equiv \Not r_{11})
             \,\And\, (r_{13} \equiv \Diamond r_{12})

        \\
        &
             \quad\,\And\, (r_{14} \equiv \Not r_{13})
      \\[3pt]
      r_{\size{\H(X'_0)}}\colon\enskip &
             r_{14}
      \\
      \noalign{\vskip 5pt}
    \end{array}$}
  \end{equation*}
  \caption{Example of invariant obtained by applying $\H$ to a $\PTL$ formula}
  \label{example-of-invariant-obtained-by-applying-h-to-a-ptl-formula-table}
\end{mytable}

It is straightforward to utilise more sophisticated methods which construct
invariants directly from formulas with other logical operators such as
logical-and and $\Box$.  In addition, it is not hard to systematically produce
invariants containing a lot fewer dependencies then the ones generated by
$\H$.  In fact, our prototype implementation of the decision procedure
described in Sect.~\ref{decision-procedure-sec} makes use of such techniques
and others as well.
Here is an invariant and initial formula produced
by the decision procedure directly from the formula $X_0$:
\begin{equation*}
  \begin{array}{ll}
     X_0\colon\enskip
       &  \Box\bigl((p\implies \Next\Diamond q)
            \,\And\, \Diamond (\Not p \Or \Next q)\bigr)
     \\[3pt]
     I'\colon\enskip
       & (r_1 \equiv \Diamond q)
         \,\And\, (r_2 \equiv \Next r_1)
         \,\And\, (r_3 \equiv \Next q)
         \,\And\, (r_4 \equiv \Diamond(\Not p \Or r_3))
     \\
       & \quad\,\And\, \bigl(r_5 \equiv\Diamond\Not((p\imp r_2) \And r_4)\bigr)
     \\[3pt]
     \init'\colon\enskip
       & \Not r_5
  \end{array}
\end{equation*}
We omit further details.

It is easy to check that $\H(X)$ contains at most one dependent variable for
each variable and operator in $X$ so the total number of dependent variables
in $\H(X)$ is bounded by $X$'s size and indeed the size of $\H(X)$ is linearly
bounded by $X$'s size.  It is also easy to check by doing induction on $X$'s
syntactic structure that $X$ is satisfiable iff the basic invariant
configuration $\Box \H(X) \,\And\, r_{\size{\H(X)}}$ is satisfiable.
Furthermore, the implication $\Box \H(X) \implies (r_{\size{\H(X)}}\equiv X)$
can be shown to be valid.  Consequently, $\Box \H(X) \,\And\,
r_{\size{\H(X)}}$ is used to represent $X$'s behaviour (modulo the dependent
variables which act as auxiliary ones).  The bounded model for the invariant
configuration (see Theorem~\ref{small-model-for-invariant-configs-thm})
satisfies $X$ as well.  \index{PTL!decision procedure using invariants} The
decision procedure described in Sect.~\ref{decision-procedure-sec} can be
utilised to check the satisfiability of arbitrary $\PTL$ formulas by reducing
them first to basic invariant configurations and then testing the associated
finite-time and infinite-time transition configurations (see
\S\ref{reduction-of-basic-configurations-subsec}).  Axiomatic completeness for
$X$ readily reduces to that for the invariant configuration $\Box \H(X)
\,\And\, r_{\size{\H(X)}}$.

\section{Some Additional Features}

\label{some-additional-features-sec}

This section describes a number of extensions to our approach.  They include
the temporal operator $\Until$ and past-time constructs and also a subset of
$\PITL$ called \emph{Fusion Logic} ($\FL$) which includes constructs of the
sort found in Propositional Dynamic Logic ($\PDL$).  In addition, the liveness
tests found in conditional liveness formulas and invariants can be generalised
to be of the form $\Dm T$, where $T$ is an $\NLone$ formula, rather than just
a state formula.  We will consider each of these issues in turn.  For the sake
of brevity, the presentation is briefer and less formal than in the previous
sections.

\subsection{The Operator $\protect\Until$}

\label{the-operator-until-subsec}

\index{until operator@$\Until$ operator}
\index{PTL!the operator $\Until$}
The operator $\Until$ is a binary operator with the syntax $X\UntilOp Y$,
where $X$ and $Y$ are $\PTL$ formulas.  Recall from Sect.~\ref{pitl-sec} that
for any interval $\sigma$ and natural number $k$ which does not exceed
$\sigma$'s interval length, $\sigma_{k:\size{\sigma}}$ denotes the suffix
subinterval obtained by deleting the first $k$ states from $\sigma$.  Here is
the semantics of $\Until$:
\begin{multline*}
   \sigma\vld X\UntilOp Y \iff
   \ifWideMargins
     \\
   \fi
   \text{for some $k\le\intlen{\sigma}$, }
    \sigma_{k:\size{\sigma}} \vld
      Y \text{ and for all } j:0\le j\lt k, \sigma_{j:\size{\sigma}} \vld X
\enspace.
\end{multline*}
Observe that the operator $\Diamond$ can be expressed in terms of $\Until$
since $\Diamond X$ is semantically equivalent to the formula $\True \Until X$.

We can alter the definition of invariants by replacing $\Diamond$-dependencies
with dependencies of the form $r\equiv (w \UntilOp w')$, where $w$ and $w'$
are state formulas.  If the $j$-th dependency $I[j]$ of an invariant $I$ is
such a dependency (called an \emph{$\Until$-dependency}), then the
corresponding conjunction $T_I[j]$ in $I$'s transition formula $T_I$ has the
form $r\equiv (w' \Or (w \And \Next r))$.  The associated conjunction $L_I[j]$
in $L_I$ is $r\imp \Dm w'$.  It is not hard to modify the material in
Sect.~\ref{invariants-and-related-formulas-sec} to ensure that finite-time and
infinite-time invariant configurations remain semantically equivalent to the
associated transition configurations.

Alternatively, we can transform an invariant with $\Until$ in it to one
without it.  Each dependency in $I$ of the form $u_k\equiv (w\Until w')$ is
replaced by the dependency $u_k\equiv \bigl(u'_k \And (w' \Or (w \And \Next
u_k))\bigr)$, where $u'_k$ is a new dependent variable with the associated
dependency $u'_k\equiv \Diamond w'$.  This approach is more hierarchical than
the first one but increases the number of dependencies used.

\subsection{Past Time}

\label{past-time-subsec}


\index{past time}
\index{PTL!past time}
Let us now consider $\PTL$ with a bounded past.  The syntax is modified to
include the two additional primitive operators $\Prev X$ (read \emph{previous
  $X$}) and $\Dpast X$ (read \emph{once $X$}).  The set of $\PTL$ formulas
including past-time constructs is denoted as $\PTLP$.  The semantics of a
$\PTL$ formula $X$ is now expressed as $(\sigma,k)\vld X$ where $k$ is any
natural number not exceeding $\intlen{\sigma}$.  For example, the semantics of
$\Prev$ and $\Dpast$ are as follows:
\begin{center}
  \begin{tabular}{L}
    (\sigma,k)\vld \Prev X \iff k\gt 0 \text{ and } (\sigma,k-1)\vld X \\
    (\sigma,k)\vld \Dpast X
      \iff \text{for some } j: 0\le j\le k, \enspace (\sigma,j)\vld X
\enspace.
  \end{tabular}
\end{center}
We define the operator $\Bpast X$ (read \emph{so-far $X$}) as $\Not\Dpast\Not
X$ and the operator $\WeakPrev X$ (read \emph{weak previous $X$}) as
$\Not\Prev\Not X$.  The operator $\First$ is defined to be $\Not\Prev\True$
and tests for the first state of an interval.  A past-time version of
$\Until$ called $\Since$ can also be included but we omit the details.

A $\PTLP$ formula $X$ is defined to satisfiable iff $(\sigma,k)\vld X$ holds
for some pair $(\sigma,k)$ with $k\le\intlen{\sigma}$.  The formula $X$ is
valid iff $(\sigma,k)\vld X$ holds for every pair $(\sigma,k)$ with
$k\le\intlen{\sigma}$.  Note that these straightforward definitions of
satisfiability and validity correspond to the so-called \emph{floating
  framework} of $\PTL$ with past time.  However, Manna and Pnueli propose
another interesting approach called the \emph{anchored
  framework}~\cite{MannaPnueli89} (also discussed
in~\cite{LichtensteinPnueli00}) which they argue is superior.  In this
framework, satisfiability and validity only examine pairs of the form
$(\sigma,0)$. There exist ways to go between the two conventions but we will
not delve into this here and instead simply assume the more traditional
floating interpretation.

We now define an analogue of the set of formulas $\NL$:
\nobreak
\begin{mydefin}[Previous Logic]
  \label{prev-logic-def}
  \index{Previous Logic (PrevL)}
  The set of $\PTL$ formulas in which the only primitive temporal operator is
  $\Prev$ is called \emph{Previous Logic} ($\PrevL$).  The subset of $\PrevL$
  with no $\Prev$ nested in another $\Prev$ is denoted as $\PrevLone$.
\end{mydefin}
We let the variables $Z$ and $Z'$ denote formulas in $\PrevLone$.  Also,
$\PrevLone_V$ denotes the set of all formulas in $\PrevLone$ only having
variables in $V$.

The following definitions extend the notation of transition configurations
to deal with past time:
\begin{mydefin}[Past-Time Transition Configurations]
  \label{tz-configuration-def}
  \index{transition configuration!past-time}
  A \emph{past-time transition configuration} is any formula of the form
  $\Bpast\!\Box (T \And Z) \And X$, where $T$ is in $\NLone_V$, $Z$ is in
  $\PrevLone_V$, and the formula $X$ is in $\PTL_V$ and is in one of the two
  categories shown below:
  \begin{center}
    \begin{tabular}{cc}
      Type of configuration & Syntax of $X$ \\\noalign{\hrule \vskip 2pt}
      Finite-time & $w \And \Finite$ \\
      Infinite-time & $w \And \Box\SDiamond L$
     \end{tabular}
  \end{center}
  Here $w$ is a state formula in $\PROP_V$ and $L$ is a conditional liveness
  formula in $\PTL_V$.
\end{mydefin}
The formula $\Bpast\!\Box (T \And Z)$ contains both $\Bpast$ and $\Box$ to
ensure that both $T$ and $Z$ are true everywhere in the interval.

The analysis of a finite-time or infinite-time past-time transition
configurations can be easily reduced to reasoning in $\PTL$ without past time.
Let us demonstrate this by first examining how to test the satisfiability of a
finite-time past-time transition configuration $\Bpast\!\Box (T \And Z) \And w
\And \Finite$.  This involves finding an interval $\sigma$ and natural number
$k\le\intlen{\sigma}$, such that $(\sigma,k)\vld \Bpast\!\Box (T \And Z) \And
w \And \Finite$ holds.  Note that this past-time transition configuration is
satisfiable iff the following formula, which shifts reasoning back to an
interval's starting state, is satisfiable:
\begin{equation}
  \label{past-time-1-eq}
  \Dpast\bigl(\Box (T \And Z) \And \First \And \Diamond w \And \Finite\bigr)
\enspace.
\end{equation}
Here we can dispense with the operator $\Bpast$ since $\Bpast\!\Box$ and
$\Box$ have the same semantics at the starting state.

Now for any $\PTLP$ formula $X$, the formula $\Dpast X$ is satisfiable iff $X$
is satisfiable.  Hence, the formula~\eqref{past-time-1-eq} is satisfiable iff
its subformula $\Box (T \And Z) \And \First \And \Diamond w \And \Finite$ is
satisfiable.  Let us now define the $\NLoneV$ formula $T'$ by replacing each
$\Prev$ construct in $Z$ by its operand and by taking each state formula in
$Z$ which does not occur in $\Prev$ and enclosing it in $\Next$.  For example,
if $Z$ is the formula $p \Or \Prev (q\And r)$, then $T'$ is $(\Next p) \Or
(q\And r)$.  Furthermore, let $w'$ be the state formula in $\PROP_V$ obtained
from $Z$ by replacing each $\Prev$ construct by $\False$.  In our example,
$w'$ is $p \Or \False$.  It can be readily checked that the following formula
relating $Z$ and $T'$ is true at any interval's initial state: $\Box Z \EQUIV
\Bm T' \And w'$.  Therefore, the original finite-time past-time transition
configuration is satisfiable iff the following formula in $\PTL$ without past
time is satisfiable:
\begin{equation}
  \label{past-time-2-eq}
  \Box (T \And (\More \imp T')) \And w' \And \Diamond w \And \Finite
\enspace.
\end{equation}
This is still not a well-formed finite-time transition configuration due to
the presence of the formula $\Diamond w$.  However, $\Diamond w$ can be
reduced by introducing a new propositional variable $r$ as shown in the next
formula:
\begin{equation}
  \label{past-time-3-eq}
  \Box \bigl(T \And (\More \imp T') \And (r\equiv (w \Or\Next r))\bigr)
    \And w' \And r \And \Finite
\enspace.
\end{equation}
The reduction of the original past-time transition configuration $\Bpast\!\Box
(T \And Z) \And w \And \Finite$ to the finite-time transition
configuration~\eqref{past-time-3-eq} systematically relates all aspects of the
analysis of the past-time transition configuration to the purely future-only
reasoning presented earlier.  This includes bounded models, decision
procedures and axiomatic completeness.

An alternative way to reduce the $\PTL$ formula~\eqref{past-time-2-eq}
involves interval-based reasoning.  We first re-express the formula in $\PTL$
as the next semantically equivalent conjunction:
\begin{equation}
  \label{past-time-4-eq}
  \Bm(T \And (\More \imp T'))
    \And w' \And \Diamond w \And \SFin T
\enspace.
\end{equation}
This makes use of the valid $\PTL$ equivalence $(\Box X \And \Finite)
\equiv(\Bm X \And \SFin X)$, for any $\PTL$ formula $X$.  However, in our case
we can omit the subformula $\More \imp T'$ in the $\SFin$ construct since the
operator $\More$ ensures that it is trivially true in the associated empty
interval.  Let $T''$ denote the subformula $T \And (\More \imp T')$.
Theorem~\ref{expressing-transitive-relexive-closure-in-ptl-thm} ensures the
semantic equivalence of $\Bm T''$ and $(\Utest T'')\Chopstar$.  Now the
formula~\eqref{past-time-4-eq} can in turn be itself re-expressed as the
following chop-formula:
\begin{equation}
  \label{past-time-5-eq}
  ((T'')\Chopstar \And w' \And \Finite);\ 
  ((T'')\Chopstar \And w \And \SFin T)
\enspace.
\end{equation}
Let $w''$ denote a state formula obtained by replacing every $\Next$ construct
in $T$ by $\False$.  Consequently, $w''$ is true exactly in states for which
$T \And \Empty$ is true.  It follows that we can test for satisfiability of
formula~\eqref{past-time-5-eq} by adapting the symbolic methods mentioned in
Sect.~\ref{decision-procedure-sec} to solve for $V$-atoms $\alpha$, $\beta$
and $\gamma$ for which the following formulas are satisfiable:
\begin{displaymath}
  \alpha \And w
  \quad
  (\Utest T'')\Chopstar \And \alpha \And \SFin \beta
  \quad
  \beta \And w'
  \quad
  (\Utest T'')\Chopstar \And \beta \And \SFin \gamma
  \quad
  \gamma \And w''
\enspace.
\end{displaymath}
Further details are omitted here.

The treatment for a infinite-time past-time transition configuration is nearly
identical to that for a finite-time one since the assumption of a bounded past
still applies and avoids the need for a past-time conditional liveness
formula.  First of all, we replace the subformula $\Finite$ by $\Box\SDiamond
L$.
\begin{equation*}
  \Box (T \And T')
    \And w' \And \Diamond w \And \Box\SDiamond L
\enspace.
\end{equation*}
The use of infinite time ensures we can omit the instance of $\More$ found in
the finite-time formulas~\eqref{past-time-2-eq} and~\eqref{past-time-3-eq}
since $T$ and $\More\imp T$ are semantically equivalent on an infinite
interval.  The formula $\Diamond w$ is itself reduced by introducing a new
propositional variable $r$ and conjoining a new implication to $L$ to obtain
the well-formed infinite-time transition configuration below:
\begin{equation*}
  \Box (T \And T')
    \And w' \And r \And \Box\SDiamond (L \And (r\imp \Dm w))
\enspace.
\end{equation*}

So far we have only considered finite- and infinite-time transition
configurations.  Invariants (and hence also invariant configurations) can be
extended to support past-time reasoning by adding two new kinds of
dependencies.  The first has the form $u\equiv Z$ and the second has the form
$u\equiv \Dpast w$.  The use of $\Dpast$ does not involve $I$'s conditional
liveness formula $L_I$ due to the assumption of a bounded past.  The
definitions of invariant configurations remain the same and the reduction of
them to past-time transition configurations is straightforward since no
dependency contains both future- and past-time temporal constructs.
Furthermore, dependencies containing the temporal operator $\Since$ (a
conventional past-time analogue of the operator $\Until$) are not much harder
to handle than $\Dpast$-dependencies.  The reduction of an arbitrary $\PTLP$
formula to an invariant with past time is also straightforward.

\subsection{Generalised Conditional Liveness Formulas and Invariants}

\label{generalised-conditional-liveness-formulas-and-invariants-subsec}

\index{conditional liveness formula!generalised}
\index{invariant!generalised}
Conditional liveness formulas and invariants require that any operand of $\Dm$
and $\Diamond$, respectively, is a state formula.  We can slightly relax this
requirement and permit arbitrary formulas in $\NLone$.  This makes invariants
more succinct since a formula such as $\SFin w$ can now be expressed using
only one dependency such as $u_k \equiv \Diamond(\Empty \And w)$ instead of
requiring two.  The formula $\Box\SDiamond w$ can be expressed with the
invariant $u_k \equiv \Diamond(w \And\Next u_k)$.  The overall analysis of
such invariants only differs slightly from that for the basic version of
invariants.  Invariants with $\Until$-dependencies (see
\S\ref{the-operator-until-subsec}) can be analogously generalised to permit
$\Until$-dependencies of the form $u_k \equiv (T\UntilOp T')$, where both $T$
and $T'$ are in $\NLone$.

Transition configurations containing generalised liveness formulas might be of
use as a notation for representing deterministic and nondeterministic
$\omega$-automata in temporal logic.  However, we need to employ
\emph{Quantified $\PTL$} ($\QPTL$) to existentially quantify over the
variables which collectively encode such an automaton's internal state.
Further details of this are omitted here.

\subsection{Fusion Logic}

\label{fusion-logic-subsec}

\index{Fusion Logic (FL)|(} Regular expressions are a standard notation for
representing regular languages.  However, within $\PITL$, it is more
appropriate to use languages based on the fusion operator rather than
conventional concatenation.  This involves a variation of regular expressions
called here \emph{fusion expressions}.  We now define a $\PITL$-based
representation of them which is in fact a special subset of $\PITL$ formulas.
This subset will then provide the basis for a generalisation of $\PTL$ called
\emph{Fusion Logic} ($\FL$) which is also itself a subset of $\PITL$.  We
originally used Fusion Logic in~\cite{Moszkowski04a} as a kind of intermediate
logic when we reduced the problem of showing axiomatic completeness of
Propositional Interval Temporal Logic ($\PITL$) with finite time to showing
axiomatic completeness for $\PTL$.  \index{Propositional Dynamic Logic (PDL)}
Fusion Logic is closely related to Propositional Dynamic Logic
($\PDL$)~\cite{FischerLadner77,FischerLadner79,KozenTiuryn90,%
  Harel84,HarelKozen2000,HarelKozen2002}.  A major reason for discussing
Fusion Logic here is because it is not hard to extend our decision procedure
for $\PTL$ with finite time to also handle more expressive interval-oriented
$\FL$ formulas by simply reducing $\FL$ formulas to lower level $\PTL$
formulas of the kinds already discussed.  This demonstrates another link
between $\PTL$ and intervals and has practical applications.

\begin{mydefin}[Fusion Expression Formulas]
  \index{fusion expression (FE) formulas}
  The set of \emph{fusion expression formulas}, denoted $\FE$, consists of
  $\PITL$ formulas with the syntax given below, where $w$ is a state formula,
  $T$ is in $\NLone$ and $E$ and $F$ themselves denote $\FE$ formulas:
  \begin{displaymath}
    w\Test \qquad  E \Or F \qquad \Utest T \qquad  E;F \qquad  E\Chopstar
    \enspace.
  \end{displaymath}
  The syntax of $\FE$ formulas is like that of programs in Propositional
  Dynamic Logic without rich tests.  However $\FE$ has a semantics based on
  sequences of states rather than binary relations.

  For any set of variables $V$, let $\FEV$ denote the set of $\FE$ formulas
  containing only variables in $V$.
\end{mydefin}
Unlike letters in conventional regular expressions, any nonmodal formula can
be used in $w\Test$.  For example, $\False\Test$ is permitted even though it
is unsatisfiable.  Consider the following $\FE$ formula:
\begin{equation*}
  \bigl((\Utest \Next p);(q\Test)\bigr)
    \Or (\Utest \Not q)\Chopstar.
\end{equation*}
This is true on an interval if either the interval has exactly two states and
$p$ and $q$ are both true in the second state or it has some arbitrary number
of states, say $k$, with $q$ false in each of the first $k-1$ states.

\begin{myremark}[Expressing concatenation]
  \label{expressing-concatentation-rem}
  \index{chop!expressing concatenation with}
  \index{concatenation!expressing using chop}
  It is important to note that the conventional concatenation of two $\FE$
  formulas $E$ and $F$ can be achieved through the use of the $\FE$ formula
  $E;(\Utest\True);F$.  Here $\Utest\True$ is itself an $\FE$ formula which is
  an alternative way to express the $\PTL$ operator $\Skip$.
  \index{chomp}
  \index{chop!chomp as variant of}
  This temporal
  operation on $E$ and $F$ is sometimes called
  ``\emph{chomp}\!'', since it is a slight variation of chop.  Hence, in the
  context of temporal logic, $\FE$ formulas can largely subsume regular
  expressions although there are slightly different conventions for such
  things as empty words.  We omit the details.
\end{myremark}

We now present the sublogic of $\PITL$ called here Fusion Logic.  In essence,
Fusion Logic augments conventional $\PTL$ with the fusion expression formulas
already introduced.
\begin{mydefin}[Fusion Logic]
  \label{fusion-logic-def}
  \index{PITL!Fusion Logic (FL)}
  \index{Fusion Logic (FL)!syntax}
  Here is the syntax of $\FL$ where $p$ is any propositional variable, $E$ is
  any $\FE$ formula and $X$ and $Y$ are themselves formulas in $\FL$:
  \begin{displaymath}
    p \qquad
    \Not X \qquad
    X \Or Y \qquad
    \Next X \qquad
    \Diamond X \qquad
    \DD{E}X.
  \end{displaymath}
  We define the new construct $\DD{E}X$ (called ``\emph{$\FL$-chop}'') and its
  dual $\BB{E}X$ (called ``\emph{$\FL$-yields}'') using the primitive $\PITL$
  constructs chop and $\Not$:
  \begin{displaymath}
    \DD{E}X \defeqv E;\!X
    \qquad \BB{E}X \defeqv \Not\DD{E}\Not X.
  \end{displaymath}
\end{mydefin}
Within an $\FL$ formula, $\Next$, $\Diamond$ and $\FL$-chop are treated as
primitive constructs.  Unlike $\PITL$, $\FL$ limits the left sides of chop to
being $\FE$ formulas.

In~\cite{Moszkowski04a}, we described an earlier version of $\FL$ having
$\Skip$ as a primitive $\FE$ formula instead of $\Utest T$.  As we noted
earlier in Remark~\ref{expressing-concatentation-rem}, the $\PTL$ formula
$\Skip$ can be expressed in $\FE$ as $\Utest \True$.  The two versions of
$\FL$ can readily be shown to be equally expressive since $\Utest T$ can be
replaced with a semantically equivalent disjunction of formulas by using of
$\Test$, $\Skip$ and chop.  For example, the $\FE$ formula $\Utest (p\imp
\Next q)$ is semantically equivalent to the $\FE$ formula $((\Not
p)\Test;\Skip) \Or (\Skip;q\Test)$.  In practice, the version described
here is much more natural and succinct.

Henriksen and Thiagarajan~\cite{HenriksenThiagarajan97,HenriksenThiagarajan99}
investigate a formalism related to Wolper's $\ETL$~\cite{Wolper81,Wolper83}
and called \emph{Dynamic Linear Time Temporal Logic} which combines $\PTL$ and
$\PDL$ in a linear-time framework with infinite time. It is similar to our
Fusion Logic and uses multiple atomic programs instead of the $\FE$ operators
$\Test$ and $\Utest$.

\begin{myremark}
  The temporal operators $\Next$ and $\Diamond$ which are primitives in $\FL$
  can actually be expressed as instances of $\FL$-chop if finite time is
  assumed:
  \begin{displaymath}
    \valid \Next X \equiv \DD{\Utest \True}X
    \qquad
    \valid \Diamond X \equiv \DD{(\Utest \True)\Chopstar}X.
  \end{displaymath}
\end{myremark}

In spite of $\FL$ being a proper subset of $\PITL$, they have the same
expressiveness.  This is discussed in~\cite{Moszkowski04a}, where a
hierarchical reduction of $\FL$ formulas to $\PTL$ formulas is also given but
is limited to dealing with finite-time intervals.  This reduction provides the
basis of a decision procedure for $\FL$ with finite-time.  We plan to describe
in future work a hierarchical reduction to transition configurations (also
restricted to finite-time).  Such transition configurations can then be tested
with the decision procedure described in Sect.~\ref{decision-procedure-sec}.
Like the first reduction in~\cite{Moszkowski04a}, this reduction can also be
used for proving the completeness of an axiom system for $\FL$ with finite
time.


\index{Fusion Logic (FL)|)}

\section{Discussion}

\label{discussion-sec}

We conclude with a look at some issues connected with $\PTL$ and $\FL$.

As noted earlier, a number of $\PTL$ decision procedures are tableau-based
algorithms.  These include ones described by Wolper~\cite{Wolper85},
Emerson~\cite{Emerson90} and Lichtenstein and
Pnueli~\cite{LichtensteinPnueli00}.  It appears that with some care a
tableau-based approach can be hierarchically reduced to our framework.  We
hope to look into this in more detail in the future.

The BDD-based techniques described in Sect.~\ref{decision-procedure-sec} can
be adapted to check in real time that an executing system is not violating
assertions expressed in $\PTL$ or $\FL$ as it runs.  Whether $\FL$ in
particular is useful for this in practice is unclear.  In addition, it would
appear that the reachability analysis necessary for our approach to work can,
as with Bounded Model Checking (BMC)~\cite{ClarkeBiere01}, employ SAT-based
techniques for $\PTL$ and $\FL$ instead of BDDs.  However, such a SAT-based
approach, unlike the BDD-based one, normally cannot exhaustively test for
unsatisfiability because in BMC there is no notion corresponding to
convergence of BDDs to the set of all atoms reachable from some starting one.
Rather BMC works by employing SAT to find at most a single solution not
exceeding some predetermined maximum bounded length which for practical
reasons is generally much less than the worst-case bounds derived from formula
syntax.  If a solution is not found, this is typically not by itself
sufficient to exclude the existence of larger satisfying intervals.

We have used versions of invariants, transition formulas and conditional
liveness formulas to analyse Propositional Dynamic Logic ($\PDL$) without the
need for Fischer-Ladner closures.  Indeed, this was the original motivation
for conditional liveness formulas.  However, at present the benefits and
novelty of utilising our approach for $\PDL$ are less compelling than for
$\PTL$.

\ifgenstyle
  \singlespacing  
\fi

\section*{Acknowledgements}

We thank Antonio Cau, Jordan Dimitrov, Rodolfo G{\'o}mez and Helge Janicke for
comments on versions of this work.  In the course of discussions, Howard
Bowman, Shmuel Katz, Maciej Koutny and Simon Thompson also made helpful
suggestions leading to improvements in the presentation of the material.  We
are especially grateful to Hussein Zedan for his patience and encouragement
during the time this research was undertaken.


\begingroup

\ifgenstyle
  \small

  \let\stdbibitem=\bibitem

  \def\bibitem#1{\itemsep=0pt %
  \parsep=0pt %
  \topsep=0pt %
  \partopsep=0pt %
  \parskip=0pt %
  \stdbibitem{#1}}

\fi

\bibliography{manna-journal-submission-strings,biblio}

\endgroup

\ifdraft
  \expandafter\ifx\csname printindex\endcsname\relax
  \else
    \printindex
  \fi  
\fi

\begin{mycomment}
  \AtEndDocument{\relax
  \typeout{****************************************************}
  \typeout{*** Delete timestamp and index in final version. ***}
  \typeout{****************************************************}
  }
\end{mycomment}


\end{document}

\section*{***Extras***}


**********************************************************************

4 Jan '06

Here is a lemma similar to Lemma~\ref{chop-coverage-lem} for later use with
chop-omega to reduce infinite-time reasoning to finite-time reasoning:
\begin{mylemma}
  \label{chopomega-coverage-lem}
  For any $V$-atom $\alpha$ and $\PITLV$ formula $A$, the following are
  equivalent:
  \begin{itemize}
  \item The formula $(\alpha \And A)\Chopomega$ is satisfiable.
  \item The formula $\alpha \And A \And \Next\SFin \alpha$ is satisfiable.
  \end{itemize}
\end{mylemma}
Like Lemma~\ref{chop-coverage-lem}, Lemma~\ref{chopomega-coverage-lem} can be
proved using $V$-intervals.  In Lemma~\ref{chopomega-coverage-lem} we use the
subformula $\Next\SFin \alpha$ instead of $\SFin \alpha$ since $\alpha \And A
\And \SFin \alpha$ might only be satisfiable on empty intervals.  The fusion
of $\omega$ copies of such an empty interval together would itself be an empty
interval and clearly not satisfy the formula $(\alpha \And A)\Chopomega$ which
requires an $\omega$-interval.

----------------------------------------------------------------------

  \emph{Right side implies left side:} The following $\PITL$ formula is valid
  by simple temporal reasoning
  \begin{equation*}
    \Valid A\Chopomega \EQUIV (A \And \Finite \And \More);A\Chopomega
\enspace.
  \end{equation*}
  From this we can readily obtain the valid implication shown below:
  \begin{equation*}
    \Valid A\Chopomega
      \Implies (A \And \Finite);\True \;\And\; (\Finite \And \More);A\Chopomega
\enspace.
  \end{equation*}
  This can then be re-expressed as the next valid implication which is
  semantically equivalent.
  \begin{equation*}
    \Valid A\Chopomega \Implies \Df A \;\And\; \SDiamond A\Chopomega
\enspace.
  \end{equation*}
  Some additional simple temporal reasoning including induction over time then
  ensures the validity of the implication $A\Chopomega \implies \Df A \And
  \Box\SDiamond\Df A$.  The assumption that $A$ is a $\Df$-fixpoint then
  yields the desired validity of the implication $A\Chopomega \implies A \And
  \Box\SDiamond A$.

**********************************************************************

14 Dec '05

The next Lemma~\ref{df-and-finite-chop-finite-equiv-df-and-finite-lem} can be
understood as stating that an interval $\sigma$ is finite and satisfies some
$\Df$-fixpoint $A$ iff some prefix of $\sigma$ (possibly $\sigma$ itself) is
finite and satisfies $A$ and the corresponding suffix of $\sigma$ is also
finite.  The lemma provides a way to ``stretch'' the scope of $A$'s
satisfiability from a prefix of $\sigma$ to include all of $\sigma$.
\begin{mylemma}
  \label{df-and-finite-chop-finite-equiv-df-and-finite-lem}
  For any $\PITL$ formula $A$ which is a $\Df$-fixpoint, the next equivalence
  is valid:
  \begin{equation*}
    (A\And\Finite);\Finite \Equiv A \And \Finite
\enspace.
  \end{equation*}
\end{mylemma}
\begin{myproof}
  \emph{Left side implies right side:} If an interval $\sigma$ satisfies the
  formula $(A\And\Finite);\Finite$ then $\sigma$ is finite and also satisfies
  the formula $(A\And\Finite);\True$ which is the same as $\Df A$.  Now $A$ is
  assumed to be a $\Df$-fixpoint so therefore the equivalence $A\equiv \Df A$
  is valid and hence $\sigma$ satisfies $A$ as well.  Hence $\sigma$ satisfies
  both $\Finite$ and $A$ and therefore also the conjunction $A \And \Finite$.
  
  \emph{Right side implies left side:} If an interval $\sigma$ satisfies the
  formula $A \And \Finite$, then $\sigma$ also satisfies the formula $(A \And
  \Finite);\Empty$.  Now any subinterval of $\sigma$ satisfying $\Empty$ is
  finite and therefore also satisfies $\Finite$.  Consequently, $\sigma$
  satisfies the formula $(A\And\Finite);\Finite$.
\end{myproof}
We will shortly use
Lemma~\ref{df-and-finite-chop-finite-equiv-df-and-finite-lem} in
Lemma~\ref{a-and-box-sdiamond-a-imp-chop-lem} to replace an instance of the
lefthand formula $(A\And\Finite);\Finite$ by the simpler righthand formula $A
\And \Finite$ (see formulas~\eqref{a-and-box-sdiamond-a-imp-chop-4-eq}
and~\eqref{a-and-box-sdiamond-a-imp-chop-5-eq} below).

\begin{mylemma}
  \label{a-and-box-sdiamond-a-imp-chop-lem}
  For any $\PITL$ formula $A$ which is a $\Df$-fixpoint, the following
  implication concerning $A \And \Box\SDiamond A$ is valid:
  \begin{equation}
    \label{a-and-box-sdiamond-a-imp-chop-1-eq}
    \Valid A \And \Box\SDiamond A
      \Implies (A \And \More \And \Finite);(A \And \Box\SDiamond A)
\enspace.
  \end{equation}
\end{mylemma}
\begin{myproof}
  Let $\sigma$ be an interval satisfying the antecedent $A \And \Box\SDiamond
  A$. Simple temporal reasoning ensures that the implication below is valid:
  \begin{equation}
    \label{a-and-box-sdiamond-a-imp-chop-2-eq}
    \Valid \Box\SDiamond A
      \Implies \Box\Diamond(A \And \Box\SDiamond A)
\enspace.
  \end{equation}
  Therefore, $\sigma$ satisfies the formula $\Box\Diamond(A \And \Box\SDiamond
  A)$.
  
  Now $A$ is assumed to be a $\Df$-fixpoint.  Also $\More$ is a $\Df$-fixpoint
  by Lemma~\ref{fixpoints-of-the-operator-df-lem} since it is defined to be
  $\Next\True$.  Therefore Lemma~\ref{fixpoints-of-the-operator-df-lem}
  ensures that the conjunction $A \And \More$ is itself a $\Df$-fixpoint.
  Consequently, the interval $\sigma$ also satisfies $\Df(A\And \More)$ which
  by the definition of $\Df$ is the same as $(A\And \More \And
  \Finite);\True$. This together with the fact the $\sigma$ satisfies the
  formula $\Box\Diamond(A \And \Box\SDiamond A)$ and some simple
  interval-based temporal reasoning yield that the interval $\sigma$ also
  satisfies the next formula:
  \begin{equation}
    \label{a-and-box-sdiamond-a-imp-chop-3-eq}
    (A\And \More \And \Finite);\Diamond(A \And \Box\SDiamond A)
\enspace.
  \end{equation}
  We now re-express the subformula $\Diamond(A \And \Box\SDiamond A)$
  in~\eqref{a-and-box-sdiamond-a-imp-chop-3-eq} using chop
  and $\Finite$ as $\Finite;(A \And \Box\SDiamond A)$. Therefore $\sigma$
  satisfies the formula shown below:
  \begin{equation}
    \label{a-and-box-sdiamond-a-imp-chop-4-eq}
    (A\And \More \And \Finite);\Finite;(A \And \Box\SDiamond A)
\enspace.
  \end{equation}
  We now wish to eliminate the second instance of $\Finite$.  The fact that $A
  \And \More$ is a $\Df$-fixpoint ensures that an interval satisfies the
  formula $(A\And \More \And \Finite);\Finite$ iff the interval satisfies the
  subformula $A\And \More \And \Finite$.  More formally,
  Lemma~\ref{df-and-finite-chop-finite-equiv-df-and-finite-lem} and the fact
  that $A \And \More$ is a $\Df$-fixpoint together yield the next valid
  equivalence:
  \begin{equation*}
    \Valid (A\And \More \And \Finite);\Finite \Equiv A\And \More \And \Finite
\enspace.
  \end{equation*}
  We then use this to re-express
  formula~\eqref{a-and-box-sdiamond-a-imp-chop-4-eq} as follows:
  \begin{equation}
    \label{a-and-box-sdiamond-a-imp-chop-5-eq}
    (A\And \More \And \Finite);(A \And \Box\SDiamond A)
\enspace.
  \end{equation}
  Therefore the interval $\sigma$ also satisfies
  formula~\eqref{a-and-box-sdiamond-a-imp-chop-5-eq}.
  Combining all of the reasoning results in ensuring that the
  implication~\eqref{a-and-box-sdiamond-a-imp-chop-1-eq} is
  indeed valid.
\end{myproof}

----------------------------------------------------------------------

  \emph{Left side implies right side:} Simple temporal reasoning establishes
  that the implication below is valid for any $\PITL$ formulas $B$ and $C$:
  \begin{equation}
    \label{df-a-and-box-sdiamond-df-a-eqv-a-chopomega-valid-2-eq}
    \Valid \Box\bigl(B \,\implies\, (C \And \More \And \Finite);B\bigr)
      \Implies (B \implies C\Chopomega)
\enspace.
  \end{equation}
  Lemma~\ref{a-and-box-sdiamond-a-imp-chop-lem} ensures that the next
  implication concerning $A \And \Box\SDiamond A$ is valid:
  \begin{equation}
    \label{df-a-and-box-sdiamond-df-a-eqv-a-chopomega-valid-3-eq}
    \Valid A \And \Box\SDiamond A
      \Implies (A \And \More \And \Finite);(A \And \Box\SDiamond A)
\enspace.
  \end{equation}
  
  Now the validity of
  implication~\eqref{df-a-and-box-sdiamond-df-a-eqv-a-chopomega-valid-3-eq}
  readily yields the validity of the next formula:
  \begin{equation}
    \label{df-a-and-box-sdiamond-df-a-eqv-a-chopomega-valid-4-eq}
    \Valid \Box\bigl((A \And \Box\SDiamond A)
      \,\implies\, (A \And \More \And \Finite);(A \And \Box\SDiamond A)\bigr)
\enspace.
  \end{equation}

  We can now achieve our main goal of showing the validity of the implication
  $A \And \Box\SDiamond A \implies A\Chopomega$ since the validity of
  implications~\eqref{df-a-and-box-sdiamond-df-a-eqv-a-chopomega-valid-2-eq}
  and~\eqref{df-a-and-box-sdiamond-df-a-eqv-a-chopomega-valid-4-eq} and modus
  ponens establish this.

**********************************************************************

8 Mar '05

A tableau-based algorithm can also be implemented which is adapted from
previous ones for $\PTL$ such as those surveyed in Wolper~\cite{Wolper85},
Emerson~\cite{Emerson90} and Lichtenstein and
Pnueli~\cite{LichtensteinPnueli00}.  We omit the details.

**********************************************************************

2 Mar '05

Another benefit of generalised conditional liveness formulas and invariants is
that it becomes easier to devise finite-time invariant configurations which
exhibit deterministic behaviour with respect to independent variables.
Here is a definition of this:
\begin{mydefin}[Deterministic Finite-Time Invariant Configurations]
  We say that a finite-time invariant configuration $\Box I \And w \And
  \Finite$ is \emph{deterministic} if the following conditions hold:
  \begin{itemize}
  \item Given a set of values for $I$'s independent variables, the initial
    formula $w$ uniquely determines the values of all dependent variables.
  \item For any set of values of $I$'s independent in two adjacent states and
    dependent variables in the first of the states, the associated transition
    formula $T_I$ uniquely determines the values of the dependent variables in
    the second state.
  \end{itemize}
\end{mydefin}

Consider the following simple invariant $I_2$ which contains an
$\Until$\!-dependency:
\begin{equation}
  \label{sample-invariant-accepting-language-1-eq}
  I_2\colon\quad
   (r_1 \;\EQUIV\; (p\not\equiv\Next p) \UntilOp r_2)
   \;\And\; (r_2 \equiv \Empty)
\enspace.
\end{equation}
If we ignore (i.e., existentially hide) the dependent variables $r_1$ and
$r_2$, then the finite-time configuration $\Box I_2 \And r_1 \And
p$ can be viewed as accepting the language with the two-letter alphabet
$\{p,\Not p\}$ and containing exactly all finite words with an odd number of
letters which start and finish with $p$ and alternate between $p$ and $\Not p$
(e.g., $p$, $p\Not pp$ and $p\Not pp\Not pp$).  However, this configuration is
not deterministic since the initial formula $r_1$ does not determine the value
of the dependent variable $r_2$.

?????? Below is a deterministic invariant $I'_2$.  The finite-time
configuration $\Box I'_2 \And r_1 \And p$ accepts the same language as the
invariant~\eqref{sample-invariant-accepting-language-1-eq} if we ignore the
dependent variables:
\begin{equation}
  I'_2\colon\quad 
    (r_1 \;\EQUIV\; (p\not\equiv\Next p) \UntilOp \Empty)
\end{equation}
Note that in the second operand $\Empty$ of the $\Until$ construct is not a
state formula but is in $\NLone$ so this invariant is only permitted if
allow the generalised syntax.
The associated transition formula is as follows:
\begin{equation}
 T_{I'_2}\colon\quad 
    (r_1 \;\EQUIV\; \Empty \Or ((p\not\equiv\Next p) \And \Next r_1))
\end{equation}

**********************************************************************

21 Jun '04/1 Mar '05

\section{Languages and Deterministic Invariants}

\label{languages-and-deterministic-invariants-sec}

*** Can this be suitably revised? *** \\

\subsection{Formal Languages and Invariants}

\begin{mydefin}[Letters Associated with an Invariant]
  Let $I$ be an invariant having $n$ independent variables.  Each of the $2^n$
  possible conjunctions containing each independent variable or its negation
  is called a \emph{letter}.
\end{mydefin}
We use $\lambda$ and $\lambda'$ to denote individual letters.  The set
$\Sigma_I$ denotes the set of all such letters.  The set $\Sigma_I^+$ denotes
the set of all finite, nonempty words (i.e., finite sequences of one or more
letters) built from elements of $\Sigma_I$.  Similar, the set
$\Sigma_I^\omega$ denotes the set of all $\omega$-words built from elements of
$\Sigma_I$.

\begin{mydefin}[Words Accepted by an Invariant]
  For any word in $\Sigma_I^+ \union \Sigma_I^\omega$, we say that an
  invariant $I$ with $\size{I}\ge 1$ \emph{accepts} the word iff there exists
  an interval which satisfies the formula $\Box I \And u_1$ and the
  behaviour of $I$'s independent variables in the interval agree with
  the word.
\end{mydefin}

\begin{mydefin}[Formal Language Accepted by an Invariant]
  For any invariant $I$ with $\size{I}\ge 1$, the notation $\LangI$ denotes
  the set of all words in $\Sigma_I^+ \union \Sigma_I^\omega$ accepted by $I$.
\end{mydefin}
Alternatively, we can speak of an invariant $I$ \emph{defining} or
\emph{recognising} the associated formal language $\LangI$.

\begin{mydefin}[Finite and $\omega$-Languages Accepted by an Invariant]
  The notation $\LangPlusI$ refers to the set of all finite words accepted by
  $I$.  In other words, $\LangPlusI \defeq \Sigma_I^+\intersection\LangI$.
  The notation $\LangOmegaI$ similarly denotes the set of all $\omega$-words
  accepted by $I$, i.e., $\LangOmegaI \defeq
  \Sigma_I^\omega\intersection\LangI$.
\end{mydefin}

\subsection{Deterministic Invariants}

\begin{mydefin}[Deterministic Invariants]
  \label{deterministic-invariants-def}
  An invariant $I$ is \emph{deterministic} iff $\size{I}\ge 1$ and $I$'s state
  and transition formulas $S$ and $T$ have the properties shown below:
  \begin{enumerate}
  \item \label{deterministic-invariants-case-label} For each letter $\lambda$,
    there exists exactly one $V$-atom $\alpha$ for which the next formula is
    satisfiable:
    \begin{equation*}
      S \And u_1 \And \lambda \And \alpha
\enspace.
    \end{equation*}
  \item For each $V$-atom $\alpha$ and letter $\lambda$ there exists exactly
    one $V$-atom $\beta$ for which the following formula is satisfiable:
    \begin{equation*}
      T \And \alpha \And \Next(S \And \lambda \And \beta)
\enspace.
    \end{equation*}
  \end{enumerate}
\end{mydefin}

The following simple invariant accepts the language with the two-letter
alphabet $\{p,\Not p\}$ and containing exactly all finite words with an odd
number of letters which start and finish with $p$ and alternate between $p$
and $\Not p$ (e.g., $p$, $p\Not pp$ and $p\Not pp\Not pp$):
\begin{equation}
  \label{sample-invariant-accepting-language-eq}
  (r_1 \equiv (r_2 \And r_3)) \And (r_2\equiv \WeakNext \Not r_2)
    \And (r_2\equiv p) \And (r_3 \equiv \Diamond(\Empty \And r_2))
\enspace.
\end{equation}

Below is a deterministic invariant which accepts the same language as the
invariant~\eqref{sample-invariant-accepting-language-eq}:
\begin{align*}
I'\colon\quad &
  (r'_1 \equiv (r'_2 \And r'_4 \And r'_5))
    \And (r'_2\equiv \WeakNext \Not r'_2)
    \And (r'_3\equiv (r'_4 \And (r'_2\equiv p))) \\
&
    \qquad \null \And (r'_3\equiv \WeakNext r'_4)
    \And (r'_4\equiv r'_4)
    \And (r'_5 \equiv \Diamond(\Empty \And r'_2 \And r'_3))
\enspace.
\end{align*}
Here are state and transition formulas of this:
\begin{align*}
S'\colon\quad &
  (r'_1 \equiv (r'_2 \And r'_4 \And r'_5))
    \And (r'_2\equiv r'_2)
    \And (r'_3\equiv (r'_4 \And (r'_2\equiv p))) \\
&
    \qquad \null \And (r'_3\equiv r'_3)
    \And (r'_4\equiv r'_4)
    \And (r'_5 \equiv r'_5) \\
T'\colon\quad &
  (r'_1 \equiv (r'_2 \And r'_4 \And r'_5))
    \And (r'_2\equiv \WeakNext \Not r'_2)
    \And (r'_3\equiv (r'_4 \And (r'_2\equiv p))) \\
&
    \qquad \null \And (r'_3\equiv \WeakNext r'_4)
    \And (r'_4\equiv r'_4)
    \And \bigl(r'_5 \equiv ((\Empty \And r'_2 \And r'_3)\Or \Next r'_5)\bigr)
\enspace.
\end{align*}

\begin{myremark}
  Formulas of the form $\Box\SDiamond w$ arise when describing in temporal
  logic both nondeterministic and deterministic $\omega$-automata which accept
  $\omega$-regular languages.  The dependency shown below can be
  used to capture this kind of behaviour:
  \begin{displaymath}
    r \equiv \Diamond\Next (w \And r)
\enspace,
  \end{displaymath}
  where $r$ is a propositional variable not occurring in $w$.  Similarly, a
  formula $\Box\Diamond \xi$ can be analogously represented as $r \equiv
  \Diamond(\xi \And \WeakNext r)$.
\end{myremark}

\begin{mylemma}[Unique Functions for Deterministic Invariants]
  \label{unique-functions-for-deterministic-invariants-lem}
  Suppose $I$ is a deterministic invariant.  Then there exists a unique unary
  function $g: \Sigma_I\rightarrow\Atoms$ and unique binary function $h:
  \Atoms\times\Sigma_I\rightarrow\Atoms$ such the next two formulas are
  satisfiable for $I$'s state and transition formulas $S$ and $T$ and any
  letter $\lambda\in\Sigma_I$ and $V$-atom $\alpha\in\Atoms$:
  \begin{eqnarray*}
    &&\sat S \And u_1 \And \lambda \And g(\lambda) \\
    &&\sat T \And \alpha
      \And \Next\bigl(S \And \lambda \And h(\alpha,\lambda)\bigr)
\enspace.
  \end{eqnarray*}
\end{mylemma}

\begin{mylemma}[Deterministic Disjunction of Two Deterministic Invariants]
  Suppose $I$ and $I'$ are two deterministic invariants with disjoint sets of
  dependent variables $u_1, \ldots, u_{\size{I}}$ and $u'_1, \ldots,
  u'_{\size{I'}}$ Let $r$ be any variable not occurring in either $I$ or $I'$.
  Take the invariant $I''$ to be $(r=(u_1 \And u'_1)) \,\And\, I \,\And\, I'$.
  It is deterministic and has the following hold:
  \begin{equation*}
    \valid \Box I'' \Implies r \EQUIV (\Box I \And \Box I' \And u_1 \And u'_1)
\enspace.
  \end{equation*}
\end{mylemma}

\begin{mylemma}[Deterministic Complement of a Deterministic Invariant]
  Suppose $I$ is a deterministic invariant with dependent variables $u_1,
  \ldots, u_{\size{I}}$ Let $r$ be any variable not occurring in $I$.  Take
  the invariant $I'$ to be $(r=\Not u_1) \,\And\, I$.  It is deterministic and
  has the following hold:
  \begin{equation*}
    \valid \Box I' \Implies r \EQUIV (\Box I \And \Not u_1)
\enspace.
  \end{equation*}
\end{mylemma}
